# Nanomaterials and nanoparticles: Sources and toxicity

**Cristina Buzea*(1), Ivan. I. Pacheco Blandino**(2), and Kevin Robbie***(1)**

*(1) Department of Physics, Queen's University, Kingston, Ontario K7L 3N6, Canada*

*(2) Gastrointestinal Diseases Research Unit & Department of Physiology, Queen's University at Kingston General Hospital, 76 Stuart St., Kingston, ON K7L 2V7, Canada*

* E-mail: cristi@physics.queensu.ca
** E-mail: pacheci1@KGH.KARI.NET
*** E-mail: robbie@physics.queensu.ca

*author to whom correspondence should be addressed*

# Abstract

This review is presented as a common foundation for scientists interested in nanoparticles, their origin, activity, and biological toxicity. It is written with the goal of rationalizing and informing public health concerns related to this sometimes-strange new science of 'nano', while raising awareness of nanomaterials' toxicity among scientists and manufacturers handling them. We show that humans have always been exposed to tiny particles via dust storms, volcanic ash, and other natural processes, and that our bodily systems are well adapted to protect us from these potentially harmful intruders. The reticuloendothelial system in particular actively neutralizes and eliminates foreign matter in the body, including viruses and non-biological particles. Particles originating from human activities have existed for millennia, e.g. smoke from combustion and lint from garments, but the recent development of industry and combustion-based engine transportation has profoundly increased anthropogenic particulate pollution. Significantly, technological advancement has also changed the character of particulate pollution, increasing the proportion of nanometer-sized particles - "nanoparticles" and expanding the variety of chemical compositions. Recent epidemiological studies have shown a strong correlation between particulate air pollution levels, respiratory and cardiovascular diseases, various cancers, and mortality. Adverse effects of nanoparticles on human health depend on individual factors such as genetics and existing disease, as well as exposure, and nanoparticle chemistry, size, shape, agglomeration state, and electromagnetic properties. Animal and human studies show that inhaled nanoparticles are less efficiently removed than larger particles by the macrophage clearance mechanisms in the lung, causing lung damage, and that nanoparticles can translocate through the circulatory, lymphatic, and nervous systems to many tissues and organs, including the brain. The key to understanding the toxicity of nanoparticles is that their minute size, smaller than cells and cellular organelles, allows them to penetrate these basic biological structures, disrupting their normal function. Examples of toxic effects include tissue inflammation, and altered cellular redox balance toward oxidation, causing abnormal function or cell death. The manipulation of matter at the scale of atoms, "nanotechnology", is creating many new materials with characteristics not always easily predicted from current knowledge. Within the near-limitless diversity of these materials, some happen to be toxic to biological systems, others are relatively benign, while others confer health benefits. Some of these materials have desirable characteristics for industrial applications, as nanostructured materials often exhibit beneficial properties, from UV absorbance in sunscreen to oil-less lubrication of motors. A rational science-based approach is needed to minimize harm caused by these materials, while supporting continued study and appropriate industrial development. As current knowledge of the toxicology of 'bulk' materials may not suffice in reliably predicting toxic forms of nanoparticles, ongoing and expanded study of 'nanotoxicity' will be necessary. For nanotechnologies with clearly associated health risks, intelligent design of materials and devices is needed to derive the benefits of these new technologies while limiting adverse health impacts. Human exposure to toxic nanoparticles can be reduced through identifying creation-exposure pathways of toxins, a study that may some day soon unravel the mysteries of diseases such as Parkinson's and Alzheimer's. Reduction in fossil fuel combustion would have a large impact on global human exposure to nanoparticles, as would limiting deforestation and desertification. While 'nanotoxicity' is a relatively new concept to science, this review reveals the result of life's long history of evolution in the presence of nanoparticles, and how the human body in particular has adapted to defend itself against nanoparticulate intruders.

# Contents

4.1.1. Particle size dependent inhalation
4.1.2. Upper airway clearance – mucociliary escalator
4.1.3. Lower airways clearance – phagocytosis and passive uptake
4.1.4. Nanoparticle size dependent phagocytosis
4.1.5. Concentration dependent phagocytosis
4.1.6. Lung burden
4.1.7. Translocation and clearance of inhaled nanoparticles
4.1.8. Adverse health effects in the respiratory tract.
4.2. Cellular interaction with nanoparticles
4.2.1. Cellular uptake
4.2.2. Oxidative stress, inflammation, and genotoxicity
4.2.3. Adverse health effects and treatment
4.2.4. "Non-invasive" terminology to be questioned
4.3. Nervous system uptake of nanoparticles
4.3.1. Neuronal uptake via olfactory nerves
4.3.2. Neuronal uptake via blood-brain-barrier
4.3.3. Adverse health effects of neuronal nanoparticles uptake and treatment
4.4. Nanoparticles translocation to the lymphatic systems
4.5. Nanoparticles translocation to the circulatory system
4.5.1. Long-term translocation
4.5.2. Short-term translocation of metals
4.5.3. Short-term translocation of non-metals
4.5.4. Nanoparticles interaction with and uptake by blood cells
4.5.5. Adverse health effects of circulatory system uptake
4.6. Liver, spleen, kidneys uptake of nanoparticles
4.6.1. Organs nanoparticles uptake
4.6.2. Adverse health effects of liver and kidney uptake
4.7. Gastro-intestinal tract uptake and clearance of nanoparticles
4.7.1. Exposure sources
4.7.2. Size and charge dependent uptake
4.7.3. Translocation
4.7.4. Adverse health effects of gastro-intestinal tract uptake
4.8. Dermal uptake of nanoparticles
4.8.1. Penetration sites
4.8.2. Translocation
4.8.3. Adverse health effects of dermal uptake
4.9. Nanoparticles uptake via injection
4.10. Nanoparticles generation by implants
4.11. Positive effects of nanoparticles
4.11.1. Nanoparticles as antioxidants
4.11.2. Anti-microbial activity

5. Physico-chemical characteristics dependent toxicity
5.1. Dose-dependent toxicity
5.2. Size-dependent toxicity
5.3. Surface area-dependent toxicity
5.4. Concentration-dependent toxicity
5.5. Particle chemistry and crystalline structure dependent toxicity

## List of abbreviations

μm - micrometer
1D - One Dimensional
2D - Two Dimensional
3D - Three Dimensional
AFM - Atomic Force Microscop(e)(y)
AQI – Air Quality Index
CNTs - Carbon NanoTubes
DNA - DeoxyriboNucleic Acid
EDS – Energy Dispersive Spectrometry
EPA - Environmental Protection Agency
GLAD – Glancing Angle Deposition
HIV – Human Immunodeficiency Virus
MISR - Multi-angle Imaging SpectroRadiometer
MODIS - MODerate-resolution Imaging Spectroradiometer
MWCNTs - Multiple-Wall Carbon NanoTubes
NASA - National Aeronautics and Space Administration
nm – nanometer
PM – Particulate Matter
ROS - Reactive Oxygen Species
SEM - Scanning Electron Microscopy
SWCNs - Single-Wall Carbon Nanotubes
TEM – Transmission Electron Microscope

## Definitions

**Aerosol** – a material that, while not gaseous itself, remains suspended in air for prolonged periods. Typical examples include dust, and fine-droplet liquid paint or hairspray.

**Aggregate/aggregation** – a material that is composed of a large number of small components which have come together as clusters, usually with branching, porous shapes. Aggregation is the process whereby the many small components form clusters, and can be driven by gravity or other forces.

**Alzheimer's disease** - a progressive, irreversible, neurodegenerative disease characterized by loss of function and death of nerve cells in several regions of the brain, leading to loss of attention, memory, and language. Its cause is unknown.

**Antibody** - a protein produced by the immune system as a response to a foreign substance, or antigen.

**Antigen** – a foreign substance that triggers the production of antibodies by the immune system.

Apoptosis – or "programmed cell death" is the process of cellular suicide that can be initiated for several reasons: when extensive cellular damage occurs, when the cell is no longer needed within the organism, and in embryonic development, among others. Apoptosis is different from cell necrosis (a form of traumatic cell death due to physical or biological injuries) in its biochemical and morphological aspects. Aberrations in apoptosis contribute to various diseases, such as cancer.

**Atomic Force Microscopy** - a scanning-probe form of surface microscopy that can image and manipulate matter at the nanometer scale.

**Autoimmune diseases** – a group of disorders where overactive functioning of the immune system results in the immune system producing antibodies or autoreactive T cells (a type of white blood cells) against its own tissue.

**Cancer** – disease characterized by rapid and uncontrolled cell division.

**Chelator** – a chemical agent that binds reversibly to a metal ion, forming a metallic complex.

**Chronic disease** – disease lasting a long time, which is ongoing or recurring, not caused by an infection and not contagious.

**Clearance** – the removal of particles or substances out of an organism, usually via urine or stool.

**Crohn's disease** – a chronic inflammatory disease of unknown cause that may affect any part of the gastrointestinal tract, most commonly the small bowel, as well as other organs. Symptoms of the disease include diarrhea, abdominal pain, and excessive weight loss.

**Cytokine** - a small protein released by cells that has a specific effect on interactions between cells, on communications between cells, or on the behavior of cells.

**Cytoplasm** - includes both the fluid (cytosol) and the organelles contained within a cell.

**Degenerative disease** - disease characterized by progressive deterioration of function or structure of tissue.

**DNA** is a nucleic acid found within the nucleus of each cell, carrying genetic information on cell growth, division, and function. DNA consists of two long strands of nucleotides twisted into a double helix and held together by hydrogen bonds. The sequence of nucleotides determines hereditary characteristics. Each cell contains an identical, complete copy of the organism's DNA, with differing cell characteristics determined by differences in gene expression.

**Endemic disease** – disease constantly present in and limited to people living in a certain location.

**Endogenous** – substances originating within, or synthesized by an organism (e.g. hormones and neurotransmitters).

**Endoplasmic reticulum** - a membrane network that extends throughout the cytoplasm and is involved in the synthesis, processing, secretion, and transport of proteins throughout the cell.

**Endothelium** - the layer of cells that line the interior surface of all parts of the circulatory system, including the heart, and blood vessels. Specialized endothelial cells perform important filtering functions in the kidney and at the blood-brain barrier.

**Enzyme** – a protein that acts as a catalyst in a biochemical reaction.

**Epidemiology** - the branch of medical sciences that studies various factors influencing the incidence, distribution, and possible control of diseases in human population.

**Etiology** – set of causes or origin of a disease.

**Exogenous** - substances originating outside an organism.

**Fibroblast** – a connective-tissue cell that secretes collagen and other components of the extracellular matrix. It migrates and proliferates during wound healing and in tissue culture.

**Gene** – a sequence of nucleotides (DNA) that defines a protein. Genes are the fundamental unit of heritability, and their collection in an individual organism (its genome) represents a code or protocol for the growth and development of that individual. Genes are arranged along the length of chromosomes, of which each species has a fixed number.

**Genotype** – the genetic constitution of an organism.

**Granuloma** – tissue resulting from aggregation of inflammation-fighting cells unable to destroy foreign substances.

**Hydrophilic** - having an affinity for water, or causing water to adhere.

**Hydrophobic** - having no affinity for water, or repelling water.

**Inflammation** - a localized protective response, produced by tissue injury that servs to destroy or arrest both the agent and the affected tissue. Blood vessel permeability locally increases, and the area becomes heavily populated with white blood cells. Signs of inflammation are redness, swelling, pain, and sometimes loss of function.

**Ischemia** - decrease in the blood supply to an organ, tissue, limb, or other part of a body caused by the narrowing or blockage of the blood vessels. Ischemia may lead to a shortage of oxygen (hypoxia) within the tissue and may result in tissue damage or tissue death.

**Lavage** - washing out or clearance of a body cavity, organ, or system.

**Lung burden** - the product of exposure rate and residency time of particulate matter inhaled into the lungs.

**Lymph** – a fluid containing white blood cells, proteins, and fats; can also carry bacteria, viruses, and cancer cells around the body. Lymph is collected from the tissues and returned to the circulatory system.

**Lymphatic system** – the network of vessels, nodes, and organs (spleen, thymus) that produce, store, and carry lymph. The lymphatic system lacks a central pump, such as the heart in the circulatory system, and must rely on muscles through.

**Lymphoedema** - a condition in which lymph nodes become enlarged and prevent lymph fluid from passing through them.

**Macrophage** - a phagocytic tissue cell of the reticuloendothelial system that is derived from the blood monocyte. The monocyte migrates from the blood into tissues where it transforms into a macrophage. Macrophages are present in most tissues. Macrophages ingest and process degenerated cells and foreign invaders, such as viruses, bacteria, and particles. The long-lived macrophages are reservoirs of HIV.

**Mesothelioma** – a rare form of cancer occurring in the lining of the lungs and chest cavity.

**Mitochondrion** - an organelle responsible for most of the oxidative metabolism in the cell. Mitochondria generate energy (in the form of ATP, adenosine triphosphate) by breaking down glucose (a type of sugar).

**Monocyte** - the largest form of a white blood cell, with a kidney-shaped nucleus; its function is the ingestion of foreign invaders, such as bacteria, tissue debris, and particles. Monocytes belong to the group of phagocytes, and mature into various macrophages in tissue.

**Murine** – pertaining to the rodent family, i.e. rats and mice.

**Nanoparticulate matter** – a collection of particles with at least one dimension smaller than 1 micron yet larger than atoms and molecules.

**Neutrophil** - an immune cell that ingests and degrades foreign organisms. Neutrophils are the most abundant type of white blood cells, and are the first to reach the site of an infection to attack foreign antigens.

**Oxidative stress** - an imbalance in favour of pro-oxidant versus antioxidant chemicals, potentially leading to damage to biomolecules.

**Parkinson's disease** - a progressive disorder of the nervous system manifested by muscle tremors and rigidity, decreased mobility and slow voluntary movements.

**Particulate matter** – airborne particles of solids and/or liquids with sizes ranging from several nanometers to several hundred microns.

**Phagocyte** - cell that ingests and kills foreign intruders via the process called phagocytosis. Three examples are: monocytes, macrophages, and neutrophils.

**$PM_{0.1}$** – particulate matter having a diameter smaller than 0.1 microns (100 nm).

**$PM_{10}$** – particulate matter having a diameter smaller than 10 microns.

**$PM_{2.5}$** – particulate matter having a diameter smaller than 2.5 microns.

**Pneumoconiosis** – lung disease due to permanent deposition of substantial amounts of particles in the lungs and by the tissue reaction to its presence. Its severity varies from relatively harmless forms of sclerosis to destructive fibrosis and scarring of the lungs.

**Protein** – molecule containing a long chain of amino acids in the order specified by a gene's DNA sequence. Proteins can be, for example, enzymes, hormones, and antibodies.

**Quantum dot** - semiconductor crystals with a diameter of a few nanometers, having many properties resembling those of atoms.

**Receptor** - A protein or large molecule on the surface of a cell that binds selectively to specific substances (ligands).

**Reperfusion** - restoration of blood flow.

**Reticuloendothelial system** - a part of the immune system that consists of phagocytic cells, including macrophages and macrophage precursors, specialized endothelial cells lining the sinusoids of the liver, spleen, and bone marrow, and reticular cells of lymphatic tissue (macrophages) and bone marrow (fibroblasts).

**Rheumatoid arthritis** - chronic, autoimmune, inflammatory disorder affecting the connective tissue lining the joints. Symptoms include pain, swelling, stiffness, and deformities. It can extend to organs.

**Scleroderma** – a degenerative, autoimmune disease of the connective tissue, characterized by the formation of fibrous tissue (collagen) which surround the joints, blood vessels and sometimes internal organs.

**Systemic lupus erythematosus** - a chronic, autoimmune disorder. Symptoms include fatigue, butterfly-shaped facial rash, inflammation of the joints, tendons, connective tissues, and organs: heart, lungs, blood vessels, brain, kidneys, and skin.

**Toxicology** - the branch of medical and biological science studying the nature, adverse effects, detection, and treatment of poisons on living organisms. A fundamental principle of toxicology is that any substance is poisonous if given in a large amount. From the study of cancer-causing substances, carcinogens, it appears that there are some materials for which there is no safe dose, no level of exposure below which they do not cause cancer.

**Transcription factor** - a protein that binds to enhancer elements in DNA to regulate the level of transcription and expression of certain genes.

**Translocation** – the process of transit of particles or substances within an organism.

**Ulcerative colitis –** a chronic disease of unknown cause characterized by inflammation of the colon producing ulcerations. Symptoms are: abdominal pain, cramps, loose discharges of pus, blood, and mucus from the bowel, and weight loss.

**Ultra fine particles** (UFP) - nanoparticles with size smaller than 100 nm.

## 1. Introduction

Every person has been exposed to nanometer sized foreign particles; we inhale them with every breath, and consume them with every drink. In truth, every organism on Earth continuously encounters nanometer-sized entities. The vast majority causes little ill effect, and go unnoticed, but occasionally an intruder will cause appreciable harm to the organism. The most advanced of the toxic intruders are viruses, composed as they are of nucleic acid-based structures that allows them to not only interfere with biological systems, but also to parasitically exploit cellular processes to replicate themselves. Among the more benign viruses are the ones causing the familiar human symptoms of the common cold or flu, which are the evident manifestations of biochemical battles occurring between these foreign intruders and our immune systems, whose nanometer sized constituents (chemicals, and proteins) usually destroy and remove the viral invaders. A growing number of recent studies show, however, that nano- and micro-organisms may play a role in many chronic diseases where infections pathogens have not been suspected, diseases that were previously attributed only to genetic factors and lifestyle. Among these diseases are: leukemia (caused by viruses from the Retrovirus and Herpes virus families) [1], cervical cancer (Papilloma virus) [2], liver cancer (Hepatitis virus) [3], gastric ulcer (Helicobacter pylori) [4], nasopharyngeal cancer (Epstein-Barr virus) [5], kidney stones (nanobacteria) [6], severe acquired respiratory syndrome SARS (Corona virus) [7], heart disease (Chlamydia pneumonia) [8], juvenile diabetes (Coxsackie virus) [9], Alzheimer's disease (Chlamydia pneumoniae) [10], pediatric obsessive-compulsive disorder (Streptococcal bacteria) [11], psychotic disorders (Borna virus) [12], and prion diseases such as mad cow disease (proteins - prions) [13].

One is tempted to think that nanoparticles (such as dust, or ash particles), while similar in size to viruses, would be more benign, as these materials lack the viruses' ability to replicate. Nevertheless, while non-replicating bodily intruders do not directly take control of cellular processes, some have been shown to sufficiently interfere with cellular function to influence basic process of cells, such as proliferation, metabolism, and death. Many diseases can be associated with dysfunction of these basic processes, the most notable being cancer (uncontrolled cells proliferation), and neurodegenerative diseases (premature cell death). In addition, several diseases with unknown cause, including autoimmune diseases, Crohn's, Alzheimer's, and Parkinson's diseases, appear to be correlated with nanoparticles exposure. Conversely, the toxic properties of some nanoparticles may be beneficial, as they are thereby able to fight disease at a cellular level, and could be used as a medical treatment, by for example targeting and destroying cancerous cells.

Very small particles, so-called nanoparticles, have the ability to enter, translocate within, and damage living organisms. This ability results primarily from their small size, which allows them to penetrate physiological barriers, and travel within the circulatory systems of a host. While natural processes have produced nanoparticles for eons, modern science has recently learned how to synthesize a bewildering array of artificial materials with structure that is engineered at the atomic scale. The smallest particles contain tens or hundreds of atoms, with dimensions at the scale of nanometers - hence nanoparticles. They are comparable in size to viruses, where the smallest have dimensions of tens of nanometers (for example, a human immunodeficiency virus, or HIV, particle is 100 nm in diameter), and which in the emerging science of nanotechnology might be called 'nano-organisms'. Like viruses, some nanoparticles can penetrate lung or dermal (skin) barriers and enter the circulatory and lymphatic systems of humans and animals, reaching most bodily tissues and organs, and potentially disrupting cellular processes and causing disease. The toxicity of each of these materials depends greatly, however, upon the particular arrangement of its many atoms. Considering all the possible variations in shape and chemistry of even the smallest nanoparticles, with only tens of atoms, yields a huge number of distinct materials with potentially very different physical, and toxicological properties. Asbestos is a good example of a toxic nanomaterial, causing lung cancer and other diseases. Asbestos exists in several forms, with slight variations in shape and chemistry yet significantly varying toxicity.

Nanometer sized particles are created in countless physical processes from erosion to combustion, with health risks ranging from lethal to benign. Industrial nanoparticle materials today constitute a tiny but significant pollution source that is, so far, literally buried beneath much larger natural sources and nanoparticle pollution incidental to other human activities, particularly automobile exhaust soot.

The misapprehension of nanotoxicity may create a general fear that all nanomaterials are toxic. The online [14] and printed media [15] are inadvertently making no distinction between nanostructured fixed structures, which are not likely to cause harm (such as computer processors), and detachable or free nanoparticles, which are likely to cause adverse health effects. While uncontained nanoparticles clearly represent a serious health threat, fixed nanostructured materials, such as thin film coatings, microchip electronics, and many other existing nanoengineered materials, are known to be virtually benign. Many synthetic nanoparticulate materials produce positive health effects, for example functionalized fullerene chemicals that act as antioxidants. The use of nanoparticles in medical diagnostics and treatment is driven by their safety, as well as utility.

In the following pages we outline existing sources of nanoparticles, both natural and man-made, and the known effects of exposure to nanoparticles. In Chapter 1 we introduce basic concepts and terminology relevant to nanoscience and nanotechnology, define concepts and terms, give examples of nanoscale systems, and introduce the basics of nanoparticles toxicity. In Chapter 2 we briefly discuss nanoparticles classifications. Chapter 3 reviews natural and anthropogenic nanoparticle sources together with their associated health effects and treatment. In Chapter 4 we present current opinions and research results related to the health implications and toxicology of nanoparticles, and we define exposure pathways, and migration or translocation mechanisms within biological systems, adverse health effects, and treatment. The mechanics and biochemistry of toxicity are discussed in Chapter 5, as well as toxicity-related risk factors, such as particle size, shape, chemistry, and surface properties. In Chapter 6 we provide an overview of current and developing applications of nanomaterials. Finally, Chapter 7 contains conclusions and reflections.

### 1.1. Nano* etymology

The prefix "nano," derived from the Greek "nanos" signifying "dwarf", is becoming increasingly common in scientific literature. "Nano" is now a popular label for much of modern science, and many "nano-" words have recently appeared in dictionaries, including: nanometer, nanoscale, nanoscience, nanotechnology, nanostructure, nanotube, nanowire, and nanorobot. Many words that are not yet widely recognized are used in respected publications, such as Science and Nature. These include nanoelectronics, nanocrystal, nanovalve, nanoantenna, nanocavity, nanoscaffolds, nanofibers, nanomagnet, nanoporous, nanoarrays, nanolithography, nanopatterning, nanoencapsulation, etc. Although the idea of nanotechnology: producing nanoscale objects and carrying out nanoscale manipulations, has been around for quite some time, the birth of the concept is usually linked to a speech by Richard Feynman at the December 1959 meeting of the American Physical Society [16] where he asked, "What would happen if we could arrange the atoms one by one the way we want them?"

The **nanometer** is a metric unit of length, and denotes one billionth of a meter or $10^{-9}$ m. Popularly, 'nano' is also used as an adjective to describe objects, systems, or phenomena with characteristics arising from nanometer-scale structure. While 'micro' has come to mean anything small, 'nano' emphasizes the atomic granularity that produces the unique phenomena observed in nanoscience. While there are some exceptional examples, most of the exciting properties of 'nano' begin to be apparent in systems smaller than 1000 nm, or 1 micrometer, 1 μm. For the purpose of this review we will describe particles with any dimension smaller than 1 micrometer as 'nanoparticles', and those somewhat larger as 'microparticles'.

Nanostructured materials did not first come into existence with the recent emergence of the field of nanotechnology. Many existing materials are structured on the micro- and nanometer scales, and many industrial processes that have been used for decades (e.g. polymer and steel manufacturing) exploit nanoscale phenomena. The most advanced nanotechnological fabrication process is microelectronic fabrication, where thin film coatings and lithography are used to create micro- and nano-sized features on computer chips. The natural world is replete with examples of systems with nanoscale structures, such as milk (a nanoscale colloid), proteins, cells, bacteria, viruses etc. Moreover, many materials that seems smooth to the naked eye have an intricate structure on the scale of nanometers (Figure 1). Thus in many ways nanomaterials are not new. Recent advances in synthesis and characterization tools, however, have fueled a boom in the study and industrial use of nano-structured materials. A new vocabulary has emerged from this research, and its important terms and concepts are defined below.

**Nanomaterials** are materials that have structural components smaller than 1 micrometer in at least one dimension. While the atomic and molecular building blocks (~0.2 nm) of matter are considered nanomaterials, examples such as bulk crystals with lattice spacing of nanometers but macroscopic dimensions overall, are commonly excluded.

**Nanoparticles** are particles with at least one dimension smaller than 1 micron and potentially as small as atomic and molecular length scales (~0.2 nm). Nanoparticles can have amorphous or crystalline form and their surfaces can act as carriers for liquid droplets or gases. To some degree, nanoparticulate matter should be considered a distinct state of matter, in addition to the solid, liquid, gaseous, and plasma states, due to its distinct properties (large surface area and quantum size effects). Examples of materials in crystalline nanoparticle form are fullerenes and carbon nanotubes, while traditional crystalline solid forms are graphite and diamond. Many authors limit the size of nanomaterials to 50 nm [17] or 100 nm [18], the choice of this upper limit being justified by the fact that some physical properties of nanoparticles approach those of bulk when their size reaches these values. However, this size threshold varies with material type and cannot be the basis for such a classification. A legitimate definition extends this upper size limit to 1 micron, the sub-micron range being classified as nano.

**Nanoparticulate matter** – refers to a collection of nanoparticles, emphasizing their collective behavior.

**Nanotechnology** can be defined as the design, synthesis, and application of materials and devices whose size and shape have been engineered at the nanoscale. It exploits unique chemical, physical, electrical, and mechanical properties that emerge when matter is structured at the nanoscale.

**Nanotoxicology** was proposed as a new branch of toxicology to address the adverse health effects caused by nanoparticles [19]. Despite suggestions that nanotoxicology should only address the toxic effects of engineered nanoparticles and structures [20] we recommend that nanotoxicology should also encompass the toxic effects of atmospheric particles, as well as the fundamentals of virology and bacteriology. While significant differences exist between the health effects of non-biological particles and viruses and bacteria, there are significant common aspects of intrusion and translocation.

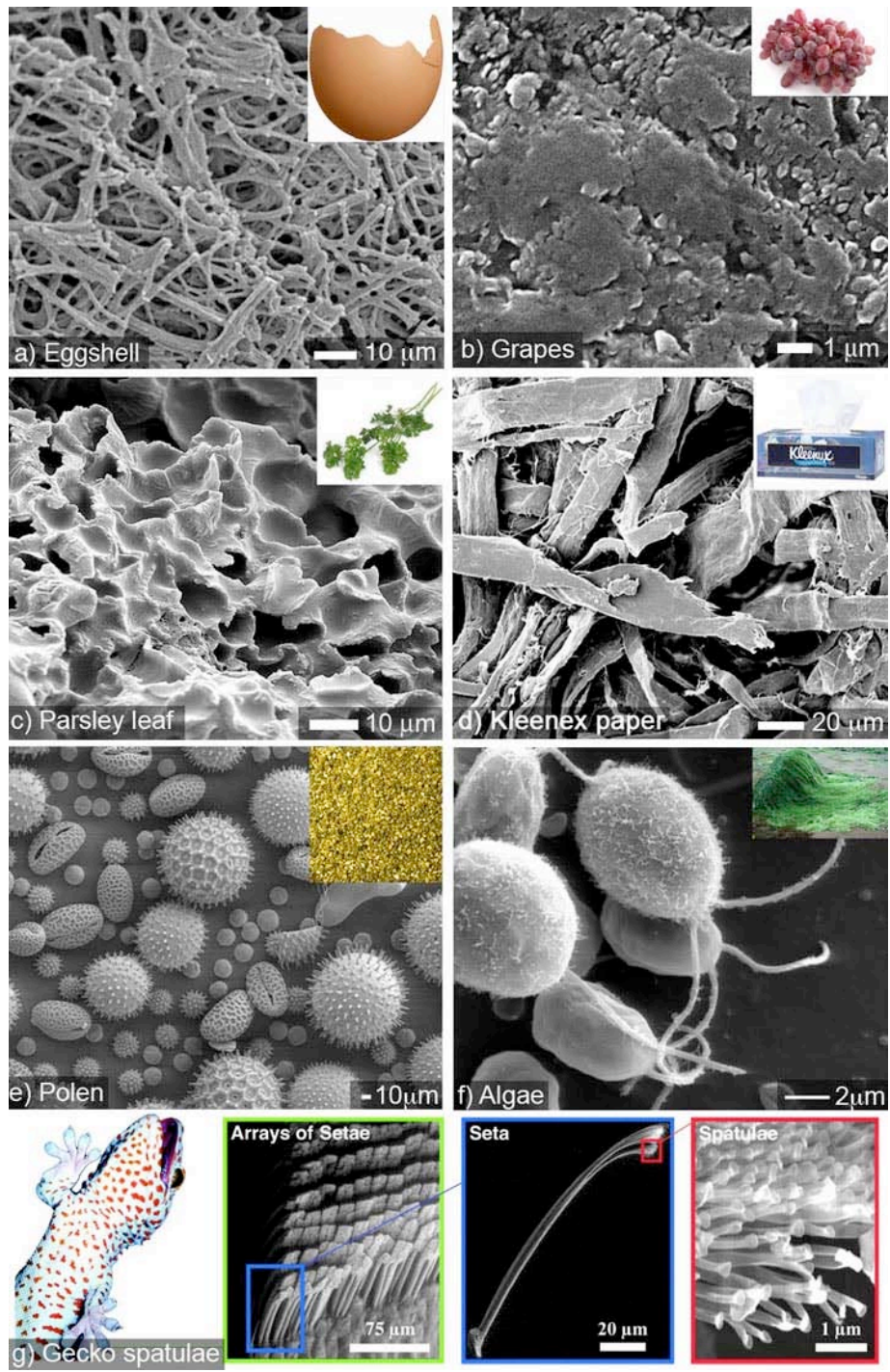

*Figure 1. SEM images showing the complexity of the world at the micro and nanoscale: (a) the inner surface of a bird's eggshell, credit: Janice Carr, Sandra L. Westmoreland, courtesy Public Health Image Library [21]; (b) the rough surface of table grape, credit: Janice Carr, courtesy Public Health Image Library [21]; (c) the textured surface of a parsley leaf, credit Janice Carr, courtesy Public Health Image Library [21]; (d) Kleenex paper, courtesy of Jim Ekstrom [22]; (e) pollen from a variety of common plants, credit Louisa Howard, Charles Daghlian, courtesy Public Health Image Library [21];*

*(f) green algae, credit Elizabeth Smith, Louisa Howard, Erin Dymek, Public Health Image Library [21]; (g) Gecko nano-adhesive system, with increasing magnification from left to right: gecko climbing vertical glass, adhesive surface microstructure, individual setae, nanostructure of spatular endings, courtesy of PNAS [23].*

The new terminology of 'nano' has united previously seemingly disparate fields, and a lexicon is needed to find and appreciate the great wealth of existing nano research, not conveniently labeled with the nano keyword.

**Health sciences epidemiology terminology.** In existing medical and toxicological terminology, nanoparticles having a diameter smaller than 100 nm are often called ultrafine particles (UFP) or ultrafine particulate matter. Ultrafine particles are labeled as a function of their size. For example, particulate matter with constituents having diameters smaller than 10 microns is abbreviated $PM_{10}$. Particulate matter having a size smaller than 100 nm is labeled as $PM_{0.1}$.

**Environmental sciences terminology.** Ambient particulate matter is categorized in three size distributions: ultrafine particles less than 0.1 μm in diameter (mainly resulting from combustion), accumulation mode particles between 0.1 and 2.5 μm in diameter (resulting from aggregation of ultrafine particles and vapors), and coarse-mode particles larger than 2.5 μm (mostly mechanically generated) [24].

**Proposed terminology.** It is important, and timely, to unify the terminology used for describing particle size in nanotechnology, health, and environmental sciences.

The materials under discussion can be classified as particles, regardless of their source. The size of these particles varies between 1 nm to several microns, and they can therefore be classified as either nanoparticles NP (any dimension smaller than 1 micron) or microparticles MP (all dimensions larger than one micron). To further specify particle size, we propose a modification of the health sciences epidemiology terminology, labeling particles by their largest dimension; for example 10 nm in diameter are labeled "$NP_{10}$", while 10 μm microparticles are labeled "$MP_{10}$".

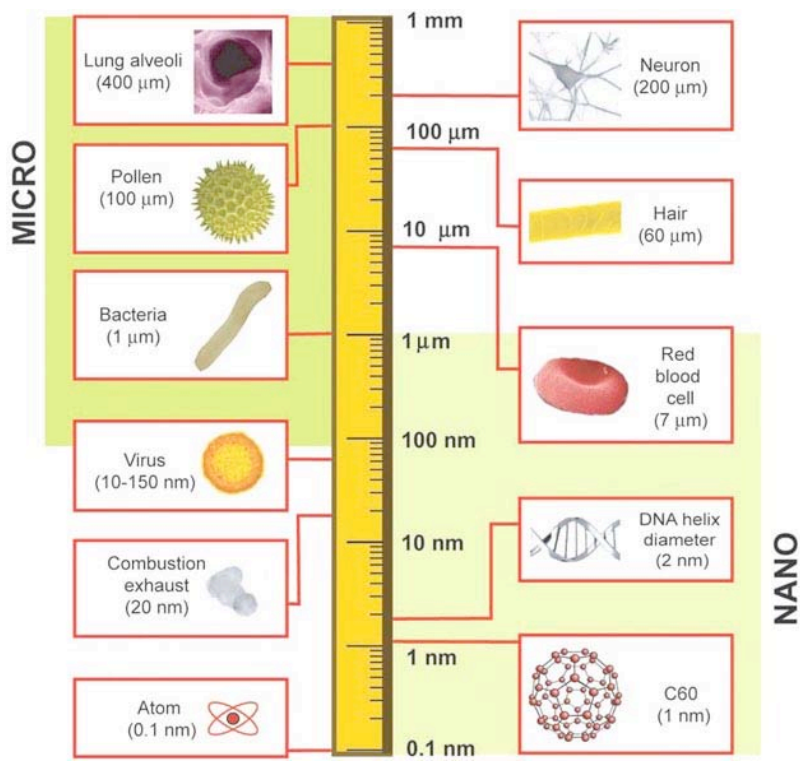

**Figure 2.** *Logarithmical length scale showing size of nanomaterials compared to biological components and definition of 'nano' and 'micro' sizes.*

Given that microparticles and nanoparticles vary in their conception by only their size, it can be difficult to fully appreciate the differences between them. To illuminate the effect of the size difference, the sizes of several natural micro- and nanostructures are shown in Figure 2, as measured from scanning and transmission microscope images [25, 26]. Generally, the sizes of nanomaterials are comparable to those of viruses, DNA, and proteins, while microparticles are comparable to cells, organelles, and larger physiological structures (Figure 2). A red blood cell is approximately 7 μm wide, a hair 60 μm, while lung alveoli are approximately 400 μm.

### 1.2. Main differences between nanomaterials and bulk materials

Two primary factors cause nanomaterials to behave significantly differently than bulk materials: surface effects (causing smooth properties scaling due to the fraction of atoms at the surface) and quantum effects (showing discontinuous behavior due to quantum confinement effects in materials with delocalized electrons) [27]. These factors affect the chemical reactivity of materials, as well as their mechanical, optical, electric, and magnetic properties.

The fraction of the atoms at the surface in nanoparticles is increased compared to microparticles or bulk. Compared to microparticles, nanoparticles have a very large surface area and high particle number per unit mass. For illustration, one carbon microparticle with a diameter of 60 μm has a mass of 0.3 μg and a surface area of 0.01 $mm^2$. The same mass of carbon in nanoparticulate form, with each particle having a diameter of 60 nm, has a surface area of 11.3 $mm^2$ and consists of 1 billion nanoparticles (Figure 3). The ratio of surface area to volume (or mass) for a particle with a diameter of 60 nm is 1000 times larger than a particle with a diameter of 60 μm (Figure 3 b). As the material in nanoparticulate form presents a much larger surface area for chemical reactions, reactivity is enhanced roughly 1000-fold. While chemical reactivity generally increases with decreasing particle size, surface coatings and other modifications can have complicating effects, even reducing reactivity with decreasing particle size in some instances.

The atoms situated at the surface have less neighbors than bulk atoms, resulting in lower binding energy per atom with decreasing particle size [27]. A consequence of reduced binding energy per atom is a melting point reduction with particle radius, following the Gibbs-Thomson equation [27]. For example, the melting temperature of 3 nm gold nanoparticles is more than 300 degrees lower than the melting temperature of bulk gold, as shown in Figure 3 c [27].

An example of a class of materials that clearly exploits quantum effects is quantum dots - synthesized nanostructures with sizes as small as a few nanometers (Figure 4). The electronic behavior of quantum dots is similar to that of individual atoms or small molecules, and quantum dots are regarded as akin to artificial atoms [28]. Notably, the confinement of the electrons in quantum dots in all three spatial directions results in a quantized energy spectrum. Another result of quantum confinement effect is the appearance of magnetic moments in nanoparticles of materials that are non-magnetic in bulk, such as gold, platinum, or palladium [27]. Magnetic moments result from several unpaired electron spins in nanoparticles formed of several hundreds atoms. Quantum confinement also results in quantified changes in the ability to accept or donate electrical charge (or electron affinity), also reflected in the catalytic ability [27]. For example, the reactivity of cationic platinum clusters in the decomposition of $N_2O$ is dictated by the number of atoms in the cluster, namely 6-9, 11, 12, 15, 20 atom-containing clusters are very reactive, while clusters with 10, 13, 14, 19 atoms have low reactivity [27].

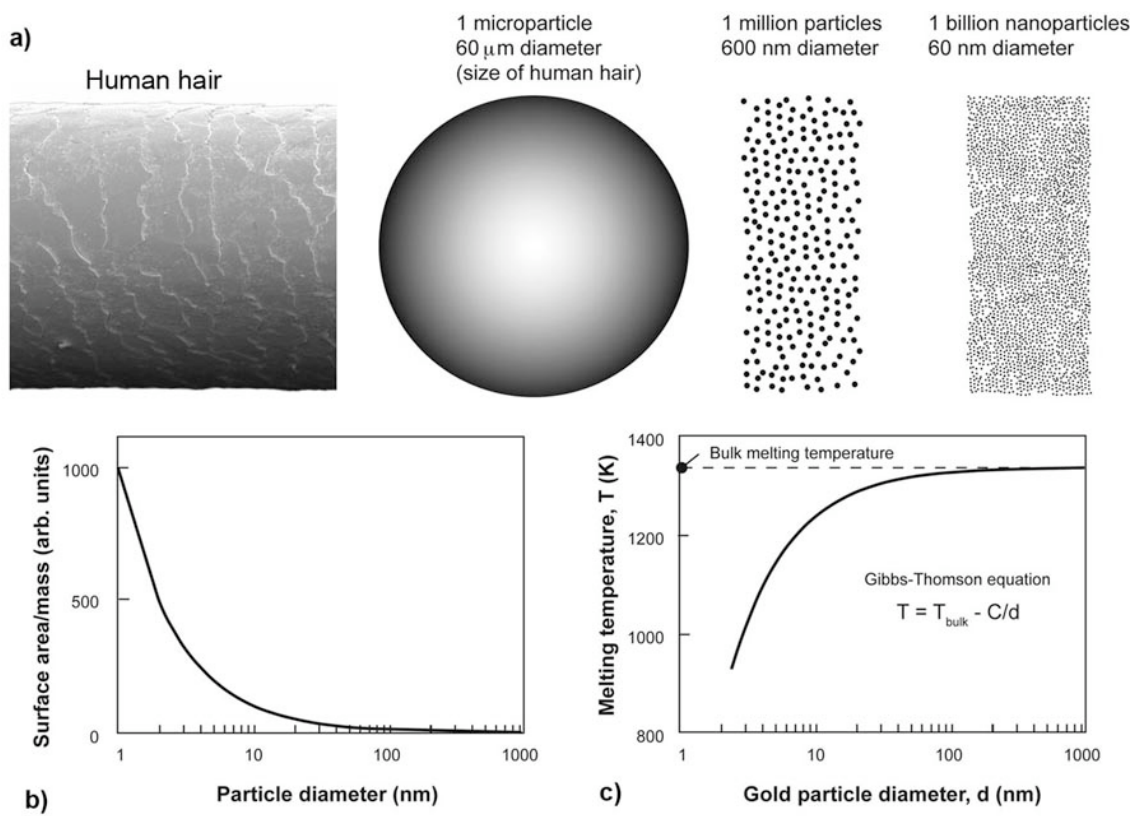

**Figure 3. (a)** *Schematics illustrating a microparticle of 60 μm diameter, about the size of a human hair - shown in the left at scale (courtesy Chelsea Elliott), and the number of nanoparticles with diameter of 600 nm and 60 nm having the same mass as one microparticle of 60 μm diameter. (b) Surface area normalized to mass versus particle diameter. (c) Gold melting temperature as a function of particle diameter, according to Gibbs-Thomson equation, shown inset; the gold bulk melting temperature is 1336 K [27].*

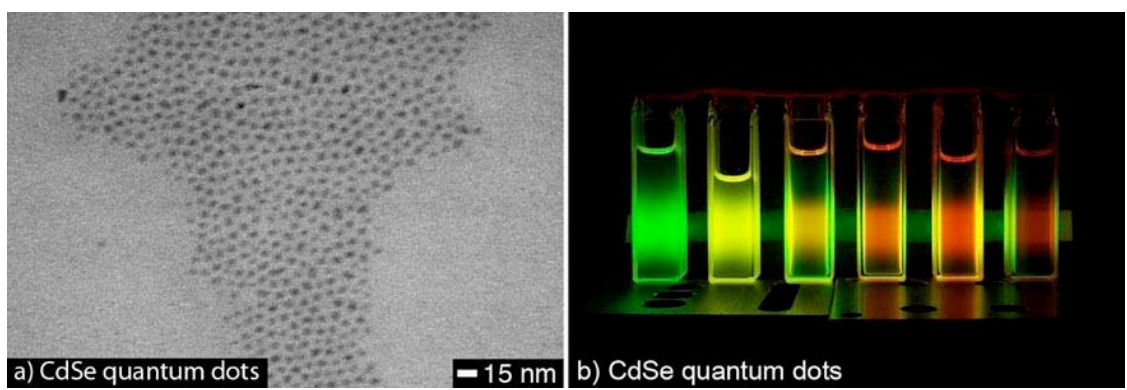

**Figure. 4. (a)** *TEM image of CdSe semiconductor nanoparticles, and **(b)** photograph of CdSe nanoparticles in solution, photo-luminescent under UV illumination. Images courtesy of Graham Rodway and Harry Ruda, University of Toronto.*

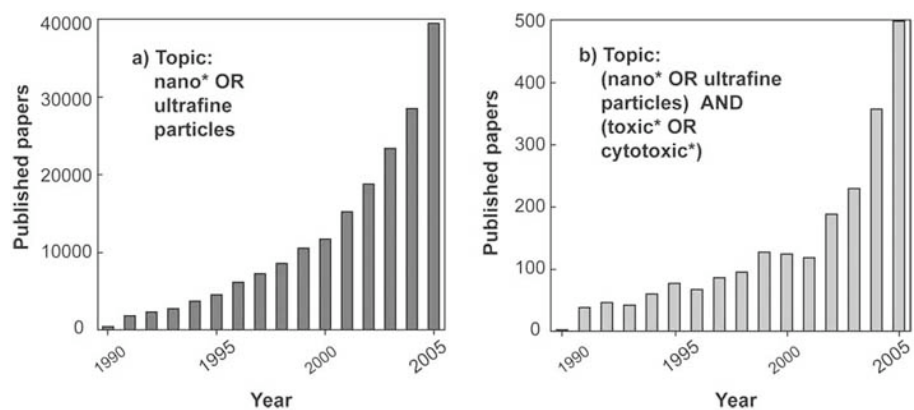

***Figure 5.*** *Statistics on scientific articles published on **(a)** nanomaterials and **(b)** their toxicity (ISI Web of Science).*

### 1.3. Nanomaterials and nanotoxicology publications statistics

The number of publications on the topic of nanomaterials has increased at an almost exponential rate since the early 1990s, reaching about 40,000 in the year 2005 (Figure 5), as indicated by a search on ISI Web of Knowledge database [29]. There is also a notable rise in the number of publications discussing their toxicity, particularly in the past two years. The total number of papers on toxicity, however, remains low compared to the total number of publications on nanomaterials, with only around 500 publications in the year 2005.

The large number of publications on nanomaterials can be explained by the fact that nanoscience and nanotechnology encompass a wide range of fields, including chemistry, physics, materials engineering, biology, medicine, and electronics. There are several reviews addressing nanotoxicology aspects, however they are intended for a narrow, specialized audience. Several are comparatively general [18], [20], [30-32], while others address selected aspects of nanoparticles toxicology, such as: health effects of air pollution [33]; epidemiological reviews of exposure to particles [34]; epidemiological studies of cardiovascular effects of airborne particles [35]; occupational aspects of nanoparticle [179]; particle inhalation, retention, and clearance [36]; pulmonary effects of inhaled particles [37], [38]; inhalation and lung cancer [39], [40]; toxicity of combustion derived particles inhalation [41]; environmental factors in neurodegenerative diseases [42]; oxidative mechanisms [43-50]; gastro-intestinal uptake of particles [51]; targeted drug delivery [52]; particle characterizations methods [53]; screening strategies and future directions of research [54]; and regulation of nanomaterials [55]. Existing reviews are either written in jargon comprehensive only to specialists in a particular field, or are, if more accessible, very succinct [32], [56]. Most nanotechnology reviews written to date focus on a specific sub-field, disregarding the vast amount of existent knowledge on the general theme of nano. In this review, we attempt to bring together a broader audience by unifying the language and experience of scientists working within these diverse fields.

### 1.4. Introduction to nanoparticles toxicity

Human skin, lungs, and the gastro-intestinal tract are in constant contact with the environment. While the skin is generally an effective barrier to foreign substances, the lungs and gastro-intestinal tract are more vulnerable. These three ways are the most likely points of entry for natural or anthropogenic nanoparticles. Injections and implants are other possible routes of exposure, primarily limited to engineered materials.

Due to their small size, nanoparticles can translocate from these entry portals into the circulatory and lymphatic systems, and ultimately to body tissues and organs. Some nanoparticles, depending on their composition and size, can produce irreversible damage to cells by oxidative stress or/and organelle injury. Figure 6 illustrates the size of an example cell and its organelles compared to nanoparticles of various sizes, making it easy to understand why nanoparticles are able to enter cells and interact with various cell components (nucleus, mitochondria, etc.).

In Figure 7 we summarize the possible adverse health effects associated with inhalation, ingestion, and contact with nanoparticles. We emphasize that not all nanoparticles produce these adverse health effects - the toxicity of nanoparticles depends on various factors, including: size, aggregation, composition, crystallinity, surface functionalization, etc. In addition, the toxicity of any nanoparticle to an organism is determined by the individual's genetic complement, which provides the biochemical toolbox by which it can adapt to and fight toxic substances. While these effects will be discussed in details in chapters 3 and 4, we summarize below the most extreme adverse health effects produced by nanoparticles in order to immediately increase the awareness of potential toxicity of some nanoparticles. Diseases associated with inhaled nanoparticles are asthma, bronchitis, emphysema, lung cancer, and neurodegenerative diseases, such as Parkinson's and Alzheimer's diseases. Nanoparticles in the gastro-intestinal tract have been linked to Crohn's disease and colon cancer. Nanoparticles that enter the circulatory system are related to occurrence of arteriosclerosis, and blood clots, arrhythmia, heart diseases, and ultimately cardiac death. Translocation to other organs, such as liver, spleen, etc, may lead to diseases of these organs as well. Exposure to some nanoparticles is associated to the occurrence of autoimmune diseases, such as: systemic lupus erythematosus, scleroderma, and rheumatoid arthritis.

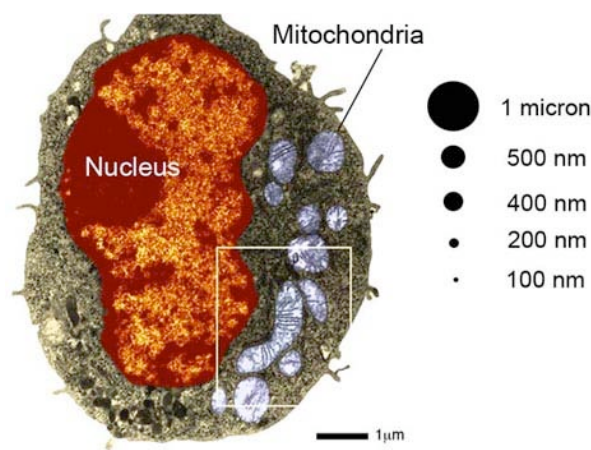

**Figure 6.** *Comparison of rat macrophage cells size to nanoparticles size (at scale). Human macrophages are up to two times larger than rat macrophages. TEM image reproduced with permission from Environmental Health Perspectives [57].*

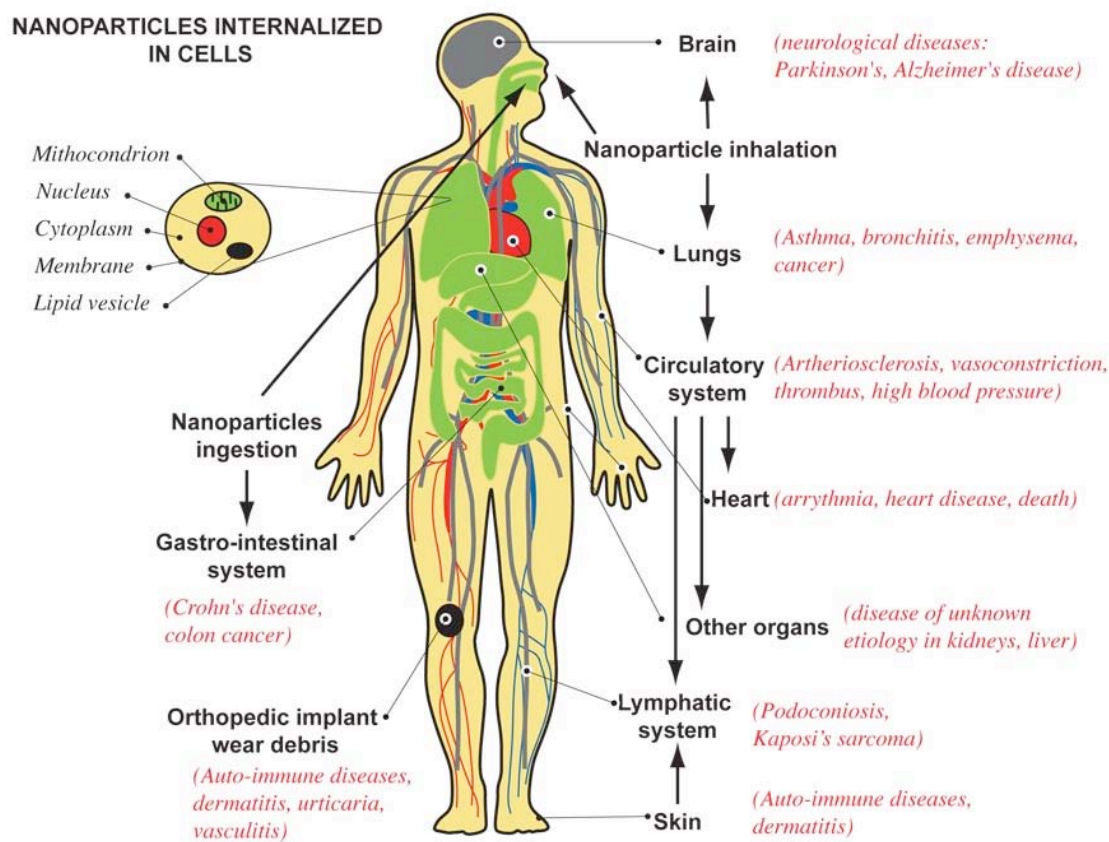

***Figure 7.*** *Schematics of human body with pathways of exposure to nanoparticles, affected organs, and associated diseases from epidemiological, in vivo and in vitro studies.*

## 2. Nanoparticle classification

Nanoparticles are generally classified based on their dimensionality, morphology, composition, uniformity, and agglomeration.

An important additional distinction should be made between *nanostructured* thin films or other fixed nanometer-scale objects (such as the circuits within computer microprocessors) and free *nanoparticles*. The motion of free nanoparticles is not constrained, and they can easily be released into the environment leading to human exposure that may pose a serious health risk. In contrast are the many objects containing nanostructured elements that are firmly attached to a larger object, where the fixed nanoparticles should pose no health risk when properly handled. An example of this important distinction is the material asbestos, which is perfectly safe in its primary state (basically a type of solid rock), but is a significant health hazard when mined or worked in such a way as to produce the carcinogenic nanometer-scale fibrous particles that become airborne (aerosol) and are therefore readily absorbed in the lungs.

It is also very important to recognize that not all nanoparticles are toxic; toxicity depends on at least chemical composition and shape in addition to simply size and particle ageing. In fact, many types of nanoparticles seem to be non-toxic [58], [59], others can be rendered non-toxic [60], while others appear to have beneficial health effects [61], [62]. An important lesson we are in the process of learning from nanoscience is that simple classifications of physical behavior (and therefore toxicity) are overly limiting and that we must study toxicology of each material and each morphology, in addition to particle ageing, to obtain accurate information to inform policy and regulatory processes.

### 2.1. Dimensionality

As shape, or morphology, of nanoparticles plays an important role in their toxicity, it is useful to classify them based on their number of dimensions (Figure 8). This is a generalization of the concept of aspect ratio.

**1D nanomaterials.** Materials with one dimension in the nanometer scale are typically thin films or surface coatings, and include the circuitry of computer chips and the antireflection and hard coatings on eyeglasses. Thin films have been developed and used for decades in various fields, such as electronics, chemistry, and engineering. Thin films can be deposited by various methods [63] and can be grown controllably to be only one atom thick, a so-called monolayer.

**2D nanomaterials.** Two-dimensional nanomaterials have two dimensions in the nanometer scale. These include 2D nanostructured films, with nanostructures firmly attached to a substrate, or nanopore filters used for small particle separation and filtration. Free particles with a large aspect ratio, with dimensions in the nanoscale range, are also considered 2D nanomaterials. Asbestos fibers are an example of 2D nanoparticles.

**3D nanomaterials.** Materials that are nanoscaled in all three dimensions are considered 3D nanomaterials. These include thin films deposited under conditions that generate atomic-scale porosity, colloids, and free nanoparticles with various morphologies [64].

### 2.2. Nanoparticle morphology

Morphological characteristics to be taken into account are: flatness, sphericity, and aspect ratio. A general classification exists between high- and low-aspect ratio particles (Figure 8). High aspect ratio nanoparticles include nanotubes and nanowires, with various shapes, such as helices, zigzags, belts, or perhaps nanowires with diameter that varies with length. Small-aspect ratio morphologies include spherical, oval, cubic, prism, helical, or pillar. Collections of many particles exist as powders, suspension, or colloids.

### 2.3. Nanoparticle composition

Nanoparticles can be composed of a single constituent material (Figure 8) or be a composite of several materials. The nanoparticles found in nature are often agglomerations of materials with various compositions, while pure single-composition materials can be easily synthesized today by a variety of methods; see chapter 3.2.6).

### 2.4. Nanoparticle uniformity and agglomeration

Based on their chemistry and electro-magnetic properties, nanoparticles can exist as dispersed aerosols, as suspensions/colloids, or in an agglomerate state (Figure 8). For example, magnetic nanoparticles tend to cluster, forming an agglomerate state, unless their surfaces are coated with a non-magnetic material. In an agglomerate state, nanoparticles may behave as larger particles, depending on the size of the agglomerate. Hence, it is evident that nanoparticle agglomeration, size and surface reactivity, along with shape and size, must be taken into account when deciding considering health and environmental regulation of new materials.

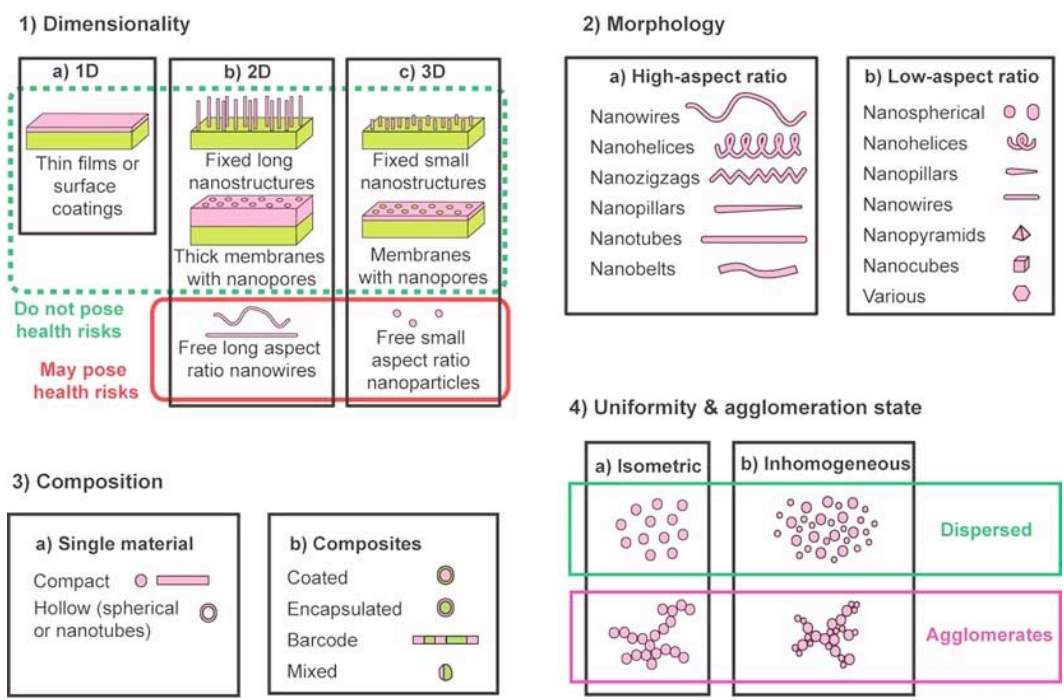

**Figure 8.** *Classification of nanostructured materials from the point of view of nanostructure dimensions, morphology, composition, uniformity and agglomeration state.*

## 3. Sources of nanoparticles and their health effects

### 3.1. Natural sources of nanoparticles

Nanoparticles are abundant in nature, as they are produced in many natural processes, including photochemical reactions, volcanic eruptions, forest fires, and simple erosion, and by plants and animals, e.g. shed skin and hair. Though we usually associate air pollution with human activities – cars, industry, and charcoal burning, natural events such as dust storms, volcanic eruptions and forest fires can produce such vast quantities of nanoparticulate matter that they profoundly affect air quality worldwide. The aerosols generated by human activities are estimated to be only about 10% of the total, the remaining 90% having a natural origin [65]. These large-scale phenomena are visible from satellites and produce particulate matter and airborne particles of dust and soot ranging from the micro- to nanoscales. Small particles suspended in the atmosphere, often known as aerosols, affect the entire planet's energy balance because they both absorb radiation from the sun and scatter it back to space [66]. It has been estimated that the most significant components of total global atmospheric aerosols are, in decreasing mass abundance: mineral aerosols primarily from soil deflation (wind erosion) with a minor component (<1%) from volcanoes (16.8 Tg), sea salt (3.6 Tg), natural and anthropogenic sulfates (3.3 Tg), products of biomass burning excluding soot (1.8 Tg), and of industrial sources including soot (1.4 Tg), natural and anthropogenic nonmethane hydrocarbons (1.3 Tg), natural and anthropogenic nitrates (0.6 Tg), and biological debris (0.5 Tg) [67] (note: 'Tg' here denotes terragram, equal to $10^{12}$ grams).

*3.1.1. Dust storms and health effects*

**Terrestrial dust storms.** Dust storms appear to be the largest single source of environmental nanoparticles. Long-range migration of both mineral dust and anthropogenic pollutants from the major continents has recently been the subject of intense investigation. Approximately 50% of troposphere atmospheric aerosol particles are minerals originating from the deserts [68]. The size of particles produced during a dust storm varies from 100 nm to several microns (Figure 9 d), with one third to a half of the dust mass being smaller than 2.5 microns [65], [68]. Particles in the range 100-200 nm can reach concentrations of 1500 particles/cm$^3$ [69].

Meteorological observations and modeling have identified ten main sources of global dust events, shown in Figure 9 e: (1) the Salton Sea, (2) Patagonia, (3) the Altipläno, (4) the Sahel region, (5) the Sahara Desert, (6) the Namibian desert lands, (7) the Indus Valley, (8) the Taklimakan Desert, (9) the Gobi Desert, and (10) the Lake Eyre basin [65].

Satellite imagery has revealed the dynamics of large-scale dust migration across continents, and demonstrated that nanoparticles generated by major environmental events in one part of the world can affect regions thousands of kilometers away, as shown in Figure 9. For example, dust storms occurring every spring in Gobi dessert strongly affect air quality in Asia and North America [70], [71]. The dust route across the Pacific can be seen in satellite images by the yellow color of the dust itself (Figure 9 a) [72]. The dust migration pattern during the 1998 trans-Pacific transport is shown in Figure 9 c, the dates representing the approximate daily location of the dust cloud [70]. During this event, the dust cloud reached the west coast of North America within 5–6 days after emission, with the region affected experiencing an intense haze and elevated particles concentrations, with an average excess 20–50 µg/m$^3$ with local peaks >100 µg/m$^3$ [70], [71].

**Extraterrestrial dust.** Nanoparticles exist widely in extraterrestrial space. Examples of dust collected from space, from the moon, and on Mars are shown in Figure 10. The extraterrestrial dust poses major environmental problems for astronauts as well as for equipment [73]. Lunar dust is very fine grained compared to typical terrestrial dust (some of the larger grains being shown in Figure 10 c), with more than 50% of particles found to be in the micron range or smaller [74]. The lunar dust contains a considerable amount of magnetic nanoparticles [75], clinging to electrostatically charged surfaces [74] such as the astronauts' space suits (Figure 10 b), rendering it nearly impossible to remove. On Mars, dust accumulating on the solar panels of the exploration robots has limited the power available to them for locomotion, sensing, and communication [76]. Aiming to mitigate the environmental effects of extraterrestrial dust on humans and machines, various research projects are directed towards the fabrication of filters or thin film coatings that repel dust [76].

**Health effects.** Terrestrial airborne dust particles can lead to a number of health problems, especially in subjects with asthma and emphysema [65]. The composition of dusts is important, as iron or other metals rich-dust can generate reactive oxygen species on the lung surface that can scar lung tissue [65]. In addition, viruses, bacteria, fungi, or chemical contaminants hitchhiking dust particles may adversely affect health and the environment (Figure 9 f). In this regard it is important to note that two hundred types of viable bacteria and fungi have been found to survive ultraviolet light exposure during intercontinental journeys from Africa to America [65].

Extraterrestrial dust brought inside the Lunar Module became airborne and irritated lungs and eyes of Apollo astronauts [77]. On longer missions to the moon or Mars, prolonged exposure could increase the risk of respiratory diseases in the astronauts, and mechanical failures of spacesuits and airlocks. Studies on rats have found that intratracheal administration of small amounts of lunar material resulted in pneumoconiosis with fibrosis formation [78] (lung disease and abnormal tissue growth).

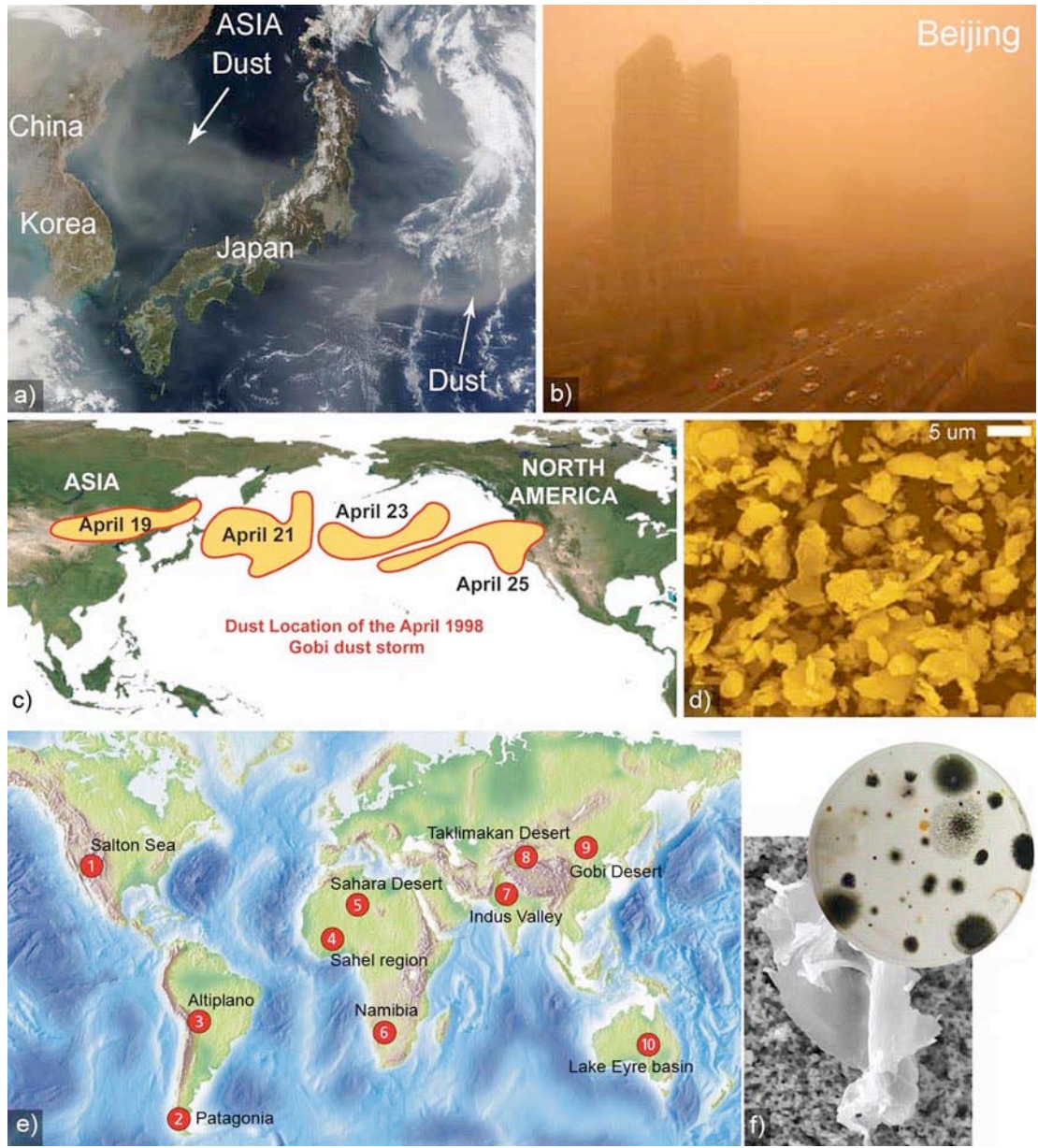

***Figure 9.*** *Sand storms visualized at macro and microscale. **(a)** Satellite image showing dust blowing off mainland China over the Sea of Japan and Pacific Ocean in April 2002, credit Jacques Descloitres, MODIS Land Rapid Response Team, NASA/GSFC [72]. **(b)** Beijing during a dust storm. **(c)** Approximate location of the dust cloud from a Gobi desert dust storm during April 1998, based on satellite images, after [70]. **(d)** Asia dust storm samples collected during the 16 March 2002 dust storm [68], courtesy of American Geophysical Union. **(e)** Ten major sources of dust in the world; **(f)** bacteria collected from African dust that reached North America, both (e) and (f) reproduced with permission from Environmental Health Perspectives [65].*

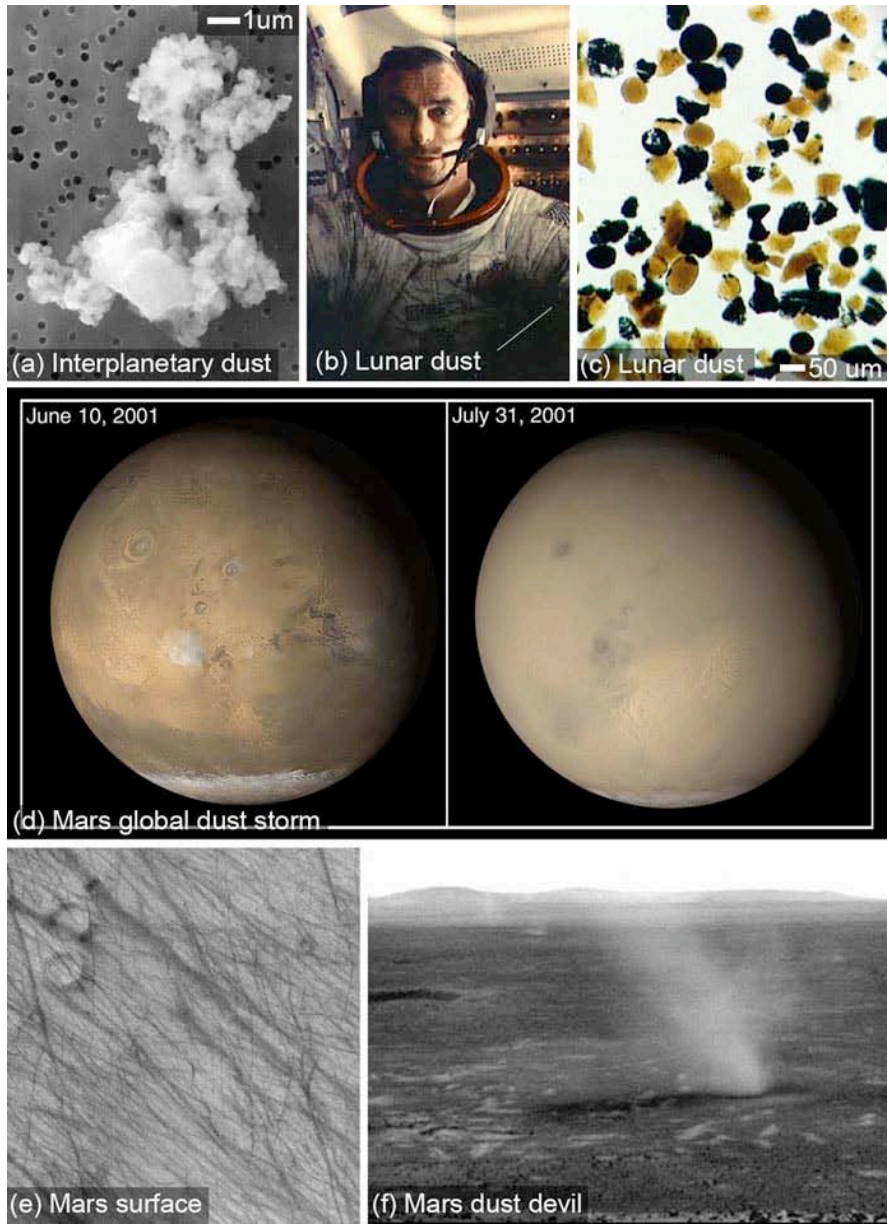

**Figure 10.** *Extraterrestrial dust. (a) SEM of interplanetary dust composed of glass, carbon and silicate mineral nanoparticles, courtesy of NASA, http://antwrp.gsfc.nasa.gov/apod/ap010813.html. (b) Lunar dust on the suit of an astronaut inside the lunar module on the lunar surface. The picture was taken after the second extravehicular activity on this mission on 12.12.1972. Image ID: AS17-145-22224, courtesy of NASA Johnson Space Center (NASA-JSC). (c) Larger lunar dust particles returned from the moon by Apollo 17 in 1973. These orange glass spheres and fragments range in size from 20 to 45 microns, courtesy of NASA Johnson Space Center (NASA-JSC) [83]. (d) Global Mars dust storm of 2001, courtesy NASA/JPL/Malin Space Science Systems [84]. (e) Mars devil-streaked surface, courtesy NASA/JPL/Malin Space Science Systems [84]. (c) Mars dust devil, courtesy NASA [84].*

### 3.1.2. Forest fires and health effects

Forest fires and grass fires have long been a part of Earth's natural history, and are primarily caused by lightning strikes or by human activity. Major fires can spread ash and smoke (Figure 11 a) over thousands of square miles (Figure 11 b, c) and lead to an increase of particulate matter (including nanoparticles) exceeding ambient air quality standards [79]. Satellite maps show a unique picture of global fire activity. Using daily global fire detection provided by MODIS on NASA's Terra satellite, the fire activity for the entire surface of the Earth has been mapped every day since February 2000 (Figure 11 d) [80]. As noticed in this figure, numerous fires occur throughout the world in the savannas of Africa, Australia, and Brazil, in North America, Europe and Asia.

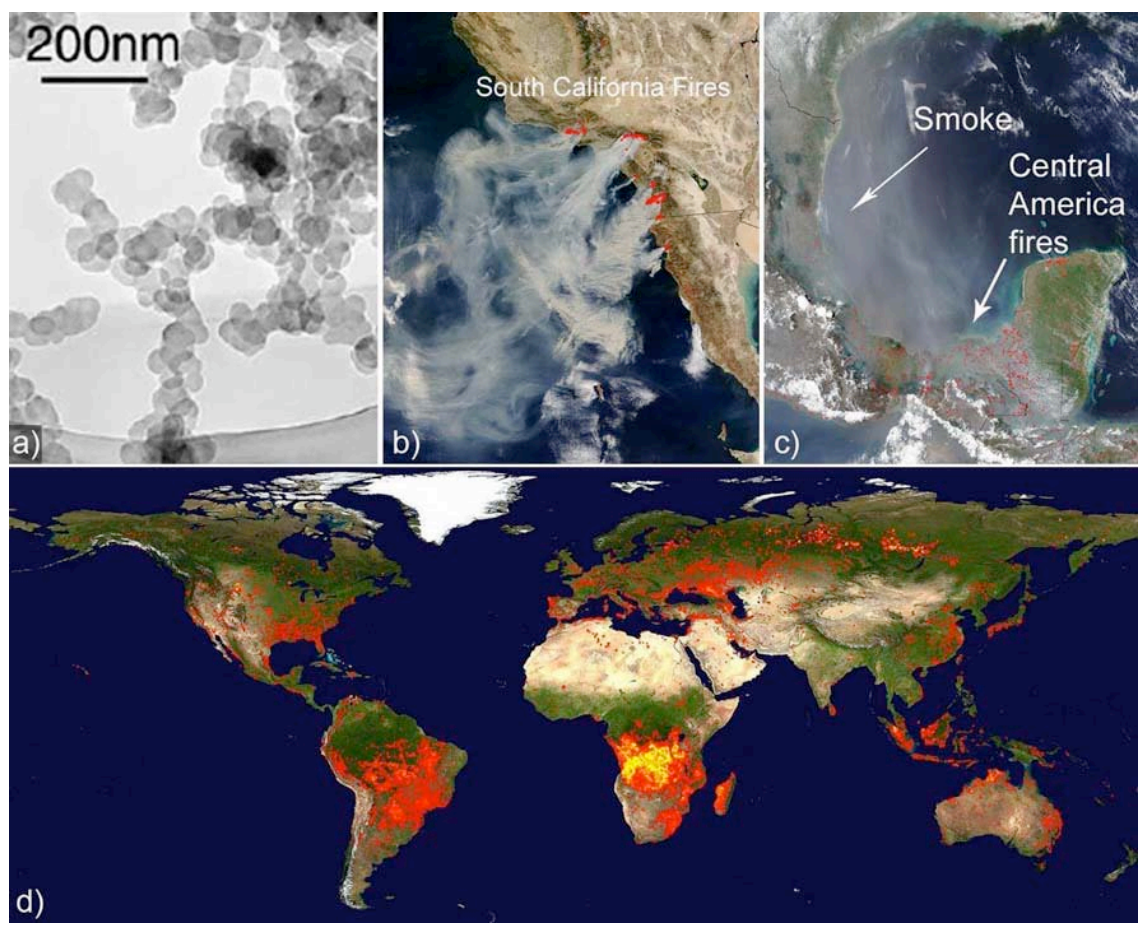

**Figure 11.** *(a) TEM image of smoke aggregates from a fire in Zambia [85], courtesy of American Geophysical Union. NASA satellite images showing smoke pollution from fires, indicated with red dots. (b) South California fires. Credit Jacques Descloitres, MODIS Rapid Response Team, NASA/GSFC [72]. (c) Smoke from fires of Central America spreads towards Golf of Mexico and U.S. in May 2003, credit Jeff Schmaltz, MODIS Rapid Response Team, NASA/GSFC. (d) Distribution of active fires detected by Terra's MODIS sensor across the planet during 10-19 July 2006, courtesy of NASA/MODIS Rapid Response Team/Scientific Visualization Studio [72].*

**Health effects.** Epidemiological studies showed that during the weeks of forest fires, medical visits increase more than 50% in the affected regions [81]. Patients with pre-existing cardio-pulmonary conditions reported worsening symptoms during smoke episodes. The usage of air cleaners was associated with less adverse health effects on the lower respiratory tract [81]. Around 75% of fire-related deaths are due to respiratory problems related to smoke inhalation and not necessarily burns. The treatment for smoke inhalation in an emergency room is usually oxygen. Due to the fact that the symptoms may be delayed until 24-36 hours after inhalation, the patient must be kept under observation for several days.

*3.1.3. Volcanoes and health effects*

When a volcano erupts, ash and gases containing particulate matter ranging from the nanoscale to microns (Figure 12 a, b, c), are propelled high into the atmosphere, sometimes reaching heights over 18000 meters. The quantity of particles released into the atmosphere is enormous; a single volcanic eruption can eject up to 30 million tons of ash [65]. Volcanic ash that reaches the upper troposphere and the stratosphere (the two lowest layers of the atmosphere) can spread worldwide and affect all areas of the Earth for years. A primary effect of upper atmospheric particulate debris is the blocking and scattering of radiation from the sun. One particularly harmful volcanic product is particles composed of heavy metals, as these are known to be toxic to humans. While some effects are seen worldwide, the highest levels of particulate matter are found in areas within tens of km from the volcano [86].

**Health effects.** Short-term effects of ash on health include: respiratory effects (nose and throat irritation, bronchitic symptoms), and eye and skin irritation. To assess the impact of long-term exposure to volcanic particulate pollution, we can look to the barefoot agricultural populations living in parts of the world containing volcanic soils, such as Africa, Mediterranean, and Central America. A large percentage of this population is affected by diseases of lympho-endothelial origin. The diseases include podoconiosis (Figure 12 d) [87-89] and Kaposi's sarcoma (Figure 12 f) [81], [90].

Podoconiosis is a noncomunicable disease producing lymphoedema (localized fluid retention) of the lower limbs. The cause of this disease is believed to be the absorption through the skin of the feet (podos) of nano and microparticles from the soil (konia) [91]. Lymphoedema occurs when the lymphatic system fails to properly collect and drain the interstitial fluid of the body, resulting in the long-term swelling of a limb or limbs (Figure 12 d). The lymphatic system is a secondary circulatory system in the body that collects fluid from several sources, primarily that lost from the circulatory system (blood), for example from damaged blood vessels in an area of inflammation (e.g. after a burn, or other injury) [82]. The lymphatic system lacks a central pump, i.e. the equivalent to the heart in the circulatory system, so it relies on a network of vessels and nodes that pumps during usual (muscle) motion of the body. If the accumulation of the interstitial fluid is faster than the pumping, then the tissue swells. In podoconiosis the effect is irreversible, and affects about 10% of the populations in volcanic tropics. Soil particles with size ranging from 400 nm up to 25 microns, were found in the dermis of the foot of individuals with podoconiosis [87], [88]. These particles were found in the macrophages, the cytoplasm of other cells, as well as in lymph node biopsies, as indicated by scanning electron microscopy. Energy dispersive x-ray analysis techniques showed compositions consistent with the elements present in black lava soil and red clay soil [89]. It is hypothesized that large quantities of small particles and chronic exposure overwhelm the normal function of the lymphatic drainage system in the patients with podoconiosis, blocking drainage of both particles and lymph fluid [87-89].

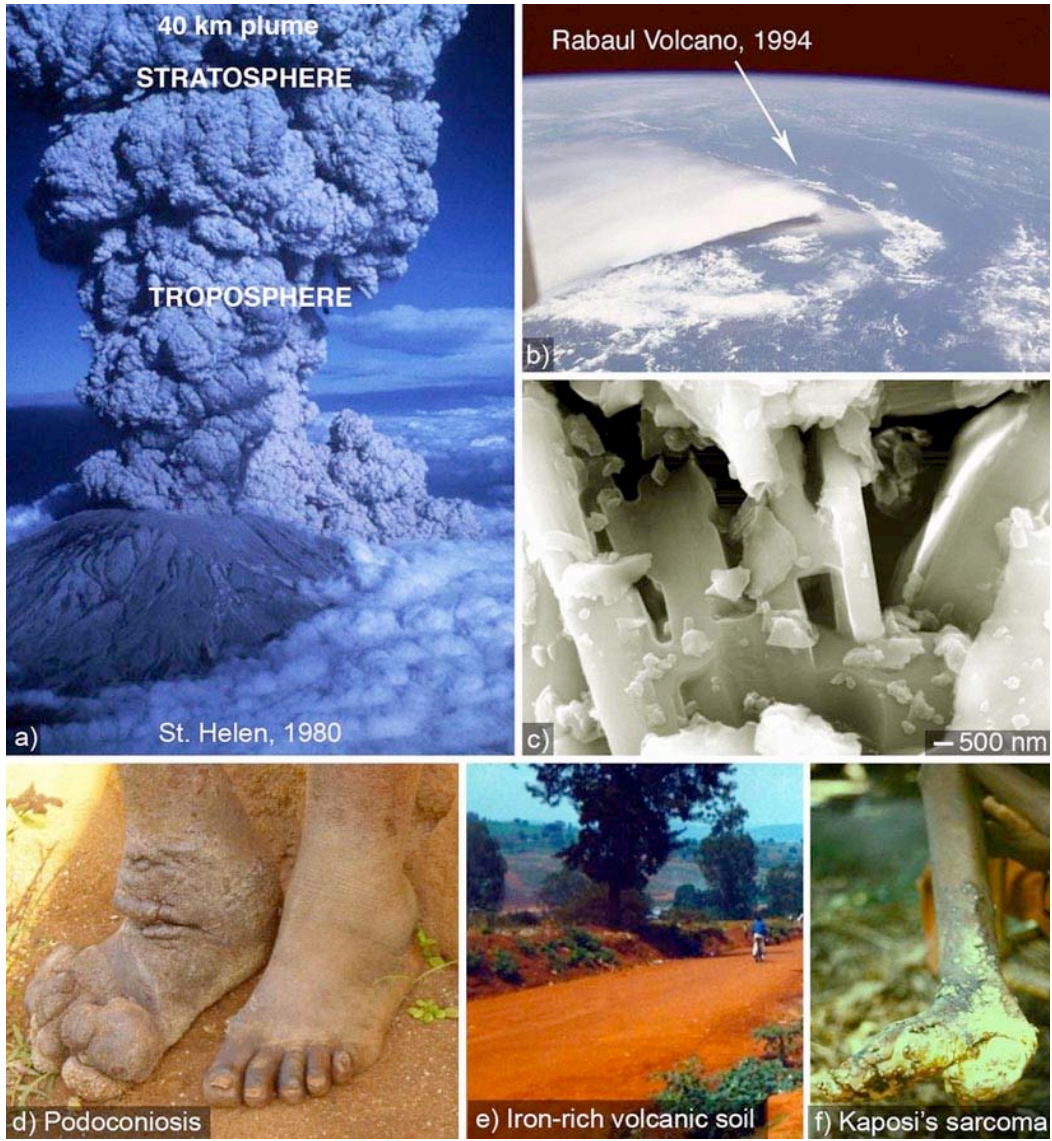

**Figure 12. (a)** *The eruption plume of St. Helen volcano, in 1980, courtesy of NASA;* **(b)** *Rabaul Eruption Plume, New Britain Island, 1994. The large scale of eruption can be compared to the Earth curvature, courtesy of Image Science and Analysis Laboratory, NASA-Johnson Space Center [95];* **(c)** *Scanning electron microscope image of volcanic ash from the first volcanic eruption of Mount St. Helens, Washington state, U.S.A. in 1980, courtesy of Chuck Daghlian, Louisa Howard [96].* **(d)** *Podoconiosis - impaired lymphatic system drainage affecting the limbs due to clogging with nano and microparticles [92], courtesy of Elsevier.* **(e)** *Volcanic iron oxide rich soil in Rwanda, [97], courtesy Biomed Central.* **(f)** *Aggressive African-endemic Kaposi's sarcoma - cancer of blood and lymph vessels - of the foot [97], courtesy Biomed Central.*

Kaposi's Sarcoma is a form of cancer affecting the blood and lymph vessels (Figure 12 f), and is also related to human herpes virus infection [81]. Endemic Kaposi's sarcoma [81], [90] is characteristic to parts of the world containing volcanic soils [81]. It was found that iron particles from the iron-rich volcanic soils (Figure 12 e) may be one of the co-factors involved in the etiology (set of causes) of Kaposi's sarcoma [81]. In chronic exposure to iron volcanic clays, ferromagnetic nanoparticles penetrate the skin of barefoot agricultural workers leading to impaired lymphatic drainage and local immunity, leaving the organism prone to infections (such as herpes virus).

**Treatment.** The treatment of podoconiosis in early stages involves elevation and elastic stockings, while in more advanced stages the only treatment is surgical. Treatment of Kaposi's Sarcoma involves iron withdrawal, and iron chelators [81]. Both these diseases, podoconiosis and Kaposi's Sarcoma, could be prevented by wearing shoes or boots (not sandals or shoes with open spaces) starting from early childhood [92].

### 3.1.4. Ocean and water evaporation and health effects

A large amount of sea salt aerosols are emitted from seas and oceans around the world [67]. These aerosols are formed by water evaporation and when wave-produced water drops are ejected into the atmosphere (Figure 13 a). Their size ranges from 100 nm to several microns. An example of sea salt nanoparticles is shown in Figure 13 b. Nanoparticles can also form in bodies of water through precipitation, as a result of temperature changes and evaporation. An example of this phenomenon is lake Michigan that rests in a limestone basin, the water containing high levels of calcium carbonate. During most of the year the calcium carbonate remains dissolved in the cold water, but at the end of summer the water temperature increases, lowering the solubility of calcium carbonate. As a result, the calcium carbonate may precipitate out of the water, forming clouds of nanometer-scale particles that appear as bright swirls when viewed from above, as shown in Figure 13 c [93].

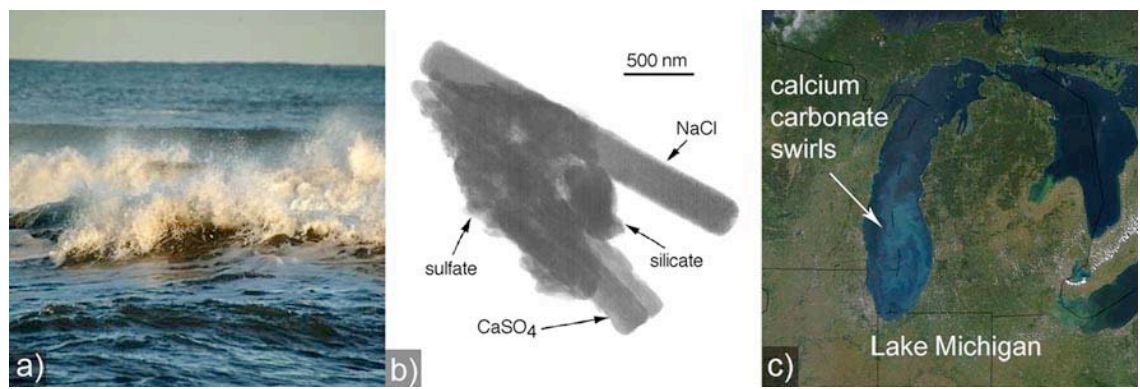

**Figure 13.** *(a) Sea spray from ocean waves, courtesy NASA; (b) TEM image of mineral dust mixture with sea salt collected from the marine troposphere [67]; (c) the pale blue swirls in Lake Michigan are probably caused by calcium carbonate (chalk) from the lake's limestone floor, credit Jeff Schmaltz, MODIS Rapid Response Team, NASA/GSFC [93].*

**Health effects.** No adverse health effects have been associated to sea salt aerosols. On the contrary, beneficial health effects have been suggested from the use of salt aerosols in the restoration of the mucociliary clearance in patients with respiratory diseases [94]. The unique microclimate of salt mines is a popular way to treat asthma, particularly in Eastern Europe. However, sea salt aerosols may transport pollutants and microorganisms that themselves may cause adverse health effects.

### 3.1.5. Organisms and health effects

Many organisms are smaller than a few microns (Figure 14 c, d), including viruses (10 nm - 400 nm), and some bacteria (30 nm - 700 μm). However, we should make a clear distinction between what we call "particles" (microparticle or nanoparticle) and nano-organisms or their components (including bacteria, viruses, cells, and their organelles). Cells, bacteria, and viruses are self-organizing, self-replicating, dissipative structures, with a shorter-lived structure than inorganic solids. Nano-organisms generally dissipate when their supply of energy is exhausted. In contrast, nanoparticles are typically inorganic solids that require no supply of energy to remain in a stable form. They interact, dissipate, or transform via chemical reactions with their environment.

Many organisms, both uni- and multicellular, produce nanoparticulate inorganic materials through intracellular and extracellular processes (Figure 14) [98]. For example, magnetite nanoparticles are synthesized by magnetotactic bacteria and used for navigation relative to the earth's magnetic field (Figure 14 a), siliceous materials are produced by diatoms (Figure 14 b), or calcium carbonate layers are produced by S-layer bacteria [98]. Magnetotactic bacteria (Figure 14 a) orient and migrate along the geomagnetic field towards favorable habitats using nanometer-size magnetic particles inside the cell. These bacteria are aquatic microorganisms inhabiting freshwater and marine environments. In figure 14 b are shown various diatom species frustules (siliceous shells). Diatoms are unicellular algae with cell walls made of silica. They are abundant in plankton communities and sediments in marine and freshwater ecosystems, where they are an important food source for other marine organisms. Some may even be found in moist soils. Diatoms are used in forensic science to confirm drowning as a cause of death and localize the site of drowning, based upon the observation of diatoms in lungs, blood, bone marrow, and organs [99]. Nanobacterium is a nano-organism that synthesizes a shell of calcium phosphate to cover itself, and resembles an inorganic particle (Figure 14 e, f). The shell ranges in size between 20 to 300 nm, and due to its porous nature it allows the flow of a slimy substance. This slime (presumably together with electrical charge) promotes the adhesion to biological tissues and the formation of colonies. Nanobacteria are very resilient, being temperature- and gamma radiation- resistant [101].

**Health effects and treatment.** Among these biological nanoparticles, diatoms might pose a health risk to workers of diatomaceous earth mining and processing [102]; biogenic magnetite is associated with neurodegenerative diseases [20], and nanobacteria shells were found in humans and animals [6], [101]. Nanobacteria are ubiquitous within living organisms, humans and animals, being identified in blood, serum, and organs [101]. These very small bacteria are suspected of being the cause (at least in part) for many diseases involving calcifications, such as: artery plaque, aortic aneurysm, heart valves, renal stone formation, chronic prostatitis, ovarian and breast tumors [6], [101], [103]. They may also be the cause of rapid kidney stone formation in astronauts on space travels, according to a NASA study, probably due to the fact that their multiplication rate in a microgravity environment increases fourfold compared to the rate under normal condition of gravity (of only about 3 days for doubling rate [104]). Definitive mechanisms relating nanobacteria to these above mentioned diseases are unknown, however there are speculations that nanobacteria colonies may act as nucleation sites for plaque or stone formation [105]. Specific therapies, such as laser irradiation [106], or antibiotics [105], have shown reduced plaque formation, and even the regression of plaques.

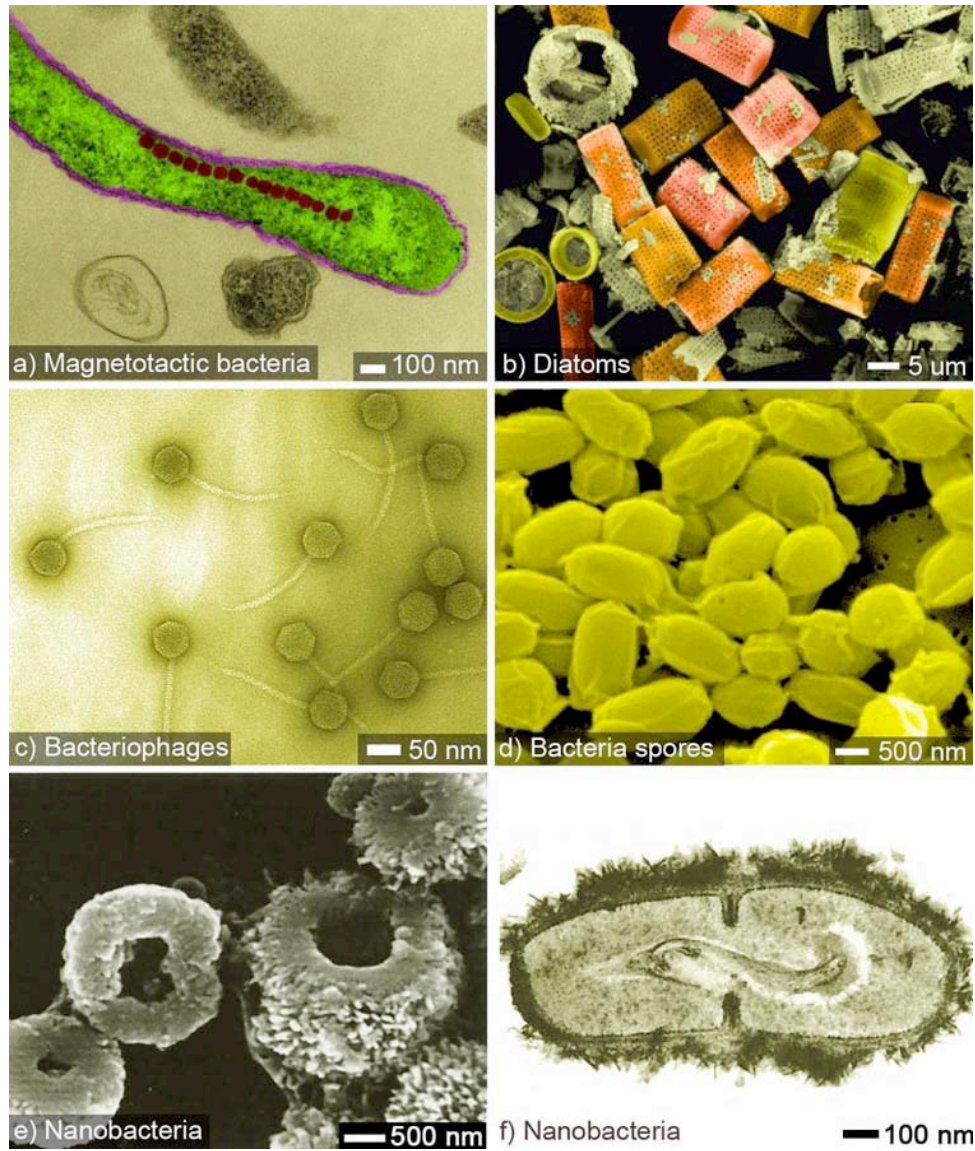

**Figure 14.** *Organisms in the nanoscale range or producing solid -state nanoscale debris. (**a**) TEM of Aquaspirillum magnetotacticum - bacterium showing magnetosomes (iron oxide granules). (**b**) SEM of diatom silica frustules or skeletons. (a and b © Dr. Dennis Kunkel/Visuals Unlimited. Reproduced with permission from Visuals Unlimited [25]. (**c**) SEM of bacteriophage - a virus that infects bacteria, courtesy of Ross Inman [100]. (**d**) SEM of Bacillus anthracis bacteria spores, that can live for many years, enabling the bacteria to survive in a dormant state until they encounter a suitable host, credit Laura Rose, courtesy of Public Health Image Library [21]. (**e**) SEM of cultured nanobacteria, (**f**) Dividing nanobacteria covered with a "hairy" apatite layer, (e and f courtesy of PNAS [6]).*

### 3.2. Anthropogenic nanomaterials

Humans have created nanomaterials for millennia, as they are byproducts of simple combustion (with sizes down to several nm) and food cooking, and more recently, chemical manufacturing, welding, ore refining & smelting, combustion in vehicle and airplane engines [107], combustion of treated pulverized sewage sludge [108], and combustion of coal and fuel oil for power generation [109]. While engineered nanoparticles have been on the market for some time and are commonly used in cosmetics, sporting goods, tires, stain-resistant clothing, sunscreens, toothpaste, food additives, etc., these nanomaterials, and new more deliberately fabricated nanoparticles, such as carbon nanotubes, constitute a small minority of environmental nanomaterials. The quantity of man-made nanoparticles ranges from well-established multi-ton per year production of carbon black (for car tires) to microgram quantities of fluorescent quantum dots (markers in biological imaging).

### 3.2.1. Diesel and engine exhaust nanoparticles and health effects

Diesel and automobile exhaust are the primary source of atmospheric nano- and microparticles in urban areas [110]. Most particles from vehicle exhaust are in the size range of 20-130 nm for diesel engines and 20-60 nm for gasoline engines (Figure 15 a) [24], [111] and are typically approximately spherical in shape. Carbon nanotubes and fibers, already a focus of ongoing toxicological studies, were recently found to be present in engine exhaust as a byproduct of diesel combustion [112] and also in the environment near gas-combustion sources [113]. The aspect ratio of these fibers is comparable to those of lung-retained asbestos, suggesting that strong carcinogens may exist in exhaust. Prior to the release of this findings [112] they were thought not to exist in the environment and their existence was attributed exclusively to engineering by material scientists. Nanoparticles constitute 20% of the particles mass but more than 90% of the number of diesel-generated particles [17]. Due to recent health concerns, particle size distribution and number concentrations studies were conducted in various cities along different continents [115].

A high number concentration of nanoparticles can be located near freeways on scales of hundreds of meters, showing that vehicular pollution is a major source of local contaminant particulate matter that includes nanoparticles (Figure 15 b).

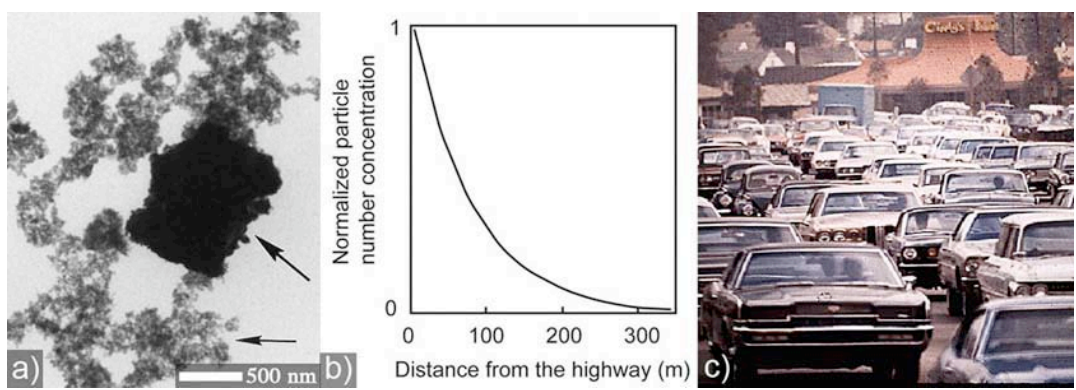

**Figure 15. (a)** *TEM showing typical engine exhaust particles consisting of carbon aggregates (small arrow) around a larger mineral particle (large arrow) [114]; (b) particle concentration decreases exponentially with downwind distance from the freeway (particles diameter between 6-220 nm) [24]; (c) Traffic in Los Angeles, courtesy EPA.*

The daily profile of nanoparticles matches that of local vehicles usage [115]. High pollution episodes or proximity to high-traffic roads can increase the mass concentration of nanoparticles by several times from typically low background levels of approximately 0.5-2 $\mu g/m^3$ [20].

**Health effects.** Research has shown some heterogeneity in the magnitude of adverse health effects of engine exhaust in different cities, probably related to the complexity and composition of particles mixtures [49]. Generally, diesel exhaust is known to be toxic as it contains high levels of polynuclear aromatic hydrocarbons (PAHs) including the known carcinogen benzo-a-pyrene (BaP) [116].

Atmospheric particle pollution from automobile exhaust seems to have a major influence on mortality, with a strong association between increased cardiopulmonary mortality and living near major roads [117], [122]. The findings of this epidemiological study are in concordance with measurements of nanoparticle concentration near highways, the concentration decreasing exponentially over several hundreds meters from the traffic [24]. Childhood cancers were also found to be strongly determined by prenatal or early postnatal exposure to oil-based combustion gases, primarily engine exhaust [118].

Professional drivers show elevated rates of myocardial infarction (heart attack) [121]. Studies done in non-smoking, healthy, young patrol officers have shown that nanoparticles from vehicular traffic may activate one or more signaling pathways that cause pro-inflammatory, pro-thrombotic and hemolytic (breakdown of red blood cells) responses [119]. It was noted that heart rate variability was significantly associated with measures of pollution. Epidemiological studies conducted on diesel locomotive drivers showed a correlation between occupational exposures to diesel engine exhaust and incidence of lung cancer in the workers [120].

These findings suggest that pollutants emitted by vehicles harm the health of many people, and that professional drivers, frequent drivers, passengers, and peoples living near major roads are at elevated risk. Results seen in these studies suggest that exposure to exhaust nanoparticles leads to increased risk of cardiovascular events over the long term.

### 3.2.2. Indoor pollution and health effects

Indoor air can be ten times more polluted than outdoor air, according to the Environmental Protection Agency (EPA) [123]. Humans and their activities generate considerable amounts of particulate matter indoors (Figure 16). Nanoparticles are generated through common indoor activities, such as: cooking, smoking, cleaning, and combustion (e.g. candles, fireplaces). Examples of indoor nanoparticles are: textile fibers, skin particles, spores, dust mites droppings, chemicals, smoke from candles, cooking, and cigarettes. A quantitative determination of nanoparticle emissions from selected indoor sources is given in Table 1 [124]. Particles have also been shown to enter buildings from outdoors through ventilation systems [24]. As humans generally spend much of their time indoors (more than 80%), indoor pollution directly affects our health.

**Health effects.** Long-term exposure to indoor cooking emissions may pose adverse health effects due to particulate matter inhalation [125]. During cooking, the level of particulate matter increases more than ten fold compared to non-cooking hours [125]. In many regions of the world, death caused from indoor smoke from solid fuels is considerable, especially in Africa and Asia (Figure 16 e). Poorly ventilated stoves using biomass fuels (wood, crop residue, dung, coal) are the main responsible for the death of an estimated 1.6 million peoples annually, from which more than a half are children under the age of five [126]. Word Health Organization estimates more than 50% of the world population uses solid fuels for cooking and heating, including biomass fuels. Wood burning is often disregarded as a source of nanoparticles and assumed to be benign to the environment simply because wood is a renewable source.

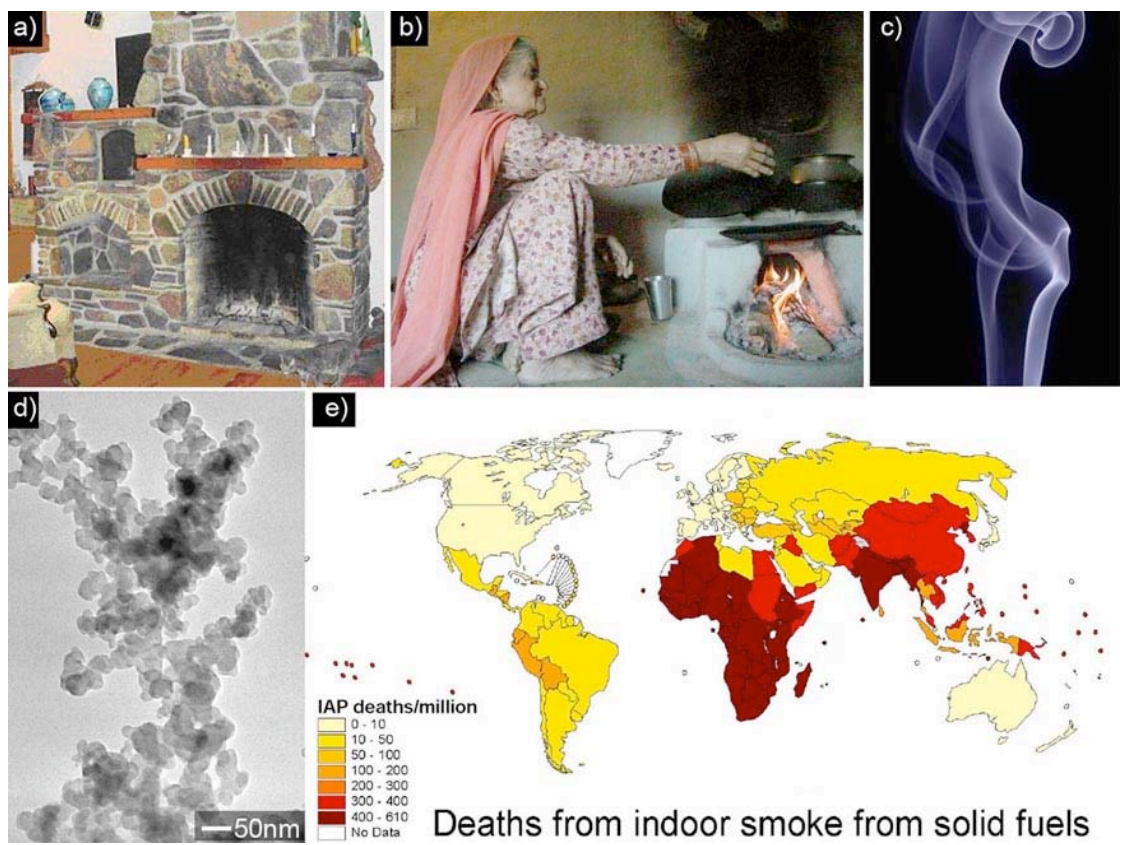

**Figure 16.** *Indoor air pollution from (**a**) heating, (**b**) cooking (Courtesy E.K. Schafhauser), (**c**) candle smoke. (**d**) TEM of soot particle from indoors pollution [127], reproduced with permission from Environmental Health Perspectives. (**e**) Death from indoor smoke from solid fuels according to World Health Organization [126].*

**Table 1.** *Measured concentrations of nanoparticles resulting from various common indoor household activities, after [124].*

| Nanoparticle source | Concentration (nanoparticles/cm$^3$) | Estimated source strength (particles/min x 10$^{11}$) |
|---|---|---|
| Pure wax candle | 241,500 | 3.65 |
| Radiator | 218,400 | 8.84 |
| Cigarette | 213,300 | 3.76 |
| Frying meat | 150,900 | 8.27 |
| Heater | 116,800 | 3.89 |
| Gas stove | 79,600 | 1.3 |
| Scented candles | 69,600 | 0.88 |
| Vacuum cleaner | 38,300 | 0.38 |
| Air freshener spray | 29,900 | 2.34 |
| Ironing a cotton sheet | 7,200 | 0.007 |

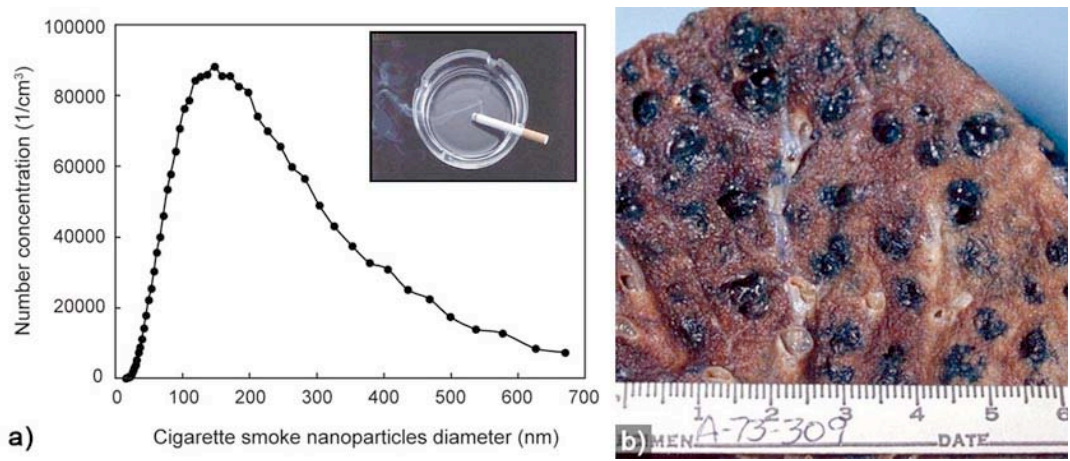

*Figure 17. (a) Measured environmental tobacco smoke particles concentration versus nanoparticle diameter. Nanoparticles are generated upon smoking one cigarette (after [128]). (b) Pathology of lung showing centrilobular emphysema characteristic of smoking. The cut surface shows multiple cavities heavily lined by black carbon deposits, content providers Dr. Edwin P. Ewing, Jr., courtesy of Public Health Image Library [21].*

*3.2.3. Cigarette smoke and health effects*

As a combustion product, tobacco smoke is composed of nanoparticles with size ranging from around 10 nm up to 700 nm, with a maximum located around 150 nm (Figure 17 a) [128]. The environmental tobacco smoke has a very complex composition, with more than 100,000 chemical components and compounds [128].

**Health effects.** Environmental tobacco smoke is known to be toxic, both due to some of its gas phases as well as nanoparticles. A plethora of studies have investigated the adverse health effects of environmental cigarette smoke. Substantial evidence shows that, in adults, first or second hand cigarette smoke is associated with an increased risk of chronic respiratory illness (Figure 17 b), including lung cancer, nasal cancer, and cardiovascular disease, as well as other malignant tumors, such as pancreatic cancer [129] and genetic alterations [130]. Children exposed to cigarette smoke show an increased risk of sudden infant death syndrome, middle ear disease, lower respiratory tract illnesses, and exacerbated asthma [129]. Cigarette smokers are more likely than nonsmokers to develop many conditions including cancers and vascular diseases [131]. It was noted that the risk of myocardial infarction decreases substantially within two years after smoking cessation, proving a reversibility of inhaled nanoparticles induced vulnerability [132].

*3.2.4. Buildings demolition and health effects*

Particulate matter concentrations can rise to very high levels when large buildings are demolished, especially the respirable ones with diameter smaller than 10 microns [133]. Older buildings are very likely to have been constructed with parts containing known toxins. Consequently, respirable asbestos fibers, lead, glass, wood, paper, and other toxic particles are often found at the site of demolition [133]. In addition, the dust cloud can travel tens of kilometers and affect the neighboring regions of the collapsed building site [133].

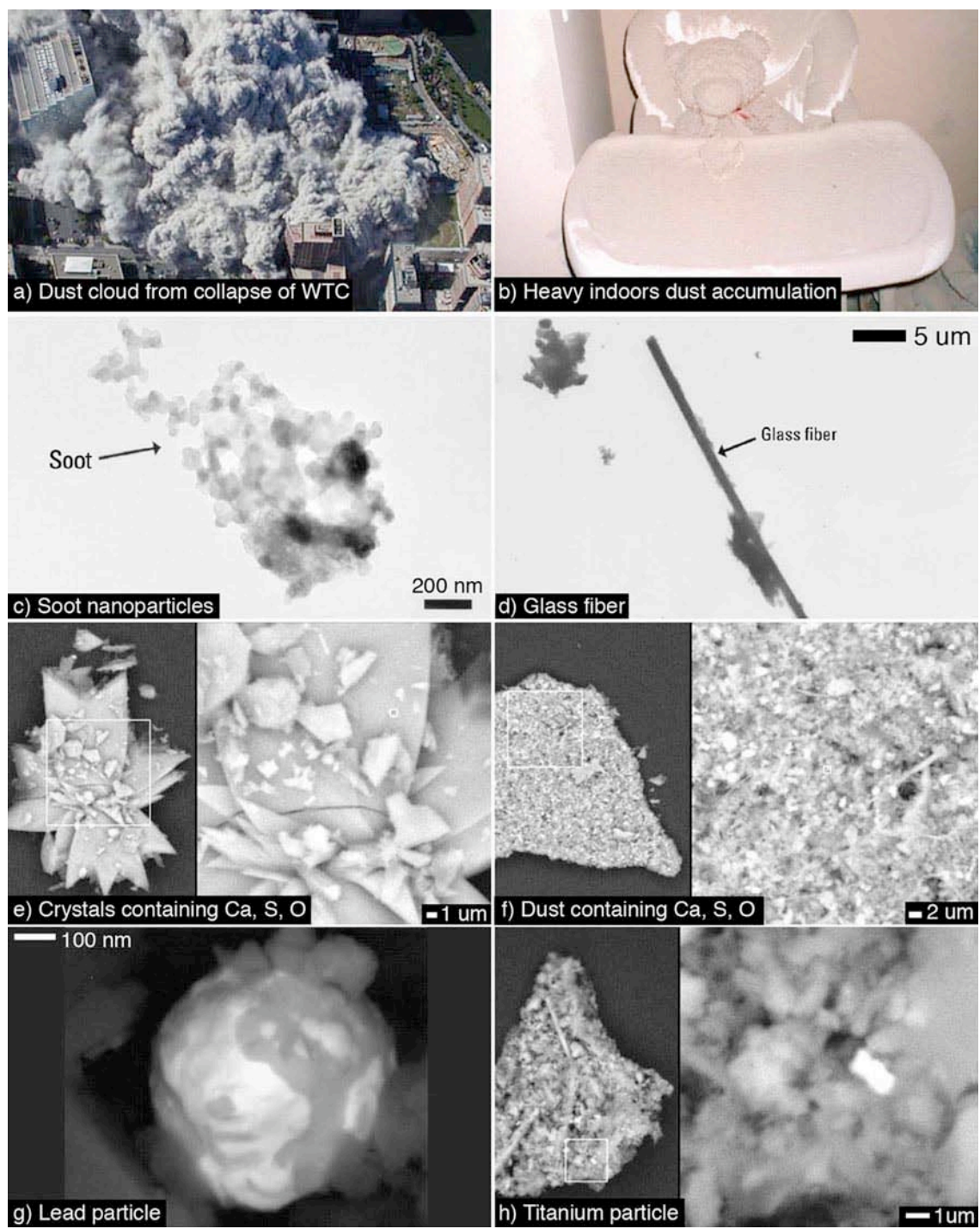

**Figure. 18.** *(a) Dust cloud from the World Trade Center collapse spreads to neighboring streets. Courtesy EPA. (b) Heavy dust accumulation in store closed to World Trade Center [135]. Particle collected from the site of collapse and neighboring streets: (c) soot [135], d) glass fiber [135], (e), (f) dust containing Ca, S and O [136], (g) lead and [g135], (h) titanium particle [136]. Images (b)-(h) courtesy Environmental Health Perspectives.*

**Health effects** of exposure to demolition particles and soot (Figure 18) are not entirely known. Early clinical and epidemiological assessments of firefighters present at the site of the environmental disaster generated by the attack on the World Trade Center on September 11, 2001, indicated exposure-related health effects, with prevalence of respiratory symptoms, especially increased cough and bronchial hyperactivity [134]. Long-term effects, however, remain to be seen.

### 3.2.5. Cosmetics, other consumer products and health effects

**Cosmetics.** The use of nanomaterials in cosmetics is not new. Black soot and mineral powders have been used as cosmetics since thousands of years ago in ancient Egypt, and some of them continue to be used today. Due to the recent development of nanotechnology, engineered nanomaterials have been embraced by the cosmetics industry for several reasons.

a) Because of their ability to penetrate deeper into the protective layers of skin than any cosmetic before, they are used as delivery agents for skin nutrients, such as synthetic peptides that instruct cells to regenerate [137].

b) Some nanoparticles have antioxidant properties [138], feature that helps maintain a youthful appearance of the skin. For example, functionalized fullerenes are now incorporated into cosmetic products, such as creams, claiming radical scavenging properties [14].

c) Due to their small size and specific optical properties, they are thought to conceal wrinkles and small creases [14]. For example, alumina nanopowder is used for optical reduction of fine lines [14].

Many cosmetic and personal care products incorporate nanomaterials. For a compilation of websites and product information see reference [140]. They include: personal care products (deodorants, soap, toothpaste, shampoo, hair conditioner), sunscreen, cosmetics (cream, foundation, face powder, lipstick, blush, eye shadow, nail polish, perfume and after-shave lotion).

There are two trends regarding the use of engineered nanoparticles in cosmetics. First, a swift application of nanotechnology advances in the cosmetic industry, in addition to re-labeling of the products that already contain nanoparticles, so that they are more appealing to the consumers [141]. Second, targeting of cosmetic companies that use nanoparticles. For the general public and uninformed journalists there is not much of a difference between the various types of nanoparticles currently used in cosmetics, such as lipid based nanoparticles, fullerenes, silicon, etc. Everything labeled "nanoparticle" is considered dangerous to some. These trends result at least in part from the lack of regulations for testing of cosmetic products before they are sold to the public [55], unlike pharmaceutical products that are required to undergo several years of research before being considered safe. Despite the fact that many of the cosmetic companies claim safety related research, their results are not always disclosed to the public.

**Other consumer products.** Many consumer products incorporate nano or microparticles. A non-comprehensive list of currently available consumer products that incorporate nanotechnology can be found in reference [14]. The authors of this list make no distinction between nanostructured fixed structures, which are not likely to cause harm (an example is their listing of computer processors), and detachable or free nanoparticles, which can cause adverse health effects.

Titanium dioxide ($TiO_2$) particles with diameter larger than 100 nm are considered biologically inert in both humans and animals [142]. Based on this understanding, titanium dioxide nanoparticles have been widely used in many products, such as white pigment, food colorant, sunscreens and cosmetic creams [19]. However, adverse effects of titanium dioxide nanoparticles have recently been uncovered [143], [276], [277], [278], [279]. New research is exploring the potential use of nanostructured titanium dioxide photocatalyst materials for sterilizing equipment of environmental microorganisms in the health care facility [144].

Silver nanoparticles are used as antibacterial/antifungal agents in a diverse range of applications: air sanitizer sprays, socks, pillows, slippers, face masks, wet wipes, detergent, soap, shampoo, toothpaste, air filters, coatings of refrigerators, vacuum cleaners, washing machines, food storage containers, cellular phones, and even in liquid condoms [14].

Coatings of nanoparticles are widely used for modifying fabrics to create stain and wrinkle free properties. In addition, one can find clothes with built-in sunscreen and moisture management technology [14]. Fabric containing bamboo-charcoal nanoparticles claims antibacterial antifungal properties [14]. They are intended for use as face cloth masks, shoes insoles. Nano-coatings are applied to wetsuits for higher performance of athletes, or self-cleaning surfaces. Textiles with 30 nm embedded nanoparticles help prevent pollen from entering gaps in the fabric [14]. Nanoparticles or nanofibers are starting to be used in water-repellent, stain resistant plush toys, stain repellent mattresses [14]. Nano-sealant sprays for fabrics or leather, and hydrophobic nanoparticle solutions adhering to concrete, wood, glass, cloth, etc., allow the surfaces to deflect water [14]. The most peculiar applications of nanofibers and nanoparticles discovered in our literature review are: nanofibers that hide hair loss, and liquid condoms [14].

**Health effects.** All the health effects of the gamut of nanoparticles used in consumer products are not yet known, though nanotoxicology has revealed adverse health effects of materials previously considered safe. For example, silver, widely used as an antibacterial agent, proves to be toxic to humans or animal cells when in nanoparticle form, its cytotoxicity being higher than that of asbestos [113]. Inhalation of silver nanoparticles leads to their migration to the olfactory bulb, where they locate in mitochondria [20], as well as translocation to circulatory system, liver, kidneys, and heart [145]. Silver nanoparticles have been found in the blood of patients with blood diseases [146] and in the colon of patients with colon cancer [147].

A controversial subject is the association between the uptake of aluminum and Alzheimer's disease. Epidemiological studies researching the connection between aluminum in antiperspirants, antacids, or drinking water and Alzheimer's disease are conflicting, some finding positive associations and others none [148]. Due to their latent evolving nature and multi-part etiology, these neurological diseases are difficult to associate with specific factors. For example, the exposure takes place much earlier than the disease occurrence, hence the subjects may not recall possible exposure, their memory being already affected by the disease. Moreover, subjects that suffer from advanced neurodegenerative diseases are not likely to participate in epidemiological studies due to their reduced ability to communicate and remember [148]. In addition, multiple factors are known to contribute to Alzheimer's disease, such as: genetics, increasing age, endocrine conditions, oxidative stress, inflammation, smoking, infections, pesticides, electromagnetic fields [148].

In general, several questions arise related to the safety of nanoparticles as consumer products. Are they biocompatible? Do the nanoparticles enter the lymphatic and circulatory systems? If not, do they accumulate in the skin and what are the long-term effects of accumulation? Do they produce inflammation? If they enter the lymphatic and circulatory system, is the amount significant? What are the long-term effects of this uptake? Related to the beneficial antioxidant properties of some nanomaterials, long-term effect need to be studied, in addition to the short-term antioxidant effect. What is the long-term fate of these nanoparticles? Are they stored in the skin? Do they enter circulation? What happens when the nanoparticles undergo chemical reactions and lose their antioxidant properties? The answers to some of these questions are known, and will be presented in the chapter dedicated to nanoparticles toxicity, however most of the remaining questions still remain unanswered.

*3.2.6. Engineered nanomaterials and health effects*

The fabrication of nanomaterials is a broad and evolving field. Nanomaterials can be synthesized by many methods including: gas phase processes (flame pyrolysis, high temperature evaporation, and plasma synthesis); vapor deposition synthesis (electron, thermal, laser beam evaporation); colloidal, or liquid phase methods in which chemical reactions in solvents lead to the formation of colloids; and mechanical processes including grinding, milling and alloying. A review of nanomaterial fabrication processes is given in [149]. A critical fact to consider with engineered nanomaterials is that they can be synthesized in almost any shape and size by materials scientists. Several examples are given in Figures 19, 20, and 21. Nanostructured materials shown in Figure 19 are firmly attached to a substrate and do not pose a health risk as long as they do not detach from the substrate. Figure 20 shows nanostructured materials where nanostructures are free and can become airborne, consequently posing a potential health risk. In Figure 21 man-made nanoparticles engineered by glancing angle deposition [150], [151] are shown together with microorganisms, such as bacteria and viruses.

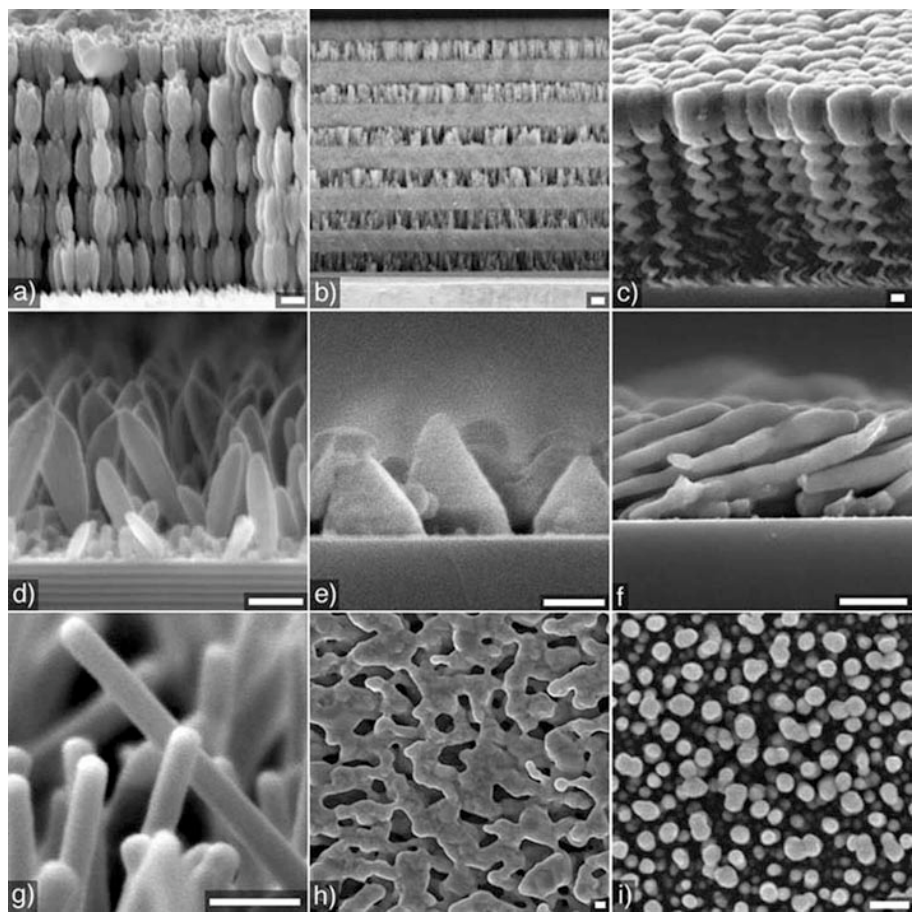

**Figure 19.** *Examples of nanostructured materials in thin film form, which are not toxic unless the nanoparticles get detached:* **(a)** *Si rugate filter [153],* **(b)** *Si 12-layered structure [154],* **(c)** *MgF$_2$ capping layered helical films [155],* **(d)** *Ti pillars [156],* **(e)** *Cu pyramids (unpublished),* **(f)** *Cu oblique columns [156],* **(g)** *ZnO nanowires, credit: Y. Lu, courtesy of National Science Foundation,* **(h)** *porous Ag [64],* **(i)** *porous Si [157]. The scale bars represent 100 nm.*

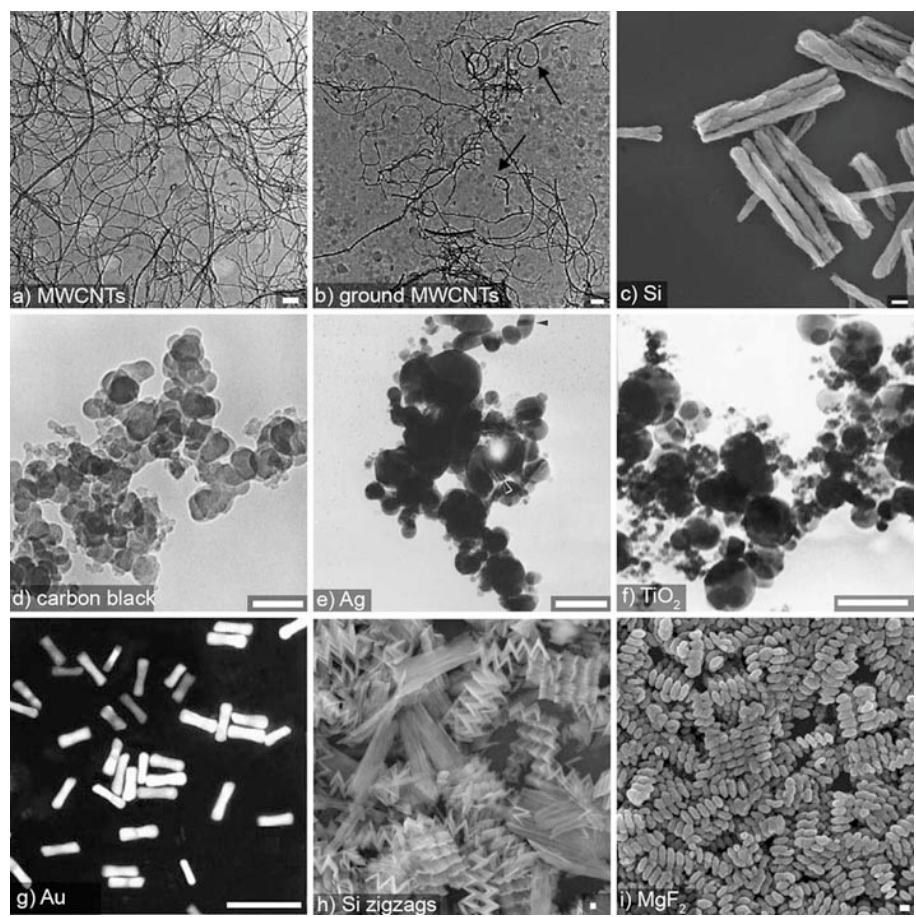

**Figure 20.** *Examples of free nanoparticles.* **(a)** *MWCNTs and* **(b)** *ground MWCNTs, [159], reproduced with permission from Elsevier.* **(c)** *Silicon rods (Kevin Robbie, unpublished).* **(d)** *Carbon black,* **(e)** *silver,* **(f)** *and titanium dioxide [113], reproduced with permission from Springer Science and Business Media.* **(g)** *Gold nanorods [160], courtesy of National Academy of Sciences of US.* **(h)** *Silicon zigzags (Kevin Robbie, unpublished).* **(i)** *Magnesium fluoride helices (Kevin Robbie, unpublished). The scale bar represents 100 nm.*

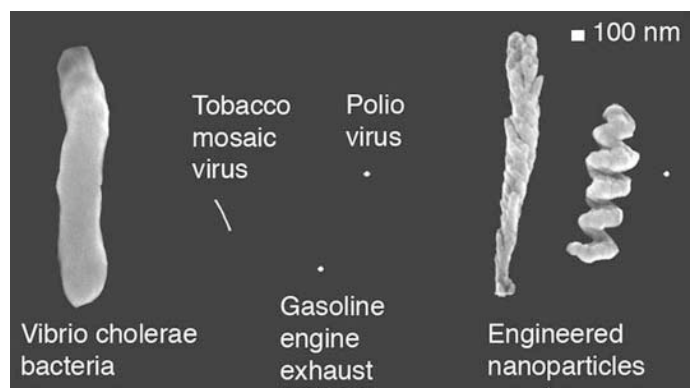

**Figure 21.** *Engineered nanoparticles [64] together with selected microorganisms, shown at equal magnification.*

**Health effects.** As a main focus of this paper, the adverse health effects of engineered nanoparticles will be discussed in Chapter 4. An important initiative by the National Institute for Occupational Safety and Health was the creation of an online nanoparticles information library that is updated with various compositions nanoparticles as well as with the known health effects of some nanoparticles [152]. As the fields of nanotechnology and nanotoxicology are developing so quickly, this is a great way to update the current knowledge on nanoparticles fabrication and toxicology.

### 3.3. Environmental and occupational exposure to toxic substances

### 3.3.1. Metals and other dusts

Small quantities of many metals, including copper, magnesium, sodium, potassium, calcium, and iron are essential for proper functioning of biological systems. At higher doses, however, metals can have toxic effects and exposure to high levels of environmental metals causes disease in humans [158]. The metals listed below in this paragraph are known to be toxic upon inhalation, ingestion or dermal exposure. Nanoparticles manufactured from these metals will have health effects not necessarily easily predicted from previous studies of non-nanoparticulate quantities of the same metals. As it could easily expose workers to these toxic materials, manufacturing of metal nanoparticles should be considered a serious occupational hazard.

The inhalation of metallic or other dusts is known to have negative health effects. The type of lung disease caused by dust inhalation depends on the nature of the material, exposure duration, and dose. The inhalation of some metal fumes (e.g. zinc, copper) may lead to metal fume fever, a influenza-like reaction [161]. Several metal dusts (e.g. platinum, nickel, chromium, cobalt) can lead to asthma [161], while inhalation of other metallic dusts can cause pulmonary fibrosis, and ultimately lung cancer. The percentage of lung cancers attributable to occupational hazards is about 15%, with exposure to metals being a major cause [161].

**Beryllium.** Beryllium alloys are used for making electrical and electronic parts, and molds for plastic. Inhalation can cause lung damage leading to a pneumonia-like syndrome called acute beryllium disease [162]. Beryllium exposure can also lead to hypersensitivity, and allergic reaction characterized by an inflammatory immune response to even tiny amounts of beryllium. Hypersensitivity can lead to chronic beryllium disease, where white blood cells accumulate around absorbed beryllium particles and form granulomas leading to anorexia, weight loss, cyanosis of the extremities, and heart enlargement [162]. Long-term exposure to beryllium causes cancer in animals and increased risk of lung cancer in humans [163].

**Lead.** Exposure to lead occurs through the air, household dust, food, and drinking water. Air-borne lead may be present in industrial emissions, such as those from smelters and refineries. Exposure to high levels of lead and its compounds can cause serious disability. At highest risk are workers involved in the manufacture of batteries, metals, and paints; the printing industry; or chronically exposed to lead dust (e.g. through sanding of surfaces coated with lead) or insecticides. Inhaled or ingested lead circulates in the blood and is deposited in bone and other tissue [158]. Following inhalation, about 50-70% of lead is absorbed into the blood, allowing it to circulate to most organs. Manifestations of lead intoxication include impairment of mental functions, visual-motor performance, memory, and attention span, as well as anemia, fatigue, lack of appetite, abdominal pain, and kidney disease, among others [158].

**Cobalt.** Diseases associated with exposure to cobalt are- asthma, acute illness (fever, anorexia, malaise, and difficulty breathing, resembling a viral illness), and interstitial pneumonitis [161], [162].

**Cadmium.** Cadmium is used in batteries, pigments, metal coatings, plastics, and is a by-product of the burning of fossil fuels and cigarettes. As a result of industrial and consumer waste, cadmium accumulates in soil at a rate increase by 1% per year [158]. Plants and feed crops growing in contaminated soil take up cadmium, leading to contamination of vegetables and animals. High-dose inhalation exposure leads to severe lung irritation, nausea, and vomiting. Long-term low-dosage exposure in humans causes lung emphysema, impairment of the immune system, central nervous system and liver damage [158]. Occupational exposure to cadmium has been linked to lung cancer in humans, some studies associating cadmium exposure with cancer of the liver, bladder and stomach, and possibly of pancreas [164].

**Aluminum.** Exposure to aluminum occurs through consumption of food and water, as well as usage of many products containing aluminum, including antacids and anti-perspirants. The use of antiperspirants combined with under-arm shaving is associated with an earlier age of breast cancer diagnosis [165]. Aluminum excess can lead to anemia, bone disease, and dementia [166]. Exposure to high levels of aluminum (and other metals, such as iron) is related to neurological disorders, such as dialysis encephalopathy, Parkinson dementia, and especially Alzheimer's disease [167]. Studies of brain plaques associated with Alzheimer's disease show abnormally high aluminum [158], but have not shown if this is a cause or effect of the disease. However, one can hypothesize that a critical mass of metabolical errors is important in producing Alzheimer's disease [168]. If aluminum can reach the brain via the olfactory bulb by passing the blood brain barrier, or via the circulatory system, then brain metabolical errors resulting from accumulations of this metal in parts of the brain could contribute to the onset of Alzheimer's disease [168]. Rats that received subcutaneous injection of aluminum glutamate show pathological signs similar to those observed in human Alzheimer's disease [169]. They show a significant increase of aluminum content in the brain (hippocampus, occipito-parietal cortex, cerebellum, striatum), and symptoms that include trembling, equilibrium instabilities, and convulsions, followed by death one hour after the injection.

**Nickel and chromium.** Nickel is used for the production of stainless steel and other nickel alloys with numerous applications. Occupational exposure to nickel via inhalation of dust and fumes is associated with cancers of lung and sinus [158]. Chromium derived from smelting has also been found to cause cancer.

**Manganese.** Manganese is both an essential nutrient and is known to have neurotoxic effects [148], [170]. At high levels, manganese exposure to contaminated water or through inhalation results in neurological impairment. Occupational exposure generally occurs only to those involved in mining and welding. An example of welding-generated nanoparticles is given in Figure 22 a. There is a clear association between manganese and neurological disease in miners exposed to $MnO_2$ dust [170]. The neurological disorder linked most closely to manganese is Parkinson's disease [148], [170]. Some welders develop Parkinson's disease much earlier in their life, usually in their mid forties, compared to the sixties in the general population [148]. Of concern for public health is the risk of neurological diseases emerging after long latencies in regions with only mildly elevated environmental manganese levels.

**Iron.** Iron is incorporated into numerous enzymes involved in cell division, DNA replication, and cellular metabolism, and it is essential for oxygen transport and gas exchange. As with manganese, low doses of iron are vital for survival. Several observations have been made linking cellular iron content to the development of cancers [97]. In studies of animals administered excessive amounts of iron, orally and by injection, an increased risk of of adenocarcinomas, colorectal tumors, hepatomas, mammary tumors, mesothelioma, renal tubular cell carcinomas, and sarcomas was observed. In humans, injection of iron compounds has been shown to cause sarcomas at the sites of deposition. Patients with hemochromatosis (genetic disease characterized by increased iron absorption) have an enhanced susceptibility to liver cancer. The accumulation of iron in brain regions with decreased function, and cell loss has been observed in many neurological diseases, such as Parkinson's disease, Alzheimer's disease, etc. [171]. Inhalation of iron dust causes a respiratory disease called pneumoconiosis [161].

**Organic dust.** Organic dusts originate from animals and/or plants and contain fragments and fibres from wood, bone, fur, skin, leather, brooms, flour, grains, tobacco, carpets, paper, etc. Organic dust from these various sources irritates the upper respiratory system, eyes, and skin, causing bronchitis, allergic reactions, asthma, conjunctivitis, and dermatitis [158].

**Silica.** Exposure to silica, or silicon dioxide ($SiO_2$), the main constituent of sand and granite, produces silicosis, a disabling pulmonary fibrosis. A controversial subject in occupational medicine is the association of silicosis with lung cancer [158]. In addition, exposure to silica is associated to autoimmune diseases including: scleroderma, rheumatoid arthritis, systemic lupus erythematosus [172].

**Coal and coal ash.** Coal dust produces pneumoconiosis in coal miners, their lungs retaining a considerable amount of dust, of up to 30 g (roughly two table spoons of dust) [173]. Epidemiological study on more than 500 chimney sweeps showed an increased number of deaths due to heart and respiratory diseases, lung, esophageal, and liver cancer [174].

**Asbestos.** Asbestos is a naturally occurring fibrous material consisting of very long chains of silicon and oxygen (polysilicate or long chain silicate). A SEM image of asbestos can be seen in Figure 22 b. Asbestos fibers have high tensile strength, flexibility and have flame retardant and insulating properties. In ancient times, asbestos was woven and used "in fabrics such as Egyptian burial cloths and Charlemagne's tablecloth, which according to legend he threw in a fire to clean." (Wikipedia, The Free Encyclopedia). Due to its desirable properties it was once used extensively in construction materials (cement, floors, roofing, pipe insulation, and fire-proofing) and in materials industry (brake pads) [175]. Asbestos exposure occurs when its handling produces small fibers, nanoparticles, that are easily carried as a suspension in both air and water where they are absorbed by inhalation and ingestion. Studies of occupational health show that exposure can cause lung cancer and mesothelioma (a rare cancer of the membranes lining the abdominal cavity and surrounding internal organs) [158]. Recent studies in a community with occupational and environmental exposure to asbestos showed increased risk of autoimmune diseases, such as: systemic lupus erythematosus, scleroderma, rheumatoid arthritis [172], [176]. These diseases affect connective tissues, skin, and organs.

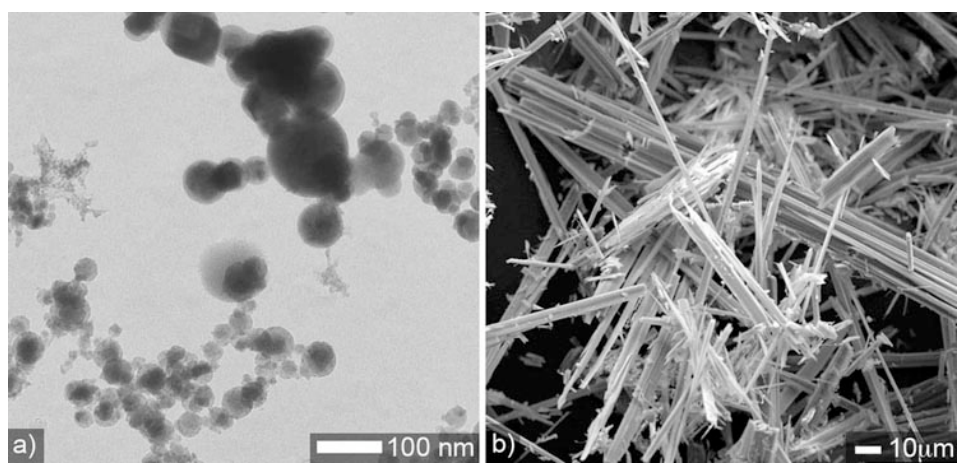

**Figure 22.** (**a**) *TEM of welding nanoparticles, courtesy of Pam Drake, National Institute for Occupational Safety and Health NIOSH.* (**b**) *Asbestos fibers, courtesy of the U.S. Geological Survey.*

**Polymer fumes.** Humans exposed to polytetrafluoroethylene (or Teflon, PTFE) and other polymer fumes develop an influenza-like syndrome (polymer fume fever). The symptoms occur several hours after exposure, and include chest pain, fever, chills, sweating, nausea and headache [177]. Severe toxic effects, like pulmonary edema, pneumonitis and death, are also possible [178].

### 3.3.2. Carcinogens and poorly soluble (durable) particles

It is clear that some types of particles cause cancer, but it is not known which characteristics of the particles are responsible for their carcinogenicity. Some particles are inherently toxic, such as metal dust, welding fume, and quartz dust, while other particles have a much lower toxicity, but still cause toxic effects under some circumstances. The latter category includes poorly soluble particles, biodurable particles without known specific toxicity that include: diesel exhaust particles, carbon black, coal-mine dust, titanium dioxide, and several others listed in Table 2 [39]. Poorly soluble particles have been shown to cause cancer in rodents, however epidemiologic studies do not clearly indicate increased cancer rates in humans exposed to these particles. The latest research on nanoparticles shows that they can exhibit more pronounced toxicity than larger microparticles, suggesting that environmental and health regulating agencies must take more consideration of particle size distribution, shape, and agglomeration when establishing regulatory exposure guidelines.

**Table 2.** *Particles with proven lung carcinogenic effects in animals and/or humans (adapted from ref. [39]). Some poorly soluble particles are shown to be carcinogenic only in rodents, while epidemiological studies do not clearly indicate human carcinogenicity. Classification in animal studies was done as follows: "+" means positive in more than one animal during inhalation studies, "-" means negative or no inhalation studies, "+/-" indicates inadequate evidence in rats, and "blank" indicates not decided yet.*

| | | Carcinogenic effect | |
|---|---|---|---|
| **Particle** | **Use/exposure** | **Rat** | **Human** |
| Particulate matter ($PM_{0.1}$, $PM_{2.5}$, $PM_{10}$) | Ambient | | Possibly carcinogenic? (unknown fraction) |
| NiO | Exhaust | + | Carcinogenic |
| Quartz (crystalline silica) | Constructions | + | Carcinogenic |
| Asbestos | Insulation, mining | + | Carcinogenic |
| Carbon black | Pigments, toner, tires | + | Possibly carcinogenic |
| Refractory ceramic fibers | insulation | + | Possibly carcinogenic |
| Wood dust | Furniture making, saw mills | +/- | Carcinogenic (some types) |
| $TiO_2$ | Pigments, sunscreens | + | |
| Diesel exhaust | Engines, cars | + | |
| Talc | Cosmetics, mining | + | |
| Volcanic fly ash | Ambient | + | |
| Coal mine dust | Mining | + | Not classifiable |
| Rockwool | Insulation | + | Not classifiable |
| Iron oxides | Pigments, paramagnetic diagnostics | +/- | |
| Graphite | Occupational | +/- | |
| Cement | Constructions, buildings | - | Not classifiable |
| Amorphous silica | Cleaning, paints, drugs | - | Not classifiable |

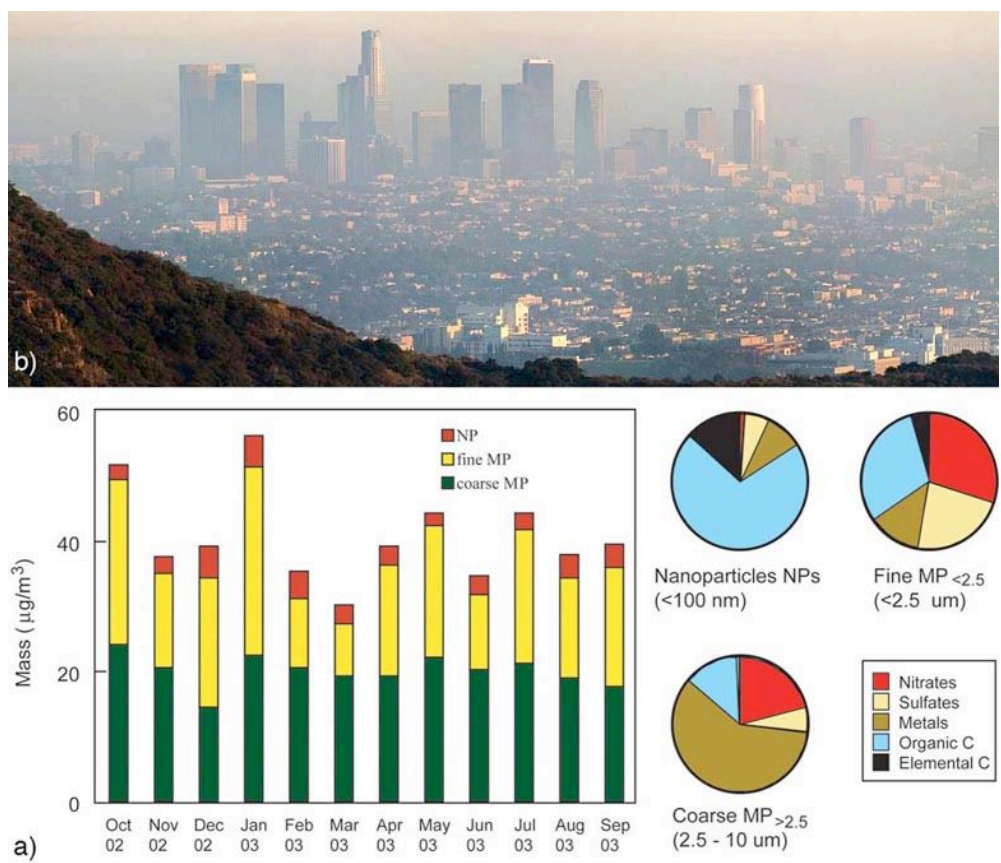

**Figure 23.** *a) Los Angeles smog. b) Size distribution and composition of particulate matter over Los Angeles during 2002-2003. NP - nanoparticles, MP – microparticles (after [123]).*

### 3.4. Aerosol pollution, monitoring, and health effects

### 3.4.1. Aerosols size and composition

Aerosol pollution is a combination of particulate matter and gaseous and liquid phases from natural and anthropogenic sources. Ambient particulate matter is generally classified according to three size distributions: nanoparticles smaller than 100 nm in diameter (mainly resulting from combustion), accumulation mode particles between 100 nm and 2.5 µm in diameter (from aggregation of smaller particles and vapors), and coarse-mode particles larger than 2.5 µm (mostly mechanically generated) [24]. These three particle categories have distinct chemical compositions (Figure 23), sources, and lifetime in the atmosphere. The larger particles, which settle faster due to gravity, are removed from the atmosphere fastest. Smaller particles are transported over greater distances and have longer lifetimes in the atmosphere [24]. Nanoparticles usually form atmospheric fractal-like dendritic aggregates similar to the soot in Figure 16 d. The polydispersity (variation in particle sizes) varies with the source, for example, primary particles in diesel aggregates ranging from 10 to 40 nm. Atmospheric measurements show that nanoparticles make up a small portion of the particulate matter mass concentration compared to microparticles [123]. However, the number concentration of nanoparticles is significantly larger than the number of microparticles.

Combustion-derived carbon particles, with traces of transition metals, make up about 50% of the mass of typical urban particulate matter, while the remaining 50% includes salts, geological dust, and organic matter [49]. As shown in this study of particulate matter in Los Angeles (Figure 23) [123], when sorted by size, we see that the particles vary considerably in composition, with the smallest nanoparticles being mostly carbon (organic and elemental), while the larger microparticles are mostly metal. In general, environmental pollution particles differ in their quantities of nitrates, sulfates, crustal materials, and carbon, with blown soil a major source in rural areas. Due to the high chemical reactivity of atmospheric nanoparticles (resulting from their high surface area), they are very likely to interact with water or other chemicals in the atmosphere to form new species. This dynamic nature of aerosol nanoparticles means that their environmental impact will be long and complex, as reactions create a cascade of products with varying effects – while some particles will be long-lived, or persistent, others may experience transformations to more or less damaging states.

*3.4.2. Aerosols concentration. Air quality index*

Nanoparticles with size smaller than 100 nm are present in large numbers in typical ambient air with a level ranging between 5,000-10,000 particles per ml, increasing during pollution episodes to 3,000,000 particles/ml [49]. Their concentration varies from region to region, as well as from season to season. Nanoparticles smaller than 100 nm make up about 70% of the total number of a ambient aerosols in urban areas, while their mass contribution is only about 1% [180]. In certain parts of the world the peak number concentration or airborne nanoparticles was found to increase over time. For example, in California, the peak concentration of nanoparticles in January 1999 ($1.45 \times 10^{11}$ particles/m$^3$) was found to be three times higher than previously measured peaks [181]. At the other extreme are modern cleanroom facilities where air particles are almost eliminated through careful design of airflow and filtering, and meticulous elimination of potential particle sources. A typical cleanroom, with Class 10 or ISO 3 particle levels has only several hundred 100 nm particles per cubic meter.

Increased awareness of the influence of particle size and shape on health impact has led the Environment Protection Agency to propose new ambient standards on fine particles smaller than 2.5 microns. The Air Quality Index (AQI) is a standard measure used by the Environmental Protection Agency for monitoring daily air quality [123]. It quantifies air pollution and predicts health effects of concern that may be experienced within a few hours or days of exposure to polluted air. The calculation of the AQI includes five major pollutants: particulate matter, ozone, carbon monoxide, sulfur dioxide, and nitrogen dioxide, all of which are regulated under the Clean Air Act. The AQI has not been standardized internationally, and other countries use different systems for describing air quality [123], [182]. The AQI values for particulate matter are shown in Table 3.

**Table 3.** *Air quality index, AQI, values for concentrations of particulate matter with diameters smaller than 2.5 (MP$_{<2.5}$) and 10 microns (MP$_{<10}$) [123].*

| AQI | MP$_{<2.5}$ ($\mu$g/m$^3$) | MP$_{<10}$ ($\mu$g/m$^3$) | Air quality descriptor |
|---|---|---|---|
| 0-50 | 0.0-15.4 | 0.54 | Good |
| 51-100 | 15.5-40.4 | 55-154 | Moderate |
| 101-150 | 40.5-65.4 | 155-254 | Unhealthy for sensitive groups |
| 151-200 | 65.5-150.4 | 255-354 | Unhealthy |
| 201-300 | 150.5-250.4 | 355-452 | Very unhealthy |

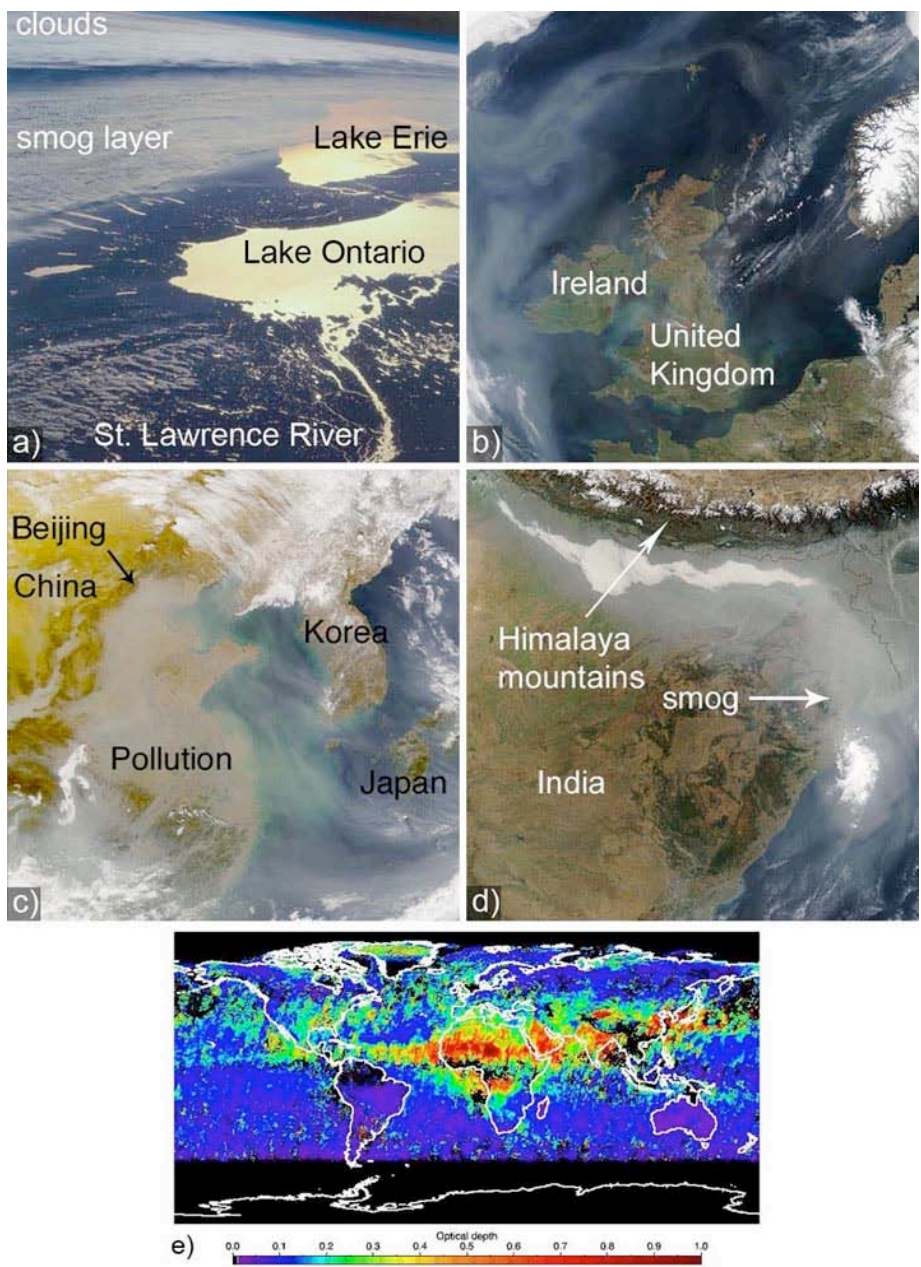

***Figure 24.*** *Images of pollution over the world. While clouds appear solid-white, pollution appears as a misty semi-transparent gray that masks the image's geographic and aquatic features. **(a)** Smog layer over upstate New York and North-Eastern Ontario, courtesy Earth Science and Image Analysis Laboratory, Johnson Space Center. **(b)** Dust from the Sahara Desert, air pollution, and smoke lingers over the Atlantic Ocean and Great Britain in April 2003 [72], credit Jacques Descloitres, MODIS Rapid Response Team, NASA/GSFC. **(c)** Pollution in China blowing east towards the Yellow Sea, Korea and Japan. Beijing, China's capital, lies under the densest portion of the aerosol pollution, credit NASA-GSFC [183]. **(d)** Pollution over India. Haze follows the course of the Ganges River in northern India, flowing eastward along the Himalaya Mountains, before turning south and spreading out in the Indian Ocean, credit Jacques Descloitres, MODIS Rapid Response Team, NASA/GSFC [183]. **(e)** Optical depth showing worldwide concentration of aerosols on June 2005, derived from data taken by MISR, NASA's*

*first Earth Observing System (EOS) spacecraft, launched on December 18, 1999. The MISR instrument orbits the Earth about 15 times each day, observes the Earth continuously from pole to pole, and every 9 days views the entire globe between 82 degrees north and 82 degrees south latitude [184].*

### 3.4.3. Satellite monitoring of aerosol concentration and size

Aerosols play an important role in the global atmosphere, directly influencing global climate and human health. Dust, smoke, and haze locally impair visibility and health in both urban and rural regions. Anthropogenic aerosol nanoparticles are especially abundant in the atmosphere, and they constitute a significant uncertainty factor in estimating the climatic change resulting from human pollution [67]. Satellite images clearly show particulate matter from both anthropogenic and natural sources in industrialized and heavily populated parts of the world (Figure 24 a-d) [183]. Atmospheric aerosols are monitored worldwide via satellites, and several years worth of measured global aerosol maps are available from NASA's MISR (Figure 24 e) [184]. Global aerosol data is measured by imaging sequential columns through the atmosphere below the satellite as it orbits the earth, in each of 4 wavelengths (blue, green, red, and near-infrared). These measures also give some indication of particle size and shape, from the variation in scene brightness over several different view angles and wavelengths. The MISR results distinguish desert dust from pollution and forest fire particles: desert dust particles and sea salt are usually larger than aerosols originating from the processes of combustion e.g. forest fires and burning of fossil fuels. MISR can help to determine ground-level pollution concentrations necessary in understanding and assessing links between pollution exposure and human health. A full assessment of the impact of pollution aerosol exposure will require records of aerosol mapping for several decades - the typical timescale of pollution-linked disease appearance.

### 3.4.4. Health effects associated to air pollution

Human exposure to inhaled ambient particles can have adverse health effects [19], [117], [158], [185], [186]. Pulmonary and cardiovascular diseases result when inhaled particles interfere with the normal function of bodily systems [48], [187], [188]. The health consequences of particle inhalation vary greatly with particle composition, concentration, etc., from benign candle wax to carcinogenic asbestos, or tobacco smoke.

As our understanding of nanoparticles has grown, so has our knowledge of disease resulting from their exposure. Until recently it was believed that particles 10 microns or smaller were responsible for disease resulting for particle pollution. But further study has shown that most of these diseases are caused by particles smaller than 100 nm, similar in size to viruses. Nanoparticles seem to be generally more toxic than microparticles, primarily do to their ability to penetrate living cells, translocate within the body, and affect the function of major organs.

**Cardiovascular diseases.** The correlation between ambient particles exposure and heart disease was accepted in the mid-nineties, when it was observed that hospital admission for cardiovascular illness increased on days with high concentrations of particles [189]. Atmospheric particle pollution from automobile exhaust seems to have a major influence on mortality, with a strong association between increased cardiopulmonary mortality and living near major roads [117]. The risk of myocardial infarction onset increases with elevated concentrations of particulate matter smaller than 2.5 μm in the day before onset and with volume of vehicular traffic. Cardiovascular diseases and effects associated with particulate pollution include: ischemic heart disease, hypertensive heart disease [190], arrhythmia, heart failure, arteriosclerosis, brachial artery vasoconstriction, and increased blood pressure in subjects with lung disease [117].

**Respiratory illnesses.** Pneumonia, bronchial asthma, chronic bronchitis, emphysema, lung cancer, acute deterioration of lung function, and hospital admissions for respiratory illnesses were all found to increase with higher levels of pollution [191], [192].

**Malignant tumors.** An epidemiological study researching the effects of chronic exposure to particulate matter smaller than 10 μm in nonsmoking subjects revealed a high incidence of lung cancer [193]. This study also showed an 8% increase in risk of lung cancer for each 10 μg/m$^3$ increase in particulate matter smaller than 2.5 μm [190]. To some surprise, levels of particulate matter smaller than 2.5 μm pollution were also found to correlate significantly with cancers of the breast, endometrium, and ovary [192], an effect that might be explained by recent studies of nanoparticles translocation to organs. Childhood cancers were also found to be strongly determined by prenatal or early postnatal exposure to oil-based combustion gases, primarily engine exhaust [118].

**Mortality and morbidity.** There is compelling evidence of correlation between particle pollution levels on a given day, and overall mortality the following day [117], [158]. Epidemiological studies have shown that the increased morbidity and mortality, correlated with increased particle pollution, are frequently the result of respiratory problems [97], but primarily due to cardiovascular diseases [158], [194]. In 1998 it was estimated that around 4000 deaths were related to atmospheric pollution in Canada. These deaths occur mainly in heavily industrialized urban centers [158].

Analysis of mortality statistics for approximately 500,000 adults residing in the United States of America covering a 16 year period of chronic exposure to air pollutants shows that cardiovascular deaths increased by 0.69% for each 10 μg/m$^3$ increase in particulate matter [117], [190]. The study found a strong correlation between a cause of death of either cardiopulmonary disease or lung cancer, and levels of particulate matter smaller than 2.5 μm [190].

Figure 25 shows the correlation of mortality rates with extreme levels of pollution during London smog episodes of the 1950's through the 1970's [20]. The exposure–response observations of daily mortality exhibit two distinct regions, with a steeper slope at lower mass concentrations and a shallower slope at higher. It has been suggested that a high concentration of aerosol nanoparticles would promote particle aggregation [145]. Aggregation of nanoparticles at high particle concentrations reduces toxicity by decreasing the reactive surface area and possibly limiting the translocation of the particles.

**Postneonatal infant mortality and birth defects.** Positive associations between exposure to particles and selected birth defects (such as atrial septal defects) were reported in studies in various countries [195], [196]. It was found that outdoor air pollution above a reference level of 12.0 μg/m$^3$ of particulate matter smaller than 10 μm contributes substantially to postneonatal infant mortality in infants born with a normal birth weight [197].

**Exacerbation of pre-existing diseases and other risks.** Certain segments of the population appear to be at greater risk to the toxic effects of particulate pollution. Patients suffering of various diseases, such as: diabetes, chronic pulmonary diseases, heart diseases, or with previous myocardial infarction are likely to suffer an increase in the severity of symptoms on days with high levels of pollutants [48], [117]. In addition, the presence of inflammation may enhance the translocation of nanoparticles into circulation [30], [255], [256], [259], [266], or via blood-brain-barrier [18], [203].

**Cumulative exposure** In addition to immediate effects, time-series studies have shown cumulative effects over weeks, associated with elevated particle concentrations [48]. Further studies are needed to assess the health effects of chronic exposure to nanoparticles.

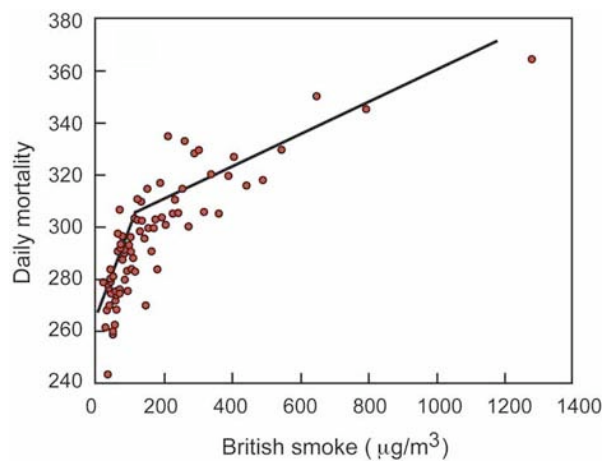

**Figure 25.** *Correlation between daily mortality rate and urban particle concentrations during the London smog episodes in the winters of 1958-1972 (data from [198]). Also shown are the regression lines for the steep and shallow slopes together with the inflection point at ~125 µg/m³ (after [20]).*

**Treatment.** Ambient particles induce oxidative stress in biological systems, either directly by introducing oxidant substances, or more indirectly by supplying soluble metals, including transition metals, that shift the redox balance of cells toward oxidation. Oxidative stress is believed to be the primary mechanism by which nanoparticles generate disease. Consequently, dietary nutrients that play a protective role in the oxidative process are suggested as potential mitigators of the toxic effects of nanoparticle pollution. Antioxidant vitamins (such as vitamin C) have a protective effect against lung diseases, and a high intake of fresh fruit and some vegetables appears to have a beneficial effect on overall lung health [199] perhaps due to reducing the toxic effects of environmental nanoparticles. Treatment of underlying health conditions also reduces the impact of air pollution [199].

## 4. Nanotoxicology - toxicology of nanoparticles

### 4.1. Respiratory tract uptake and clearance

#### 4.1.1. Particle size dependent inhalation

After inhalation, nanoparticles deposit throughout the entire respiratory tract, starting from nose and pharynx, down to the lungs [37], [200]. Lungs consist of airways, that transport air in and out, and alveoli, which are gas exchange surfaces, as shown in Figure 26 c, d. Human lungs have an internal surface area between 75-140 m² and about 300 million alveoli [30]. Due to their large surface area, the lung is the primary entry portal for inhaled particles.

Spherically-shaped solid material with particle diameters smaller than 10 microns can reach the gas exchange surfaces (Figure 26 d) [30], [37]. Larger diameter particles tend to be deposited further up in the respiratory tract as a result of gravitational settling, impaction, and interception [201]. Many larger-diameter fibers are deposited at "saddle points" in the branching respiratory tree. Smaller diameter particles are more affected by diffusion and these can collect in the smaller airways and alveoli. Fibres having a small diameter may penetrate deep into the lung, though very long aspect ratio fibres will remain in the upper airways [30].

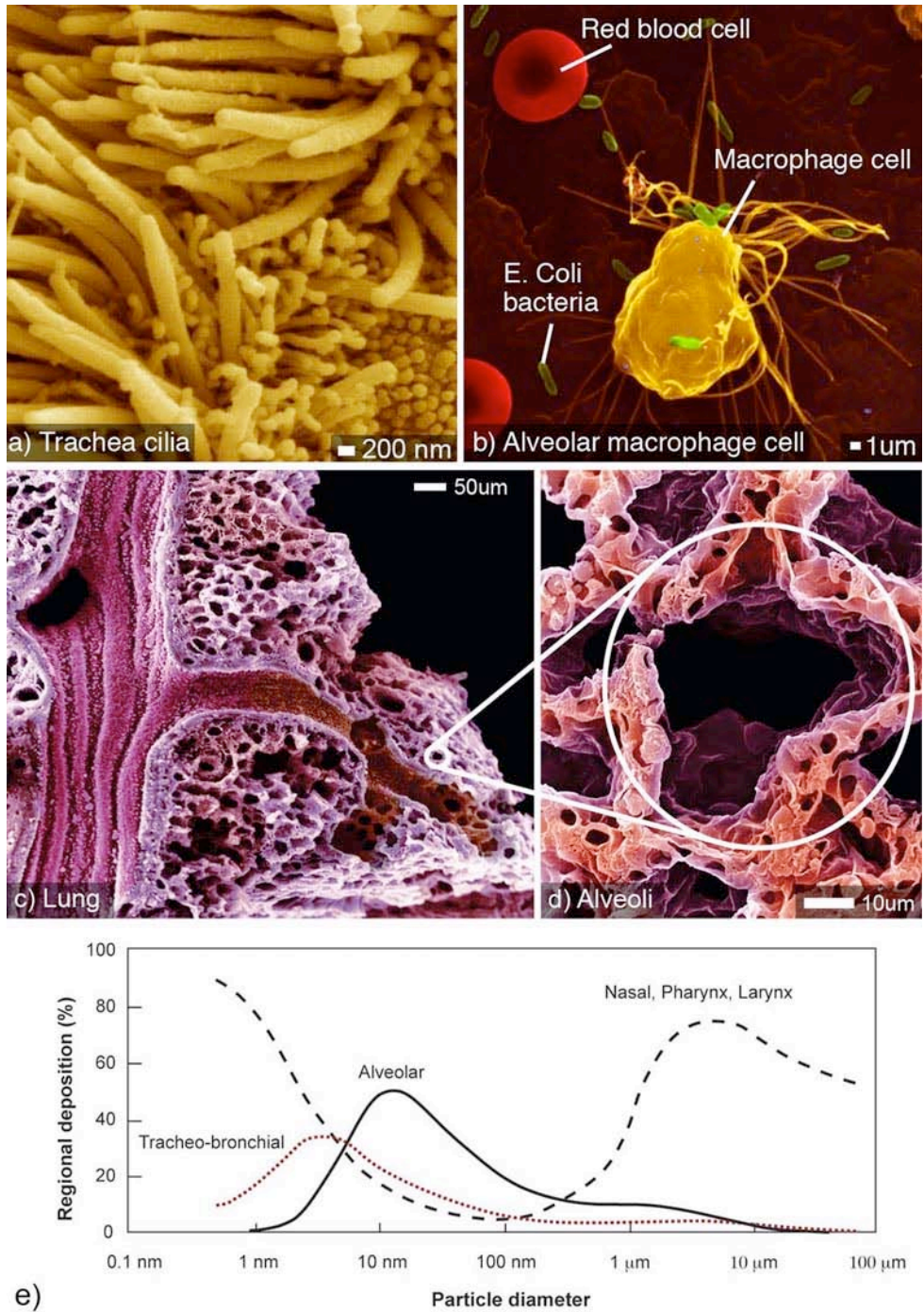

**Figure 26. (a)** *SEM image of lung trachea epithelium, showing cilia (mucociliary escalator), courtesy Louisa Howard.* **(b)** *Human alveolar macrophage (center, yellow) phagocytosis of Escherichia coli (multiple ovoids, green), together with a red blood cell (red). © Dr. Dennis Kunkel/Visuals Unlimited, reproduced with permission from Visuals Unlimited [25].* **(c), (d)** *Alveoli in the lung. © Dr. David M. Phillips/Visuals Unlimited, reproduced with permission from Visuals Unlimited [25].* **(e)** *Deposition of inhaled particles in the human respiratory tract versus the particle diameter, after [37].*

As shown in Figure 26 e, the nasopharyngeal region captures mainly microparticles and nanoparticles smaller than 10 nm, while the lungs will receive mainly nanoparticles with diameter between 10 – 20 nm [37].

### 4.1.2. Upper airway clearance – mucociliary escalator

Pulmonary retention and clearance of particles has been under study for many years. The nineteen-fifties were marked by a great interest in pneumoconiosis and studies of the effects of inhalation of radioactive particles, while in the nineties studies of occupational and environmental particles generated a considerable amount of knowledge regarding the adverse health effects of nano and microparticles in the respiratory tract [36].

The clearance of deposited particles in the respiratory tract is by physical translocation to other sites, and chemical clearance. Chemical dissolution in the upper or lower respiratory tract occurs for biosoluble particles in the intra-cellular or extra-cellular fluids, and will not make the subject of further discussions in this review. Non-soluble particles will undergo a different much slower clearance mechanism that we will discuss further in detail later. For relatively insoluble particles, the elimination process is very slow in comparison to soluble nanoparticles [145].

In the upper airways, particle clearance is performed mainly by the mucociliary escalator [36]. The first contact of inhaled nanoparticles in the respiratory tract is with the lining fluid, composed of phospholipids and proteins [202]. This contact leads to particle wetting and displacement towards the epithelium by surface forces from the liquid-air interface [203]. When in contact with esophageal epithelial cells, nanoparticles uptake by these cells is possible in the presence of pre-existent inflammation [204]. The cilia of the bronchial epithelial cells (Figure 26 a) move the covering mucous layer, including particles, away from the lungs and into the pharynx, a process generally requiring up to several hours. The nanoparticles that are cleared from the lung via the mucociliary escalator enter the gastro-intestinal tract [145], [206]. The clearance from the gastro-intestinal tract will be discussed in chapter 4.7. The mucus layer contains protective antioxidants, which can become depleted when a large number of oxidative compounds are inhaled [202].

### 4.1.3. Lower airways clearance – phagocytosis and passive uptake

**Phagocytosis.** Particles smaller than 10 microns can reach the lower airways [207]. Particle clearance from the lungs alveoli occurs primarily through macrophage phagocytosis. Macrophages are cells that act as vehicles for the physical removal of particles from alveoli to the mucociliary escalator or across the alveolar epithelium to the lymph nodes in the lung or to those closely associated with the lungs [203]. When the lung is subject to prolonged exposure, white blood cells from the circulatory system (neutrophils) are recruited to help.

Phagocytes engulf and break down pathogenic microorganisms, damaged or apoptotic cells, and inert particles [209]. In addition to the "professional cleaners", phagocytes (neutrophils and monocyte/ macrophages, see Figure 26 b), most cells also have some phagocytic ability [209]. The main difference between the phagocytic ability of professional and non-professional phagocytes is related to the presence of dedicated receptors able to recognize molecules pertaining to pathogens, molecules very different from those found in the human body [209]. Phagocytosis is a very complex mechanism due to the diversity of receptors, its understanding requiring thorough knowledge of chemical processes at molecular level. Many phagocytic receptors serve a dual function, adhesion and particle internalization [209]. The phagocytosis of particles is more effective if the particles are labeled with special molecules (such as antibodies or complement molecules) able to speed-up phagocytosis, a labeling process called opsonisation. Opsonins are present in the lung-lining fluid [207]. Hydrophobic particles will be readily coated by opsonins and subsequently available for phagocytosis [208]. Coating of particles with hydrophilic polymers, such as polyethylene glycol, diminishes the opsonisation of particles, consequently decreasing the probability of being phagocytized [208]. However, unopsonized particles are nevertheless eventually phagocytized by macrophages [210].

Phagocytosis takes up to several hours and involves several steps:

1) First, specific receptors on the phagocyte membrane bind with specific molecules (ligands) localized on the surface of particle [209], [210]. Older studies suggest that the opsonisation with complement protein 5a may be responsible for the chemotactic (pertaining to the movement of a cell in a direction corresponding to a concentration gradient of a chemical substance) signal of nanoparticles [20], while newer studies propose the electric charge may play a role in activating the scavenger-type receptors for certain type of nanoparticles (such as: titanium dioxide, iron oxide, quartz) [211]. For uncharged nanoparticles, such as carbon based (diesel exhaust), some authors suggest that toll-like receptors are responsible for the recognition of these nanoparticles (as well as bacteria, virus, and fungi) [212].

2) After the binding of the phagocyte receptor with a ligand, the cytoskeleton (a network of protein filaments) of the phagocyte rearranges, resulting in pseudopod formation, and ultimately leading to internalization of the particle with the formation of a phagocytic vesicle (phagosome) [213].

3) The phagosome fuses with a lysosome (an organelle containing digesting enzymes), forming a phagolysosome. The fusion process can take from 30 minutes up to several hours, depending on the chemical interaction between the surface of the particle and the phagosome membrane [209]. Lysosoms release protease (which break down proteins) and NADPH oxidase (oxygen radicals) [213]. This process assists in the chemical dissolution of the particle [213]. Depending on the type of receptor used in the detection of the particle, macrophages may also release intercellular chemical messengers alerting the immune system that an infection is present.

4) If the particle is digested by lysosome enzymes, the residues are removed by exocytosis (release of chemical substances into the environment). If not, phagocytosis is followed by gradual movement of macrophages with internalized particles towards the mucociliary escalator, a process that can last up to 700 days in humans [20]. If the macrophage is unable to digest the particle and the particle produces damage to phagosomal membrane due to peroxidation, the oxidative compounds will likely interact with macrophage's cytoskeleton, and lead to reduced cell motility, impaired phagocytosis, macrophage death [214], and ultimately reduced clearance of particles from the lung [215]. Macrophage death can lead to release of oxidative lysosome compounds outside the cells. If particles cannot be cleared they can kill successive macrophages attempting to clear them, and create a source of oxidative compounds, and inflammation with macrophage debris accumulation (pus). Oxidative stress is associated to various diseases, such as cancer, neurodegenerative, and cardiovascular diseases.

This mechanism of alveolar clearance is not perfect, as it allows smaller nanoparticles to penetrate the alveolar epithelium and reach the interstitial space [20]. From the interstitial space nanoparticles may enter the circulatory and lymphatic systems and reach other sites throughout the body [24], [145].

Phagocytosis occurs in different areas of the body, phagocytes present in lungs, spleen, liver, etc., having different names, according their location, such as alveolar macrophages, splenic macrophages, Kupfer cells, respectively [82].

*4.1.4. Nanoparticle size dependent phagocytosis*

Human alveolar macrophages measure between 14 to 21 μm, while rat alveolar macrophages measure between 10 to 13 μm [38]. Macrophages can engulf particles of a size comparable to their own dimensions, but are significantly less effective with particles that are much larger or smaller. Experimental data show that, compared with larger particles, nanoparticles smaller than 100-200 nm are more capable of evading alveolar macrophages phagocytosis [203], entering pulmonary interstitial sites, and interacting with epithelial cells to get access to the circulatory and lymphatic systems [20], [145].

There are contradictory reports related to the phagocytosis of nanoparticles smaller than 100 nm. *In vitro* studies show that nanoparticles activate, and are phagocytized by alveolar macrophages [20]. However, macrophage lavage recovery studies show that nanoparticles smaller than 100 nm are not efficiently phagocytized in comparison with particles between 1-3 μm [20], [145]. A twelve-week inhalation study in rats showed that 20 nm nanoparticles of titanium dioxide are characterized by longer retention time in the lungs and increased translocation to interstitial sites than larger nanoparticles (250 nm) of the same material [216]. Small nanoparticles that evade the alveolar macrophages penetrate the alveolar epithelium, resulting in a slower clearance rate from the lung and possibly later translocation to the circulatory and lymphatic system.

*4.1.5. Concentration dependent phagocytosis*

At high concentrations, nanoparticles tend to cluster, forming aggregates often larger than 100 nm. Larger nanoparticles (>100 nm) can be readily phagocytized by alveolar macrophages [145], [217]. Results of studies involving inhalation or intratracheal instillation of high concentrations of nanoparticle (silver, iron, India ink, or titanium dioxide) smaller than 100 nm, which aggregate in larger particles, suggest that most nanoparticles are indeed stopped by alveolar macrophages [145]. Rat studies based on inhalation of low concentrations of 15 nm diameter silver nanoparticles showed that soon after inhalation (30 minutes), nanoparticles are distributed in the blood and brain, and subsequently to organs, such as heart, kidney, while the lungs rapidly cleared of the nanoparticles [145]. Hence, minute concentrations of nanoparticles with size smaller than 100 nm can have a higher probability of translocating to the circulatory system and organs (and produce damage) than high concentrations of the same particles, which are likely to form aggregates, and which will be stopped from translocation by macrophage phagocytosis.

*4.1.6. Lung burden*

Insoluble particle burden in the lung can induce a range of toxicological responses differing from those due to soluble particles [39]. Particles that are soluble or partly soluble (for example, cement) will dissolve in the aqueous fluid lining the epithelium (and pass into the circulatory and lymphatic systems), while the insoluble ones (such as carbon black) must be removed through other mechanisms such as the mucociliary escalator. Particles that are not soluble or degradable in the lung will rapidly accumulate upon continued exposure, as shown in Figure 27 for carbon black, asbestos, multiwall carbon nanotubes, and grounded carbon nanotubes [159]. If the macrophage clearance capacity is exceeded, then the lung defense mechanisms are overwhelmed, resulting in injury to the lung tissue.

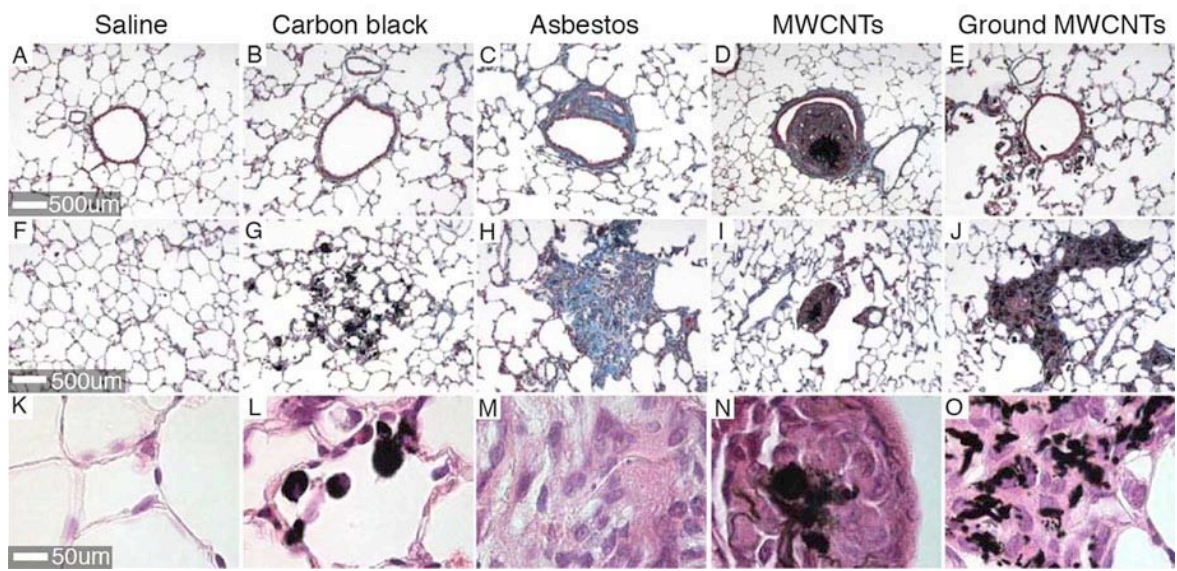

**Figure 27.** *Rat lung lesions induced by nanoparticles of: (B,G,L) carbon black, (C,H,M) asbestos, (D,I,N) multiwall carbon nanotubes, and (F,J,O) grounded nanotubes compared to saline solution (A,F,K) [159]. Reproduced with permission from Elsevier.*

The adverse effect of inhaled nanoparticles on the lungs depends on the lung burden (determined by the rate of particle deposition and clearance) and on the residence time of the nanoparticles in the lung [39], [203]. For example, carbon nanotubes are not eliminated from the lungs or very slowly eliminated (81% found in rat lungs after 60 days) [159]. The persistent presence within the alveoli of inhaled particles (Figure 27), especially those with mutagenic potential, increases the risk of lung cancer [39].

*4.1.7. Translocation and clearance of inhaled nanoparticles*

Inhaled nanoparticles are shown to reach the nervous system via the olfactory nerves [18], [20], [218], and/or blood-brain-barrier [18], [203]. Nanoparticles that reach the lung are predominantly cleared via: mucociliary escalator into the gastrointestinal tract (and then eliminated in the feces) [206], lymphatic system [219], and circulatory systems [20]. From the lymphatic and circulatory systems, nanoparticles may be distributed to organs, including kidneys from where partial or total clearance may occur.

*4.1.8. Adverse health effects in the respiratory tract.*

**Adverse health effects.** Recent research has lead to changes in terminology and brought about the realization that no particles are completely inert, and that even low concentrations of particles can have negative health effects [36]. The adverse health effects of nanoparticles depend on the residence time in the respiratory tract [203]. Smaller particles have a higher toxicity than larger particles of the same composition and crystalline structure, and they generate a consistently higher inflammatory reaction in the lungs [216]. Smaller nanoparticles are correlated with adverse reactions such as: impaired macrophage clearance, inflammation, accumulation of particles, and epithelial cell proliferation, followed by fibrosis, emphysema, and the appearance of tumors [36], [39], [216], [220-222]. Particle uptake and potential health effects may be dependent on genetic susceptibility and health status [206].

Recent research has demonstrated that nanoparticles inhalation can affect the immune system defense ability to combat infections [223]. Nanoparticles of various compositions are able to modulate the intrinsic defensive function of macrophages, affecting their reactivity to infections. It was found that several types of nanoparticles (such as $ZrO_2$) enhance the expression of some viral receptors, making macrophages exposed to nanoparticles hyper-reactive to viral infections and leading to excessive inflammation [223]. On the other hand, exposure to other nanoparticles ($SiO_2$, $TiO_2$) leads to a decrease in the expression of some other viral and bacterial receptors, leading to lower resistance to some viruses or bacteria.

**Adaptability.** Organisms are capable of adapting to specific environmental stresses. Recent studies suggest that pre-exposure to low concentrations of nanoparticles stimulates the phagocytic activity of cells, while high concentration of nanoparticles impairs this activity [224], [225]. At the same time, genotype is an important factor in adaptability [226].

**Treatment.** Treatments for nanoparticles inhalation include those that act to enhance mucociliary clearance, and those that reduce the effects of oxidation and inflammation. Mucociliary clearance can be enhanced two fold by inhalation of increasing concentrations of saline solutions [227]. The saline solution acts as an osmotic agent increasing the volume of airway surface liquid. Anti-inflammatory medicine (sodium cromoglycate) was found to strongly reduce airway inflammation caused by diesel exhaust nanoparticles [117]. Sodium cromoglycate works by reducing allergic responses (inhibits the release of mediators from mast cells - cells responsible for the symptoms of allergy). Antioxidant vitamins (particularly vitamin C) [199], rosmarinic acid [46], and a high intake of fresh fruit and some vegetables have a protective effect against lung diseases [199].

In order to better understand the adverse health effects and possible treatment of inhaled nanoparticles, the next chapter explores the biological interaction of nanoparticles at a cellular level.

## 4.2. Cellular interaction with nanoparticles

### 4.2.1. Cellular uptake

Like nanoorganisms (viruses), nanoparticles are able to enter cells and interact with subcellular structures. Cellular uptake, subcellular localization, and ability to catalyze oxidative products depend on nanoparticle chemistry, size, and shape [228]. The mechanism by which nanoparticles penetrate cells without specific receptors on their outer surface is assumed to be a passive uptake or adhesive interaction. This uptake may be initiated by Van der Waals forces, electrostatic charges, steric interactions, or interfacial tension effects, and does not result in the formation of vesicles [203], [229]. (Steric interactions occur when nanoparticles have molecules with size, geometries, bondings, and charges optimized for the interaction with the receptors.) After this type of uptake, the nanoparticles are not necessarily located within a phagosome (which offers some protection to the rest of the cellular organelles from the chemical interaction with the nanoparticle). For example $C_{60}$ molecules enter cells and can be found along the nuclear membrane, and within the nucleus [214]. This type of uptake and free movement within the cell makes them very dangerous by having direct access to cytoplasm proteins and organelles. Upon non-phagocytic uptake, nanoparticles can be found in various locations inside cell, such as the outer-cell membrane [133], [230], cytoplasm [133], [230], mitochondria [57], [228], lipid vesicles [116], [230], along the nuclear membrane [133], or within the nucleus [228], [230]. Depending on their localization inside the cell, the nanoparticles can damage organelles or DNA, or ultimately cause cell death.

Nanoparticles are internalized not only by professional phagocytes such as alveolar macrophages [30], [145], [228], but by various types of cells, including endothelial cells [230], pulmonary epithelium [142], [231-236], gastrointestinal epithelium [204], red blood cells [203], [237], platelets [238], and nerve cells [239].

Particle internalization location depends on nanoparticle size. For example, environmental particles with size between 2.5 - 10 μm were found to collect in large cytoplasmic vacuoles (Figure 28 c, d), while smaller nanoparticles (<100 nm) localize in organelles, such as mitochondria (Figure 28 e, f), leading to disruption of mitochondrial architecture [57]. Very small nanoparticles, such as $C_{60}$ molecules with a diameter of 0.7 nm, are able to penetrate cells via a different mechanism than phagocytosis, probably through ion channels or via pores in the cell membrane [214].

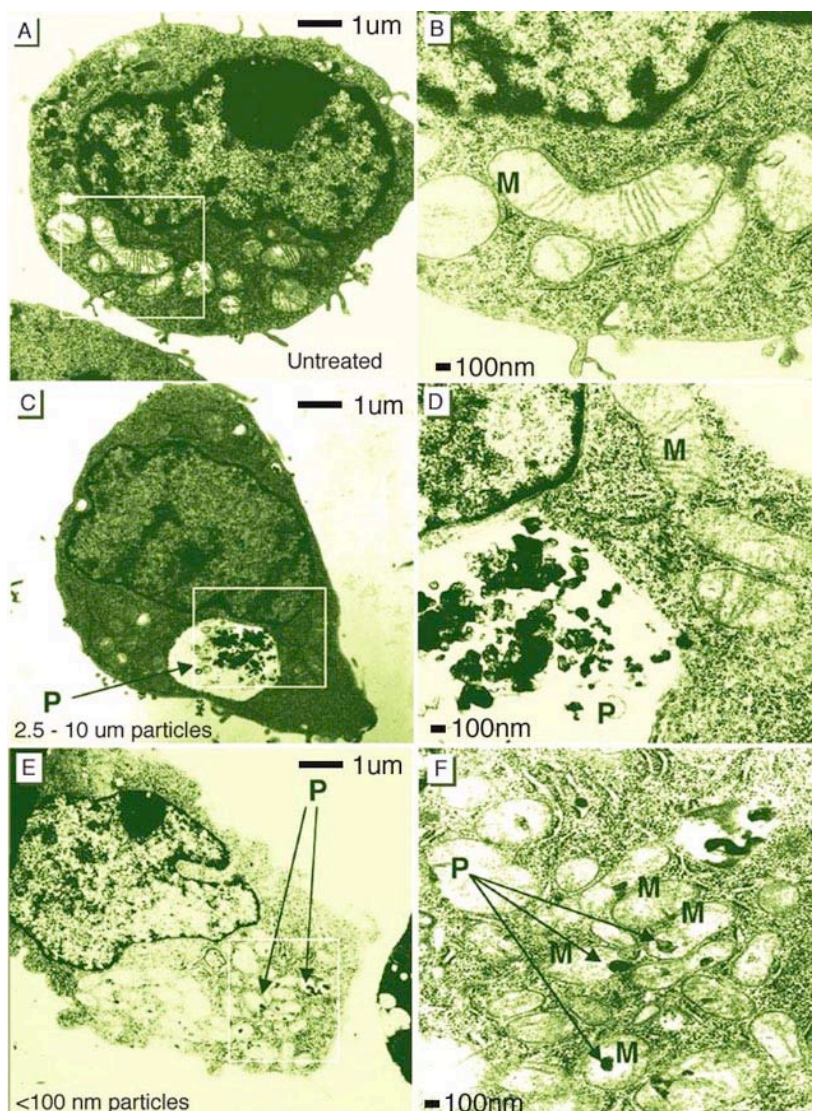

**Figure 28.** *TEM images showing effects of environmental particles size (P) on murine macrophage cells RAW 264.7 treated with various size particles: (a,b) untreated, (c,d) 2.5 - 10 μm size particles, (e,f) particles smaller than 100 nm. M denotes mitochondria [57]. Reproduced with permission from Environmental Health Perspectives.*

Uptake location is likely to depend on material type, however current research does not provide sufficient information to drawing conclusions on this subject.

### 4.2.2. Oxidative stress, inflammation, and genotoxicity

While the exact mechanism whereby nanoparticles induce pro-inflammatory effects is not known, it has been suggested that they create reactive oxygen species (ROS), and thereby modulate intracellular calcium concentrations, activate transcription factors, and induce cytokine production [240]. Below we outline in a very simplified and schematic depiction the current understanding of these very complex cellular mechanisms.

**Oxidative stress generation.** Both *in vivo* and *in vitro* studies have shown that nanoparticles of various compositions (fullerenes, carbon nanotubes, quantum dots, and automobile exhaust) create reactive oxygen species [20]. Reactive oxygen species have been shown to damage cells by peroxidizing lipids, altering proteins, disrupting DNA, interfering with signaling functions, and modulating gene transcription [240].

Oxidative stress is a response to cell injury, and can also occur as an effect of cell respiration, metabolism, ischemia/reperfusion, inflammation, and metabolism of foreign compounds [46].

The oxidative stress induced by nanoparticles may have several sources [46]:

(i) Reactive oxygen species can be generated directly from the surface of particles when both oxidants and free radicals are present on the surface of the particles. Many compounds hitch-hiking on the surface of nanoparticles (usually present in ambient air) are capable of inducing oxidative damage, including ozone ($O_3$) and $NO_2$.

(ii) Transition metals (iron, copper, chromium, vanadium, etc.) nanoparticle can generate reactive oxygen species acting as catalysts in Fenton type reactions [46]. For example, the reduction of hydrogen peroxide ($H_2O_2$) with ferrous iron ($Fe^{2+}$)

$$\overset{\bullet}{O_2^-} + H_2O_2 \xrightarrow{Fe} \overset{\bullet}{O}H + OH^- + O_2$$

results in the formation of hydroxyl radical ($\overset{\bullet}{O}H$) that is extremely reactive, attacking biological molecules situated within diffusion range [46].

(iii) Altered functions of mitochondrion. As shown in several studies, small nanoparticles are able to enter mitochondria [57], [228] and produce physical damage, contributing to oxidative stress [24].

(iv) Activation of inflammatory cells, such as alveolar macrophages and neutrophils, which can be induced by phagocytosis of nanoparticles, can lead to generation of reactive oxygen species and reactive nitrogen species [46], [241]. Alveolar macrophages participate in the initiation of inflammation in the lung (see paragraph 4.1.3).

Nanoparticles have been shown to generate more free radicals and reactive oxygen species than larger particles, likely due to their higher surface area [24], [231], [242].

**Inflammation.** Inflammation is the normal response of the body to injury. When generated in moderation, inflammation stimulates the regeneration of healthy tissue, however when in excess, it can lead to disease [49]. *In vitro* and *in vivo* experiments demonstrate that exposure to small nanoparticles is associated with inflammation, with particle size and composition being the most important factors [46]. Inflammation is controlled by a complex series of intracellular and extracellular events. The oxidative stress results in the release of pro-inflammatory mediators or cytokines - intercellular chemical messengers alerting the immune system when an infection is present [215], [241]. Some nanoparticles can produce cell death via mitochondrial damage without inflammation [228].

**Antioxidants.** The oxidative stress also results in the release of antioxidants - proteins that act to remove the oxidative stress [46], [49]. In addition to the antioxidants released as a response to the oxidative process, nanoparticles may interact with metal-sequestering proteins and antioxidants (from body fluids and intracellularly), that will likely modify the surface properties of the nanoparticle to some extent, rendering them less toxic [46].

**DNA damage.** Generation of reactive oxygen species to the point that they overwhelm the antioxidant defense system (shifting the redox balance of the cell) can result in oxidation, and therefore destruction, of cellular biomolecules, such as DNA, leading to heritable mutations [46], [203]. For example, the chemical modification of histones (or binding proteins that support the supercoiled structure of DNA) opens the coiled DNA and allows its alteration [49]. Epidemiological, *in vitro* and *in vivo* studies show that nanoparticles of various materials (diesel, carbon black, welding fumes, transition metals) are genotoxic in humans or rats [41]. Oxidative DNA damage markers showed higher levels on workdays for bus drivers from central areas compared to bus drivers from suburban/rural areas of Copenhagen [46]. Nasal biopsies from children living in Mexico City showed greater DNA damage compared to children living on less polluted areas [46].

A general schematic, of the molecular events by which nanoparticles exert their toxic effects at the cellular level, is given in Figure 29. In summary, nanoparticles can directly generate reactive oxygen species on the their surfaces or by activation of macrophages [20], [46], [241]. Overall, the generation of oxidative species leads to increased inflammation [215], [241], and increased antioxidant production [49]. The activation of macrophages leads to modulation in intracellular calcium concentration, that in turn activates further the reactive oxygen species production, which in turn enhances further calcium signaling by oxidation of calcium pumps in the endoplasmic reticulum, leading to calcium depletion [46], [240], [243]. Intracellular calcium modulation results in impaired motility and reduced macrophage phagocytosis [46]. Non-phagocytized nanoparticles are likely to access and interact with epithelial cells [46], thus enhancing inflammation. Ultimately, the interaction of nanoparticles with cells may lead to DNA modifications, cell injury, and disease [49].

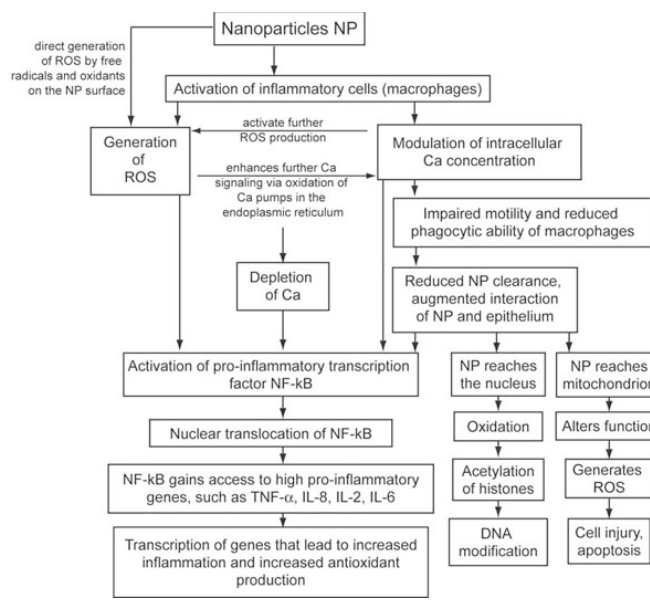

**Figure 29.** *Schematics of the molecular events by which nanoparticles exert their toxic effects at the cellular level.*

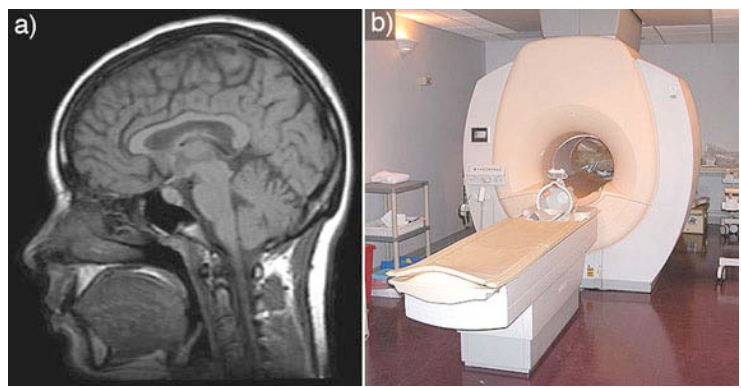

**Figure 30. (a)** *Head MRI image. Courtesy of United States National Library of Medicine, National Institute of Health. (b) MRI. Courtesy of National Institute of Neurological Disorders and Stroke.*

### 4.2.3. Adverse health effects and treatment

Nanoparticles, due to their small size, can influence basic cellular processes, such as proliferation, metabolism, and death. Many diseases can be associated with dysfunction of these basic processes. For example, cancer results from uncontrolled cell proliferation, while neurodegenerative diseases are caused in part by premature cell death [148]. Oxidative stress has been implicated in many diseases, including cardiovascular and neurological disease, pancreatitis, and cancer [46]. Severe inflammation is assumed to be the initiating step [50] in the appearance of autoimmune diseases (systemic lupus erythematosus, scleroderma, and rheumatoid arthritis) associated with exposure to some nanoparticles, such as silica and asbestos [172], [176].

Regarding the treatment of adverse health effects caused by nanoparticles cytotoxicity, antioxidants [24], [49], [61], [62], [199], anti-inflammatory drugs [117], [244], and metal chelators [240], [245] show promising effects. It has been reported that rats that underwent instillation of nanoparticles into the lungs together with an antioxidant (nacystelin) showed inflammation reduced by up to 60% in comparison to those exposed to nanoparticles alone [49]. Antioxidant therapy has been found to protect against the development of hypertension, arteriosclerosis, cardiomyopathies, and coronary heart disease [24], providing further evidence of the link between the oxidative stress response and cardiovascular effects. The adverse health effects of transition metals can be diminished by metal chelators [240].

### 4.2.4. "Non-invasive" terminology to be questioned

The process of nanoparticle uptake by cells is clinically used today in targeted drug delivery and cell imaging (Figure 30). The safety of these techniques, however, depends on cellular uptake of nanoparticles without affecting normal cellular function. Cellular imaging techniques are currently named "non-invasive" techniques [246], [247], which means non-penetrating, however they should perhaps be relabeled as "minimally-invasive", given that the nanoparticles enter the cells and are likely to affect cellular functions. Iron oxide and other magnetic nanoparticles have been used for many years as magnetic resonance imaging (MRI) contrast agents. Depending on their size and coating, MRI nanoparticles can localize in liver, spleen, lymph nodes, etc [246]. Some nanoparticles were found to be teratogenic (causing birth defects) in rats and rabbits [246]. Minor side effects of contrasting agents are nausea, vomiting, hives, and headache [248]. More serious adverse reactions involving life-threatening cardiovascular and respiratory reactions are possible in patients with respiratory disorders [248].

### 4.3. Nervous system uptake of nanoparticles

The nervous system is composed of the brain, spinal cord, and nerves that connect the brain and spinal cord to the rest of the body. In addition to nanoparticle uptake due to inhalation (discussed below), nervous system uptake may occur via other pathways (such as dermal). Uptake via olfactory nerves and the blood-brain-barrier are the most studied pathways.

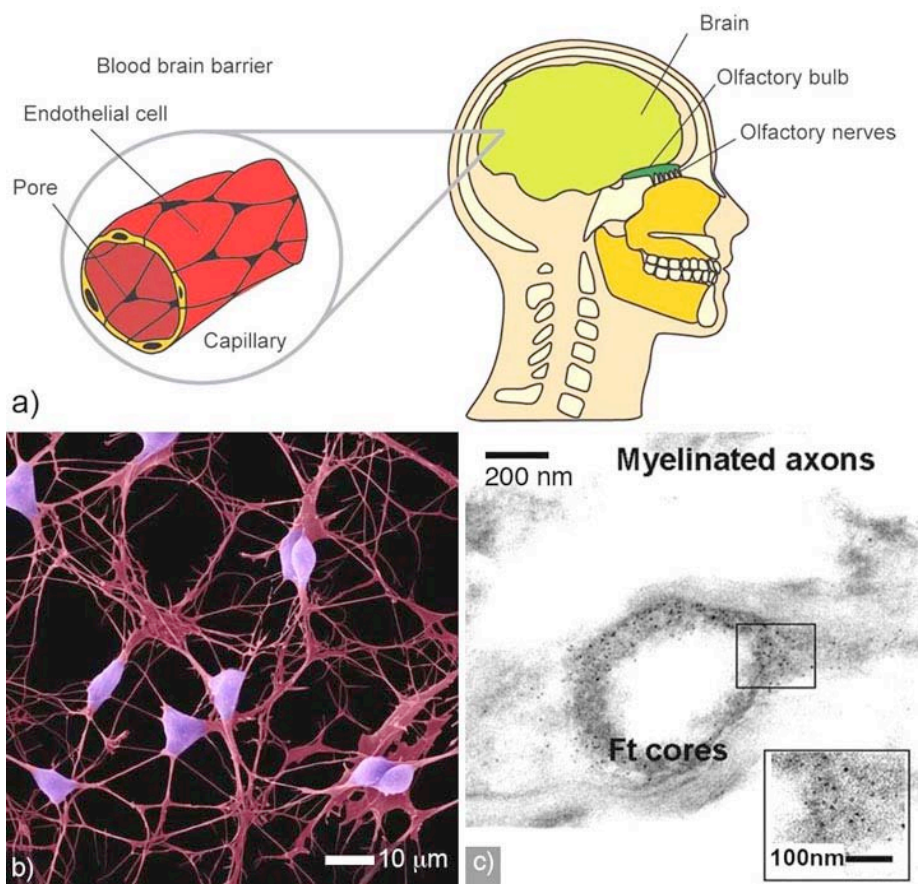

**Figure 31. (a)** *Schematics of the nanoparticles neuronal uptake via olfactory bulb and blood-brain barrier.* **(b)** *Cortical neurons (nerve cells) growing in culture. Neurons have a large cell body with several long processes extending from it, usually one thick axon and several thinner dendrites. The axon carries nerve impulses away from the neuron. Its branching ends make contacts with other neurons and with muscles or glands. © Dr. Dennis Kunkel / Visuals Unlimited. Reproduced with permission from Visuals Unlimited [25].* **(c)** *TEM images of iron accumulation in the brain of neurologically affected patients. Iron is stored in ferritin, Ft, a protein involved in excess iron storage [171]. Reproduced with permission from Elsevier.*

*4.3.1. Neuronal uptake via olfactory nerves*

Neuronal uptake (Figure 31 a) of inhaled nanoparticles may take place via the olfactory nerves [18], [20], [218] or/and blood-brain-barrier [18], [203].

The nasal and tracheo-bronchial regions have many sensory nerve endings [20]. As demonstrated several decades ago with polio viruses (30 nm) and silver coated gold nanoparticles (50 nm) in monkeys, intranasally instilled viruses and particles migrate to the olfactory nerves and bulb with an axonal transport velocity of about 2.5 mm/hr [20]. The silver coated gold nanoparticles that reached the olfactory bulb were preferentially located in mitochondria [20], raising a major concern of their toxicity. More recent studies confirm the uptake of inhaled nanoparticles from olfactory mucosa via the olfactory nerves in the olfactory bulb [18], [20], [218], [239]. For example, rat inhalation studies with 30 nm magnesium oxide [200] and 20-30 nm carbon [239] nanoparticles indicate that nanoparticles translocate to the olfactory bulb [200]. If inhalation occurred via one nostril only, the accumulation was observed only in the side of the open nostril [20]. Experiments show that microparticles with diameter larger than a micron do not cross the olfactory nerve, as expected from the geometrical restrictions imposed by the diameter of the olfactory axons of only 100-200 nm (Figure 31 a) [20]. Translocation of nanoparticles into deeper brain structures may be possible [145], as suggested by the movement of viruses through neurons [20].

*4.3.2. Neuronal uptake via blood-brain-barrier*

The passage of nanoparticle to the nervous system is also possible via the blood-brain-barrier (Figure 31 a). The blood-brain-barrier is a physical barrier with negative electrostatic charge between the blood vessels and brain [249], selectively restricting the access of certain substances [18]. This anionic barrier is believed to stop most anionic molecules, while the cationic molecules increase the permeability of the blood-brain-barrier by charge neutralization [18]. This route has been extensively studied for the purpose of drug delivery to the brain [18], [249]. Regarding the passage of nanoparticles, the blood-brain-barrier permeability is dependent upon the charge of nanoparticles [249]. It allows a larger number of cationic nanoparticles to pass compared to neutral or anionic particles, due to the disruption of its integrity [249]. As shown by magnetic resonance imaging (MRI) with magnetic nanoparticles, the blood-brain-barrier in healthy subjects stops some proteins and viruses present in the brain vascular system from translocating to the brain [18]. However, subjects with specific circulatory diseases (like hypertension) [18], brain inflammation [18], respiratory tract inflammation (increased levels of cytokines that cross blood-brain-barrier and induce inflammation) [203] may have increased blood-brain-barrier permeability, which will allow nanoparticles access to the nervous system.

*4.3.3. Adverse health effects of neuronal nanoparticles uptake and treatment*

Experimental evidence suggests that the initiation and promotion of neurodegenerative diseases, such as Alzheimer's disease, Parkinson's disease, Pick's disease, are associated with oxidative stress and accumulation of high concentrations of metals (like copper, aluminum, zinc, but especially iron) in brain regions associated with function loss and cell damage [245], [250]. Iron is necessary in many cellular functions, especially in the brain, where it participates in many neuronal processes. In excess, however, iron is toxic to cells. The brain continuously accumulates iron, resulting in increased stored iron amounts with the age. In order to prevent its toxicity, organisms developed a way to store excess iron in proteins called ferritin (Ft). Dysfunction of ferritin resulting from excessive accumulation of iron (Figure 31 b) may lead to oxidative stress and myelin (the electrically insulating coatings of axons) breakdown [171]. Metal homeostasis imbalance and neuronal loss are both present in neurodegenerative diseases. (Homeostasis is a dynamic equilibrium balancing act necessary for a proper function of a living system) It is not known if the presence of metals in brain of subjects with neurodegenerative diseases is due to nanoparticles themselves translocating to the brain or their soluble compounds [41].

Despite the fact that the etiology of neurodegenerative diseases is unknown, environmental factors are believed to play a crucial factor in their progress, being able to trigger pro-inflammatory responses in the brain tissue [42], [250]. Recent studies on DNA damage in nasal and brain tissues of canines exposed to air pollutants shows evidence of chronic brain inflammation, neuronal dysfunction, and similar pathological findings with those of early stages of Alzheimer's disease [203], [251]. Autopsy reports on humans suggest similar results [203]. Significant oxidative damage was found in the brain of largemouth bass after exposure to $C_{60}$ [252]. Rat inhalation studies with stainless steel welding fumes showed that manganese accumulates in blood, liver, and brain [41]. Epidemiological studies show a clear association between inhalation of dust containing manganese and neurological diseases in miners [170] and welders [148]. Some welders develop Parkinson's disease much earlier in their life, usually in their mid forties, compared to the sixties in the general population [148]. Brain inflammation appears to be a cumulative process, and the long-term health effects may not be observed for decades [203]. Currently there are 1.5 millions peoples suffering from Alzheimer's in the United States of America [203] and an estimated 18 millions worldwide [253].

**Treatment.** Antioxidants and metal chelators are treatment options for the adverse health effects caused by the neuronal uptake of nanoparticles. In the therapy of neurodegenerative diseases, metals chelators transported across the blood-brain-barrier seem to be a very promising approach [245]. Functionalized fullerenes [61] and nanoparticles made of compounds holding oxygen vacancies show great antioxidant properties [62]. Fullerols, or poly-hydroxylated fullerenes, are excellent antioxidants with high solubility and ability to cross the blood-brain barrier, showing promising results as neuroprotective agents [61]. $CeO_2$, and $Y_2O_3$ nanoparticles have strong antioxidant properties on rodent nervous system cells [62]. Cerium oxide tends to be non-stoichiometric, Ce atoms having a dual oxidation state, +3 or +4, leading to oxygen vacancies. Dual oxidation state confers $CeO_2$ and probably $Y_2O_3$ nanoparticles antioxidant properties that promote cell survival under conditions of oxidative stress. It appears that the antioxidant properties depend upon the structure of the particle but they are independent of its size within 6-1000 nm.

### 4.4. Nanoparticles translocation to the lymphatic systems

Translocation of nanoparticles to lymph nodes is a topic of intense investigation today for drug delivery and tumor imaging [219]. Progression of many cancers (lung, esophageal, mesothelioma, etc.) is seen in the spread of tumor cells to local lymph nodes [219]. The detection and targeted drug delivery to these sites are the steps involved in the therapeutic treatment of cancer. Several studies show that interstitially injected particles pass preferentially through the lymphatic system and not the circulatory system, probably due to permeability differences [219]. After entering the lymphatic system, they locate in the lymph nodes [219]. The free nanoparticles reaching the lymph nodes are ingested by resident macrophages [254]. Nanoparticles that are able to enter the circulatory system can also gain access to the interstitium and from there are drained through the lymphatic system to the lymph nodes as free nanoparticles and/or inside macrophages [219], [254].

The adverse health effects of nanoparticle uptake by lymphatic system are not sufficiently explored. However, one can hypothesize that oxidative stress created by certain types of nanoparticles could lead to damage of lymphocytes (type of white blood cell), lymph nodes, and/or spleen.

### 4.5. Nanoparticles translocation to the circulatory system

Inhalation or instillation studies in healthy animals show that metallic nanoparticles with size smaller than 30 nm pass rapidly into the circulatory system [20], [41], [145], [219], [229], while non-metallic nanoparticles with size between 4 and 200 nm pass very little or not at all [255-259]. In contrast, subjects suffering from respiratory and circulatory diseases have higher capillary permeability, allowing fast translocation of metallic or non-metallic nanoparticle into circulation [255], [256], [259].

#### 4.5.1. Long-term translocation

Nanoparticles, unlike larger particles, are able to translocate across the respiratory epithelium after being deposited in the lungs [20], [229]. Once they have crossed the respiratory epithelium, they may persist in the interstitium for years, or they may enter the lymphatic system [219] and circulatory system [145]. From the circulatory system long-term translocation to organs (such as the liver, heart, spleen, bladder, kidney, bone marrow) is possible, depending on the duration of exposure [20]. Smaller particles (20 nm) are cleared faster from the lung than larger particles (100 nm), probably because small nanoparticles are not efficiently phagocytized by macrophages and are able to enter more rapid the circulatory and/or lymphatic systems [145].

#### 4.5.2. Short-term translocation of metals

Evidence of rapid translocation of metal nanoparticles from lungs into the circulation and to organs has been provided by animal studies. These results show the location of nanoparticles with diameters of 30 nm (Au) [20], 22 nm (TiO$_2$) [229] in pulmonary capillaries; 15 nm (Ag) [145], and welding fumes [41] in blood, liver, kidney, spleen, brain, and heart. Animal studies on rats with inhalation of titanium dioxide nanoparticles (22 nm diameter) show that they can translocate to the heart and can be found in the heart connective tissue (fibroblasts) [229]. Within 30 minutes post exposure, large quantities of intratracheally instilled gold nanoparticles (30 nm) have been found in platelets inside of pulmonary capillaries of rats [20], motivating the hypothesis that nanoparticles may induce aggregation of platelets, leading to the formation of blood clots. Figure 32 a shows an electron microscope image of a capillary with red blood cells.

#### 4.5.3. Short-term translocation of non-metals.

There is no conclusive evidence showing fast translocation of carbon-based nanomaterials into systemic circulation. Short-term translocation of radiolabeled nanoparticles from lungs to the organs is currently the subject of debate as a significant fraction of radioactive labels detaches from their labeled nanoparticles, so radioactivity observed throughout the body may not indicate the actual translocation of nanoparticles, but of radiolabels. Technetium's short lived isotope $^{99m}$Tc, with an atomic diameter of about 0.37 nm, is used in labeling nanoparticles that are subsequently injected or inhaled by subjects. In many cases the radiolabel can separate from the nanoparticles and follow a different translocation route. In the presence of oxygen, the radioactive label can transform into pertechnetate ($^{99m}$TcO$_4^-$) having a slightly larger diameter of roughly 0.5 nm. Most studies show very little or no translocation of radiolabeled polystyrene nanoparticles with diameters of 56 nm and 200 nm [255], or carbon nanoparticles with diameters of 5 nm [256], 4-20 nm [257], 35 nm [259], 100 nm [258], while others show a rapid and substantial translocation into circulation for particles sized 5-10 nm [194], 20-30 nm [239].

While the short-term extra-pulmonary translocation into circulation in healthy subjects is still under debate, there seems to be agreement on the fact that nanoparticle fast translocation into circulation may be enhanced by pulmonary inflammation [255], [256], [259], and increased microvascular permeability [255]. Subjects suffering from respiratory or blood diseases may have an increased susceptibility of nanoparticles translocation from lungs to circulation and organs.

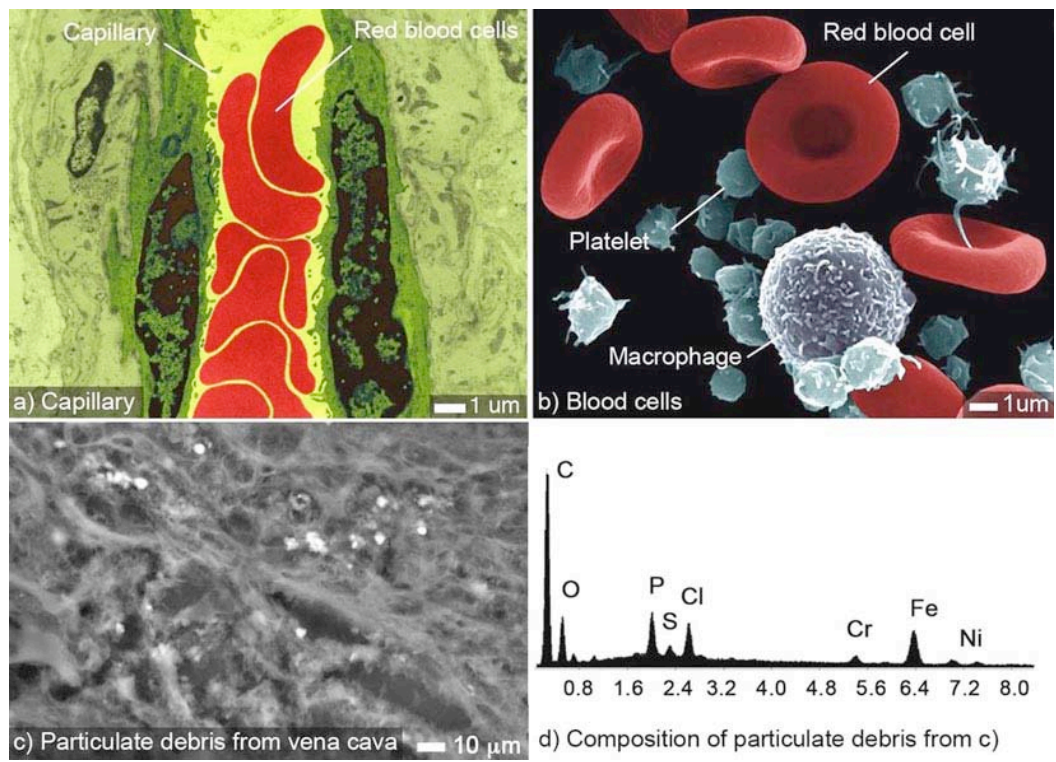

**Figure 32. (a)** *Red blood cells in a capillary.* **(b)** *Platelets, red, and white blood cells. © Dr. Dennis Kunkel/Visuals Unlimited, reproduced with permission from Visuals Unlimited [25].* **(c)** *Particulate debris entrapped inside the tissue formed around a vena cava filter removed after 156 days in a patient with blood disease [146].* **(d)** *EDS spectrum showing the composition of the debris shown in (c), identified as stainless steel [146]. Figures c, d, reproduced from [146] with kind permission of Springer Science and Business Media.*

*4.5.4. Nanoparticles interaction with and uptake by blood cells*

There are three main types of cells in the blood (Figure 32 b): red cells in charge of oxygen transport; white cells responsible for fighting infections; and platelets that help prevent bleeding by forming blood clots. The uptake of nanoparticles by each type of blood cells is essentially different.

Nanoparticle uptake by red blood cells (that do not have phagocytic abilities, due to the lack of phagocytic receptors) is entirely dictated by size [203], while the nanoparticle charge or material type plays little importance [237].

In contrast, nanoparticle charge plays an essential role in their uptake by platelets and their influence on blood clot formation [238]. Uncharged polystyrene particles do not have an effect on blood clots formation. Negatively charged nanoparticles significantly inhibit thrombi formation, while positively charge nanoparticles enhance platelet aggregation and thrombosis [238]. The interaction between platelets and positively charged particles seems to be due to the net negative charge that platelets carry on their surface [238]. The positively charged nanoparticles interact with negatively charged platelets and reduce their surface charge, making them more prone to aggregation. Until now it was thought that blood clots can be formed due to three main causes: when the blood flow is obstructed or slowed down, when the vascular endothelial cells are damaged, or due to the blood chemistry. However, it seems possible, in the view of recent findings that nanoparticles may act as nucleating centers for blood clots [146], [260]. It is important to note that pulmonary instillation of large nanoparticles (400 nm) caused pulmonary inflammation of similar intensity to that caused by 60 nm particles, but did not lead to peripheral thrombosis [30]. The fact that the larger particles failed to produce a thrombotic effect suggests that pulmonary inflammation itself is insufficient to cause peripheral thrombosis [30], and that thrombi formation occur via direct activation of platelets [117], [238].

Microscopic and energy dispersive spectrometry (EDS) analysis of blood clots from patients with blood disorders revealed the presence of foreign nanoparticles, as shown in Figure 32 c, d. The blood clots were collected after half a year of wear of vena cava filters implanted in order to prevent pulmonary embolism in patients affected by blood disorders [146]. Most notably, patients with the same type of blood disorder show fibrous tissue clots embedding nanoparticle with different composition: gold, silver, cobalt, titanium, antimony, tungsten, nickel, zinc, mercury, barium, iron, chromium, nickel, silicon, glass, talc, stainless steel. The common denominator of the particles is their size, ranging from tens of nanometers to few microns [146].

The uptake of nanoparticles by macrophages (a type of white cell) has already been discussed.

*4.5.5. Adverse health effects of circulatory system uptake*

**Thrombosis.** Translocation of nanoparticles into the circulatory system was correlated with the appearance of thrombi (or blood clots) [117], [238]. The time frame of this process is very short, thrombosis occurring during the first hour after exposure. Hamster studies of tracheally or intravenously instilled nanoparticles of charged polystyrene (60 nm) [117] and diesel exhaust particles (20-50 nm) [117], [261] significantly increased arterial or venous thrombus formation during the first hour after administration. There is a clear dose-dependent response correlating the quantity of pollutant administered and the observed thrombus sizes [30], [117]. Prothrombotic effects persisted 24 h after instillation [30].

If inhaled nanoparticles were to be found in red blood cells located in pulmonary capillaries [203], one would expect adverse health effects as blood-related diseases, such as anemia, due to reduced oxygen transport capacity of the red blood cells.

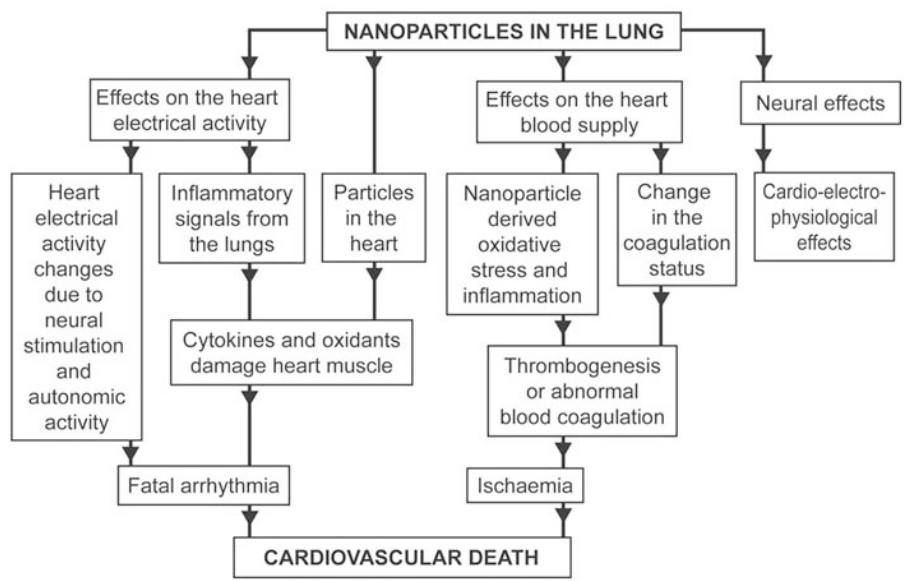

**Figure 33.** *Diagram of hypothetical mechanisms and pathways that link nanoparticles in the lung with adverse cardiovascular effects (modified after [49] and [117]).*

**Cardiovascular malfunction.** It is clear from clinical and experimental evidence that inhalation of nano and microparticles can cause cardiovascular effects [262]. Despite the fact that there is an intuitive relationship between inhaled nanoparticles and adverse respiratory effects, the causal link between particles in the lung and cardiovascular effects is not entirely understood [49]. It was thought that the pulmonary inflammation caused by the particles triggers a systemic release of cytokines, resulting in adverse cardiovascular effects. However, recent studies on animals [145], [263], and humans [194] have shown that nanoparticles diffuse from the lungs into the systemic circulation, and then are transported to the organs, demonstrating that cardiovascular effects of instilled or inhaled nanoparticles can arise directly from the presence of nanoparticles within the organism [188]. Proposed mechanisms of cardiovascular effects are summarized in Figure 33 [49], [117].

### 4.6. Liver, spleen, kidneys uptake of nanoparticles

#### 4.6.1. Organs nanoparticles uptake

Endothelial cells (cells that line the vascular system) form a physical barrier for particles, having very tight junctions, typically smaller than 2 nm [264]. Nevertheless larger values, from 50 nm [51] up to 100 nm [264] have been reported, depending on the organ or tissue. A very tight endothelial junction is present in the brain, often called the blood brain barrier. However, experiments performed on rats injected with ferritin macromolecules (with size around 10 nm) into the cerebrospinal fluid, demonstrated passage of ferritin into deep brain tissue. In certain organs, such as liver, the endothelium is fenestrated with pores of up to 100 nm, allowing easier passage of larger particles (Figure 34 a). In the presence of inflammation the permeability of the endothelium is increased, allowing a larger passage of particle.

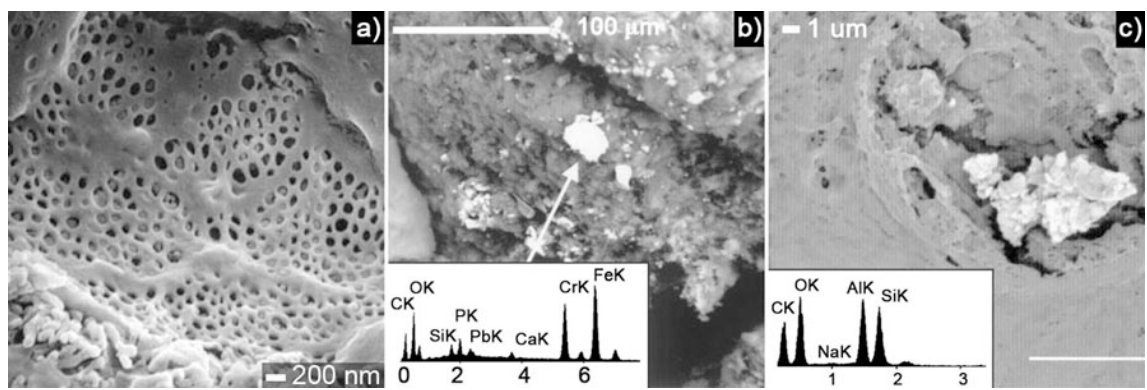

***Figure 34. (a)** Mouse liver with fenestrated hepatic endothelial cells [264]. SEM and EDS spectrum of particles found in patients with diseased **(b)** liver and **(c)** kidneys [265]. Reproduced from [265] with permission from Elsevier.*

Micro and nanoparticle debris was detected by scanning electron microscopy in organs and blood of patients with: orthopedic implants [265], drug addiction [265], worn dental prostheses [266], blood diseases [146], colon cancer, Crohn's disease, ulcerative colitis [147], and with diseases of unknown etiology [265]. Coal workers autopsies reveal an increased amount of particles in the liver and spleen compared to non-coal workers [41]. The workers with pronounced lung diseases have more nanoparticles in their organs than healthier ones [41]. The pathway of exposure most likely involves the translocation from lungs to circulation of the inhaled nanoparticles, followed by uptake by the organs.

Rat inhalation studies with stainless steel welding fumes showed that manganese accumulates in blood and liver [41]. Rat inhalation studies with 4-10 nm silver nanoparticles show that within 30 minutes the nanoparticles enter the circulatory system, and after a day can be found in the liver, kidney and heart, until subsequently cleared from these organs after a week [145]. Clearance from the liver can occur via biliary secretion into the small intestine [267].

A case study shows that the wear of dental bridges leads to the accumulation of wear nanoparticles in liver and kidneys [266]. The most probable absorption pathway was assumed to be via intestinal absorption [266]. Scanning electron microscopy and energy-dispersive microanalytical techniques identified the chemical compositions of particles in the liver and kidney biopsies, as well in stool, as the same as the porcelain from dental prostheses. The maximum size of particles found in the liver (20 microns) was larger than in the kidneys (below 6 microns), suggesting that particles are absorbed by intestinal mucosa, translocate to liver before reaching the circulatory system and kidneys. After the removal of dental bridges, particles in stool are no longer observed.

*4.6.2. Adverse health effects of liver and kidney uptake*

Up to now there is little knowledge (or discussion) on the effect of nanoparticles on organs such as liver, kidneys, spleen, etc. However, one can speculate that as long as there is translocation to and accumulation of nanoparticles in these organs, potentially adverse reactions and cytotoxicity may lead to disease.

Diseases with unknown origins have been correlated with the presence of micro- and nanoparticles in kidneys and liver (Figure 34 b, c) [265]. For comparison, the liver and kidneys of healthy subjects did not show any debris. Particles debris has been found also in the liver of patients with worn orthopedic prosthesis [265].

Dental prosthesis debris internalized by intestinal absorption can lead to severe health conditions, including fever, enlarged spleen and liver, suppression of bile flow, and acute renal failure [266]. These symptoms appeared about a year after the application of dental porcelain bridges. After the removal of dental bridges, and subsequent treatment with steroids, the clinical symptoms declined [266].

### 4.7. Gastro-intestinal tract uptake and clearance of nanoparticles

#### 4.7.1. Exposure sources

Endogenous sources of nanoparticles in the gastro-intestinal tract are derived from intestinal calcium and phosphate secretion [268]. Exogenous sources are particles from food (such as colorants – titanium oxide), pharmaceuticals, water, or cosmetics (toothpaste, lipstick) [268], dental prosthesis debris [266], and inhaled particles [145]. The dietary consumption of nanoparticles in developed countries is estimated around $10^{12}$ particles/person per day [252]. They consist mainly of $TiO_2$ and mixed silicates. The use of specific products, such as salad dressing containing a nanoparticle $TiO_2$ whitening agent, can lead to an increase by more than 40 fold of the daily average intake [252]. These nanoparticles do not degrade in time and accumulate in macrophages. A database of foods and pharmaceuticals containing nanoparticles can be found in reference [268]. A portion of the particles cleared by the mucociliary escalator can be subsequently ingested into the gastro-intestinal tract. Also, a small fraction of inhaled nanoparticles was found to pass into the gastrointestinal tract [145].

#### 4.7.2. Size and charge dependent uptake

The gastro-intestinal tract is a complex barrier-exchange system, and is the most important route for macromolecules to enter the body. The epithelium of the small and large intestines is in close contact with ingested material, which is absorbed by the villi (Figure 35).

The uptake of nano- and micro-particles have been the focus of many investigations, the earliest dating from mid seventeen century, while more recently entire issues of scientific journals have been devoted to the subject [51]. The extent of particles absorption in the gastro-intestinal tract is affected by size, surface chemistry and charge, length of administration, and dose [30].

The absorption of particles in the gastro-intestinal tract depends on their size, the uptake diminishing for larger particles [269]. A study of polystyrene particles with size between 50 nm and 3 μm indicated that the uptake decreases with increasing particle size from 6.6% for 50 nm, 5.8% for 100 nm nanoparticles, 0.8% for 1μm, to 0% for 3 μm particles.

The time required for nanoparticles to cross the colonic mucus layer depends on the particle size, with smaller particles crossing faster than larger ones: 14 nm diameter latex nanoparticles cross within 2 minutes, 415 nm within 30 minutes, and 1000 nm particles do not pass this barrier [30]. Particles that penetrate the mucus reach the enterocytes and are able to translocate further [30]. Enterocyes are a type of epithelial cell of the superficial layer of the small and large intestine tissue, which aid in the absorption of nutrients. When in contact with the sub-mucosal tissue, nanoparticles can enter the lymphatic system and capillaries, and then are able to reach various organs [30]

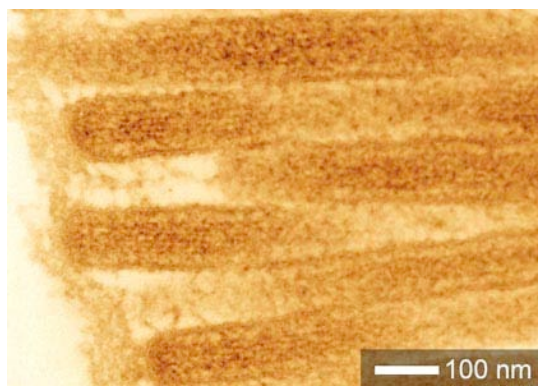

**Figure 35.** *TEM image of a thin section cut through a segment of human small intestine epithelial cell. One notices densely packed microvilli, each microvillus being approximately 1 micron long by 100 nm in diameter. Courtesy of Chuck Daghlian, Louisa Howard, Katherine Connollly [96].*

Diseases, such as diabetes, may lead to higher absorption of particles in the gastrointestinal tract [30]. For example, rats with experimentally induced diabetes had a 100 fold increase in absorption of 2 μm polystyrene particles [30] relative to non-diabetic rats. Also inflammation may lead to the uptake and translocation of larger particles of up to 20 μm [266].

The kinetics of particles in the gastro-intestinal tract depend strongly on the charge of the particles, positively charged latex particles are trapped in the negatively charged mucus while negatively charged latex nanoparticles diffused across the mucus layer and became available for interaction with epithelial cells [30].

### 4.7.3. Translocation

Varying the characteristics of nanoparticles, such as size, surface charge, attachment of ligands, or surfactant coatings, offers the possibility for site-specific targeting of different regions of the gastro-intestinal tract. The fast transit of material through the intestinal tract (on the order of hours), together with the continuous renewal of epithelium, led to the hypothesis that nanomaterials will not remain there for indefinite periods [30]. Most of the studies of ingested nanoparticles have shown that they are eliminated rapidly: 98% in the feces within 48 hours and most of the remainder via urine [20]. However, other studies indicate that certain nanoparticles can translocate to blood, spleen, liver, bone marrow, [269], lymph nodes, kidneys, lungs, and brain, and can also be found in the stomach and small intestine [270]. Oral uptake of polystyrene spheres of various sizes (50 nm – 3 μm) by rats resulted in a systemic distribution to liver, spleen, blood, and bone marrow [269]. Particles larger than 100 nm did not reach the bone marrow, while those larger than 300 nm were absent from blood [269]. In the study no particles were detected in heart or lung tissue. Studies using iridium did not show significant uptake, while titanium oxide nanoparticles were found in the blood and liver [20]. For several days following oral inoculation of mice with a relatively biologically inert nanometer-sized plant virus (cowpea mosaic virus), the virus was found in a wide variety of tissues throughout the body, including the spleen, kidney, liver, lung, stomach, small intestine, lymph nodes, brain, and bone marrow [270].

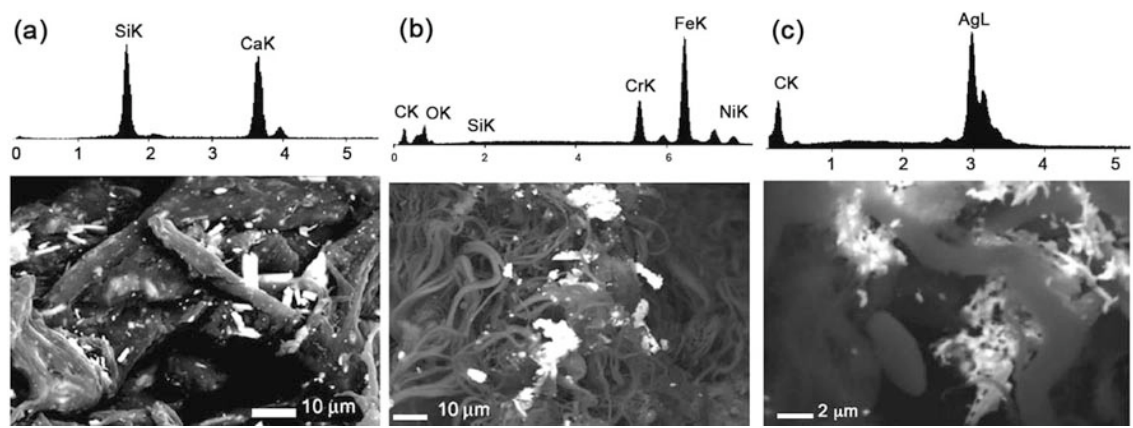

**Figure 36.** EDS spectrum and SEMs of particles of different size and morphology in patients with colon cancer [147]. Particles are composed mainly of **(a)** calcium and silicon, **(b)** stainless steel, **(c)** silver. Reproduced from [147] with permission from Elsevier.

The exact order of translocation from the gastro-intestinal tract to organs and blood is not known, however a case study of dental prosthesis porcelain debris internalized by intestinal absorption suggests that intestinal absorption of particles is followed by liver clearance before they reach the general circulation and the kidneys [266].

### 4.7.4. Adverse health effects of gastro-intestinal tract uptake

**Reaction reduced toxicity.** In the intestinal tract there is a complex mix of compounds, enzymes, food, bacteria, etc., that can interact with ingested particles and sometimes reduce their toxicity [30]. It was reported that particles *in vitro* are less cytotoxic in a medium with high protein content.

**Crohn's disease, ulcerative colitis, cancer.** Nanoparticles have been constantly found in colon tissue of subjects affected by cancer, Crohn's disease, and ulcerative colitis (Figure 36), while in healthy subjects nanoparticles were absent [147]. The nanoparticles present in diseased subjects had various chemical compositions and are not considered toxic in bulk form. Microscopic and energy dispersive spectroscopy analysis of colon mucosa indicated the presence of carbon, ceramic filosilicates, gypsum, sulphur, calcium, silicon, stainless steel, silver, and zirconium [147]. The size of debris varied from 50 nm to 100 microns, the smaller the particle the further is able to penetrate. The particles were found at the interface between healthy and cancerous tissue. Based on these findings it was suggested that the gastrointestinal barrier is not efficient for particles smaller than 20 microns [266].

Crohn's disease affects primarily people in developed countries, and occurs in both the native population and in immigrants from under-developed countries. It affects 1 in 1000 people [244]. Crohn's disease is believed to be caused by genetic predisposition together with environmental factors [244]. Recently it was suggested that there is an association between high levels of dietary nanoparticles (100 nm-1μm) and Crohn's disease [244]. Exogenous nanoparticles were found in macrophages accumulated in lymphoid tissue of the human gut, the lymphoid aggregates being the earliest sign of lesions in Crohn's disease [244]. Microscopy studies showed that macrophages located in lymphoid tissue uptake nanoparticles of: spherical anatase ($TiO_2$) with size ranging between 100-200 nm from food additives; flaky-like aluminosilicates 100-400 nm typical of natural clay; and environmental silicates 100-700 nm with various morphologies [271]. A diet low in exogenous particles seems to alleviate the symptoms of Crohn's disease [244].

This analysis is still controversial, with some proposing that an abnormal response to dietary nanoparticles may be the cause of this disease, and not an excess intake [268]. More precisely, some members of the population may have a genetic predisposition where they are more affected by the intake of nanoparticles, and therefore develop Crohn's disease [252]. Some evidence suggests that dietary nanoparticles may exacerbate inflammation in Crohn's disease [268]. These studies measured the intake of dietary particle, but did not analyze the levels of outdoor and indoor nanoparticle pollution at the subjects' residences. As was described previously, significant quantities of nanoparticles are cleared by the mucociliary escalator and subsequently swallowed, ultimately reaching the gastro-intestinal tract.

**Treatment.** The diseases associated with gastro-intestinal uptake of nanoparticles (such as Crohn's disease and ulcerative colitis) have no cure and often requires surgical intervention. Treatments aim to keep the disease in remission and consist of anti-inflammatory drugs and specially formulated liquid meals [244]. If dietary nanoparticles are conclusively shown to cause these chronic diseases, their use in foods should be avoided or strictly regulated.

### 4.8. Dermal uptake of nanoparticles

#### 4.8.1. Penetration sites

The skin is composed of three layers - epidermis, dermis and subcutaneous (Figure 37 a). The outer portion of the epidermis, called *stratum corneum* - is a 10 $\mu$m thick keratinized layer of dead cells and is difficult to pass for ionic compounds and water soluble molecules [30]. The surface of epidermis is highly microstructured, as seen in Figure 37, having a scaly appearance as well as pores for sweat, sebaceous glands, and hair follicle sites.

As with many subjects involving nanoparticles, dermal penetration is still controversial [18]. Several studies show that nanoparticles are able to penetrate the stratum corneum [18], [20], [87-89], [272], [273]. Nanoparticle penetration through the skin typically occurs at hair follicles [272], and flexed [273] and broken skin [20]. Intracellular nanoparticles penetration is also possible, as demonstrated by cell culture experiments [274]. MWCNTs are internalized by human epidermal keratinocytes (the major cell type of the epidermis) in cytoplasmic vacuoles and induce the release of pro-inflammatory mediators [274]. Spherical particles with diameter between 750 nm and 6 microns selectively penetrate the skin at hair follicles with a maximum penetration depth of more than 2400 microns (2.4 mm) [272]. Broken skin facilitates the entry of a wide range of larger particles (500 nm - 7 $\mu$m) [20]. While stationary unbroken skin has been shown to be impervious to penetration, nanoparticles have been observed to penetrate when the skin is flexed. Thus mechanical deformation is capable of transporting particles through the stratum corneum and into the epidermis and dermis.

A current area under discussion is whether or not nanoparticles of $TiO_2$ found in commercially available sunscreens penetrate the skin [275]. For example, the application of a sunscreen containing 8% nanoparticles (10-15nm) onto the skin of humans showed no penetration, while oil-in-water emulsions showed penetration, higher penetration being present in hairy skin at the hair follicles site or pores [275]. The quantity of nanoparticles that penetrate is very small, with less than 1% of the total amount in the applied sunscreen being found in a given hair follicle [276].

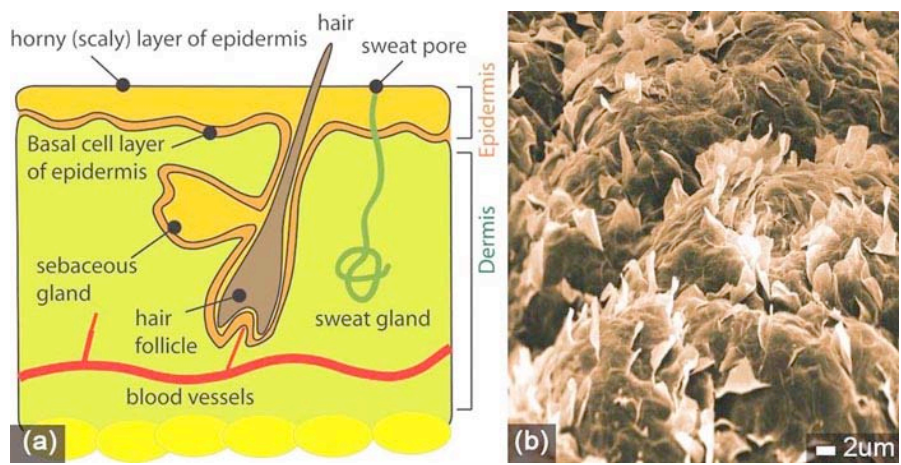

**Figure 37. (a)** *Schematics of cross-section in the skin. (b) The surface of human skin epidermis. © Dr. David M. Phillips/Visuals Unlimited, reproduced with permission from Visuals Unlimited [25].*

### 4.8.2. Translocation

The dermis has a rich supply of blood and macrophages, lymph vessels, dendritic cells, and nerve endings [20]. Therefore, the particles that cross through the stratum corneum and into the epidermis and dermis are potentially available for recognition by the immune system.

Translocation of nanoparticles through skin into the lymphatic system is demonstrated by soil particles found in lymph nodes of patients with podoconiosis [87-89].

Neuronal transport of small nanoparticles along sensory skin nerves may be possible, in a similar way to the proven path for herpes virus [20].

### 4.8.3. Adverse health effects of dermal uptake

Many manufacturing processes pose an occupational health hazard by exposing workers to nanoparticles and small fibers, as suggested from the intracellular uptake of MWCNTs by human epidermal keratinocytes [274]. This can explain beryllium sensitization in workers wearing inhalation protective equipment exposed to nanoparticulate beryllium [273]. Also, this may be relevant for latex sensitivity and other materials that provoke dermatologic responses.

**Soil particles.** Lymphatic system uptake of nanoparticles via the dermis is shown to cause podoconiosis (Figure 12 d) [87-89] and Kaposi's sarcoma (Figure 12 f) [81], [90], diseases discussed in Chapter 3.1.3.

**Titanium dioxide.** Currently, a controversial subject is the toxicity of titanium dioxide from cosmetics [139]. There are concerns about the toxicity of titanium dioxide - commonly used as a physical sunscreen since it reflects and scatters UVB (290-320 nm) and UVA (320-400 nm) light rays – the skin-damaging portion of the solar spectrum. $TiO_2$ also absorbs a substantial amount of UV radiation, however, which in aqueous media leads to the production of reactive oxygen species, including superoxide anion radicals, hydrogen peroxide, free hydroxyl radicals, and singlet oxygens. These reactive oxygen species can cause substantial damage to DNA [142]. Titanium dioxide particles under UV light irradiation have been shown to suppress tumor growth in cultured human bladder cancer cells via reactive oxygen species [277]. Sun-illuminated titanium dioxide particles in sunscreen were observed to catalyze DNA damage both *in vitro* and *in vivo* [143], [278]. Reports regarding the toxicity of titanium dioxide nanoparticles in the absence of UV radiation are contradictory. Nanoparticles were seen to have no inflammatory effect or genotoxicity in rats (when introduced by instillation) [279]. However, several other studies reported that titanium dioxide caused chronic pulmonary inflammation in rats (again by instillation) [280], and *in vitro* had a pro-inflammatory effect in cultured human endothelial cells [281].

**Silver.** It is known that silver has a beneficial antibacterial effect when used as a wound dressing, reducing inflammation and facilitating healing in the early phases [282], [283]. However, there are contradictory studies on silver nanoparticles and ions cytotoxicity from laboratories around the world. Silver is known to have a lethal effect on bacteria, but the same property that makes it antibacterial may render it toxic to human cells. Concentrations of silver that are lethal for bacteria are also lethal for both keratinocytes and fibroblasts [282].

### 4.9. Nanoparticles uptake via injection

Injection is the administration of a fluid into the subcutaneous tissue, muscle, blood vessels, or body cavities. Injection of nanoparticles has been studied in drug delivery.

The translocation of nanoparticles following injection depends on the site of injection: intravenously injected nanoparticles quickly spread throughout the circulatory system, with subsequent translocation to organs; intradermal injection leads to lymph nodes uptake; while intramuscular injection is followed by neuronal and lymphatic system uptake [20]. For example, the injection of magnetic nanoparticles smaller than 100 nm into the tongues and facial muscles of mice resulted in synaptic uptake [20].

Nanoparticles injected intravenously are retained longer in the body than ingested ones. For example, 90% of injected functionalized fullerenes are retained after one week of exposure [20].

Intravenously injected nanoparticles (quantum dots, fullerenes, polystyrene, plant virus) with size ranging from 10-240 nm show localization in different organs, such as liver, spleen, bone marrow, lymph nodes [20], small intestine, brain, lungs [270]. Talc particles introduced by injection are found in the liver of intravenous drug users [265]. The distribution of particles in the body is a function of their surface characteristics and their size. Coating nanoparticles with various types and concentrations of surfactants before injection significantly affects their distribution in the body [284]. For example, coating with polyethylene glycol or other substances almost completely prevents hepatic and splenic localization [20], [284]. Another example is the modification of nanoparticles surface with cationic compounds that facilitate arterial uptake by up to 10 fold [285].

The adverse health effects of injected nanoparticles are a function of particle chemistry and charge. A common side effect of injecting nanoparticles intravenously is hypersensitivity, a reaction that occurs in a large number of recipients and is probably due to the complement activation [286].

### 4.10. Nanoparticles generation by implants

Nanoparticle debris produced by wear and corrosion of implants is transported to region beyond the implant [265]. Implants release metal ions and wear particles and, after several years of wear, in some cases the concentration of metals in blood exceeds the biological exposure indices recommended for occupational exposure [287].

Materials considered chemically inert in bulk form (like ceramic porcelain and alumina), or in other terms biocompatible, are used for implants and prostheses [265]. However, nanoparticles with the same composition have been observed in liver and kidneys of diseased patients with implants and prostheses. It was suggested that the concept of biocompatibility should be revised in view of these findings [265].

Patients with orthopedic implants have a statistically significant rise in the incidence of autoimmune diseases, perhaps due to the particulate wear debris generated by the implant, which is associated with electrochemical processes that may activate the immune system [288]. Immunological responses and aseptic inflammation in patients with total hip replacement are a response to wear particles [289]. Exposure to orthopedic wear-debris leads to inflammatory initiated bone resorption, implant failure, dermatitis, urticaria, and vasculitis [288], [290]

### 4.11. Positive effects of nanoparticles

#### 4.11.1. Nanoparticles as antioxidants

Fullerene derivatives [61] and nanoparticles made of compounds holding oxygen vacancies ($CeO_2$, and $Y_2O_3$) [62] have demonstrated neuroprotective properties and antiapoptotic activity. Fullerene derivatives have been shown to prevent apoptosis in hepatic, kidney, and neuronal cells, a fact attributed to their antioxidant properties [61]. The decrease of apoptotic cell death is related to the neutralization of reactive oxygen species both *in vitro* and *in vivo*. Neurodegenerative disorders, such as Parkinson's and Alzheimer's diseases present hyper-production of oxygen and nitric oxide radical species [61]. As described previously, oxidative stress by oxygen radicals induces cellular instability by a cascade of events, leading to cell death. The use of fullerenes as radical sponges (or scavengers) has been shown to decrease neuronal death [61]. Functionalized fullerenes can react with oxygen species that attack lipids, proteins, and DNA, conferring neuroprotective properties. In particular, poly-hydroxylated fullerenes (fullerols) [$C_{60}(OH)_n$] are excellent antioxidants and offer exceptional neuroprotective properties, having high solubility and ability to cross the blood-brain barrier [61].

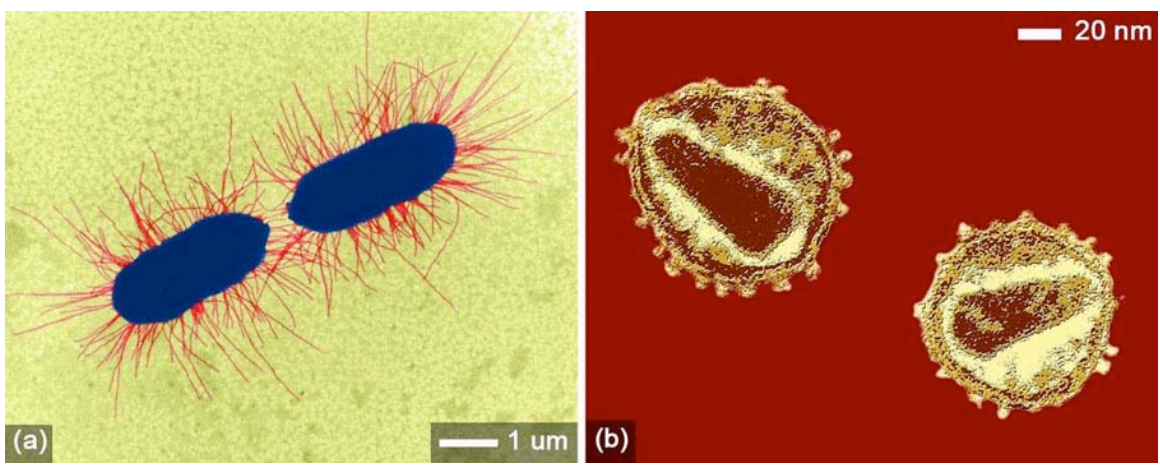

**Figure 38.** *(a) E. Coli bacteria, just after division, showing fimbriae on the cell surface. © Dr. Dennis Kunkel / Visuals Unlimited, reproduced with permission [25]. (b) The HIV Virus. © Dr. Hans Gelderblom/Visuals Unlimited, reproduced with permission [25].*

*4.11.2. Anti-microbial activity*

Several types of nanoparticle are known to have an antimicrobial effect, such as: silver [14], titanium dioxide [291], fullerenes [61], zinc oxide [292], and magnesium oxide [286].

Antimicrobial activity of fullerenes was observed on various bacteria, such as *E. Coli* (Figure 38 b), *Salmonella, Streptococcus spp.* [61]. The bactericide action is probably due to inhibition of energy metabolism once the bacteria have internalized the nanoparticles. Zinc oxide nanoparticles are bactericidal, disrupting membrane permeability and being internalized by Escherichia coli bacteria [292]. Silver nanoparticles and ions are broad spectrum antimicrobial agents [293]. Their antibacterial action results from destabilization of the outer membrane of bacteria, and depletion of the levels of adenosine triphosphate, a molecule that is the principal form of energy immediately usable by the cell.

Fullerenes have also been shown to have an anti-HIV activity, probably due to a good geometrical fit of a $C_{60}$ sphere into the active site (diameter of about 1 nm) on the fundamental enzyme (HIV protease) necessary for HIV virus (Figure 38 a) survival, leading to strong Van der Waals interactions between the enzyme and fullerene [61]. It has been demonstrated that silver nanoparticles undergo a size dependent interaction with HIV-1 virus, with nanoparticles exclusively in the range of 1–10 nm attached to the virus [294]. Due to this interaction, silver nanoparticles inhibit the virus from binding to host cells, as demonstrated *in vitro*.

## 5. Physico-chemical characteristics dependent toxicity

From previous knowledge of toxicological properties of fibrous particles (such as asbestos), it is believed that the most important parameters in determining the adverse health effects of nanoparticles are dose, dimension, and durability (the three D's) [38]. However, recent studies show different correlations between various physico-chemical properties of nanoparticles and the associated health effects, raising some uncertainties as to which are the most important parameters in deciding their toxicity: mass, number, size, bulk or surface chemistry, aggregation, or all together. In the following we will emphasize what we believe are the most important nanoparticle characteristics associated with their toxicity.

### 5.1. Dose-dependent toxicity

Dose is defined as the *amount* or *quantity* of substance that will reach a biological system. The dose is directly related to exposure or the concentration of substance in the relevant medium (air, food, water) multiplied by the duration of contact.

Generally, the negative health effects of nanoparticles do not correlate with nanoparticle mass dose (see Figure 39) [20], [49]. Comparing the health effects of inhaled $TiO_2$ nanoparticles with different sizes, it is remarkable that the low dose (10 mg/m$^3$) exposure to 20 nm diameter particles resulted in a greater lung tumor incidence than the high dose (250 mg/m$^3$) exposure of 300 nm diameter particles [30]. The measure that correlates with the effects is the surface area and not the mass dose (Figure 39 a) [20], [216], [295].

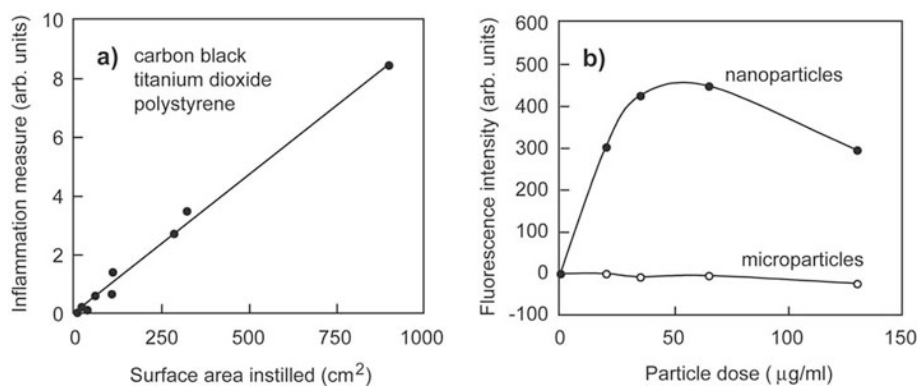

**Figure 39.** *(a) Inflammation generated by instillation of low-toxicity particles (carbon black, titanium dioxide and polystyrene) with the dose expressed as surface area, after [49] (b) Indication of oxidation induced fluorescence for nanoparticles and microparticles versus mass dose, after [49].*

### 5.2. Size-dependent toxicity

In the last decade, toxicological studies have demonstrated that small nanoparticles (<100 nm) cause adverse respiratory health effects, typically causing more inflammation than larger particles made from the same material [20], [49], [142], [216], [242], [296]. Rat inhalation [216] and instillation [20] of titanium oxide particles with two sizes, 20 nm and 250 nm diameter, having the same crystalline structure show that smaller particles led to a persistently high inflammatory reaction in the lungs compared to larger size particles. In the postexposure period (up to 1 year) it was observed that the smaller particles had (1) a significantly prolonged retention, (2) increased translocation to the pulmonary interstitium and pulmonary persistence of nanoparticles, (3) greater epithelial effects (such as type II cell proliferation), (4) impairment of alveolar macrophages function [216].

### 5.3. Surface area-dependent toxicity

For the same mass of particles with the same chemical composition and crystalline structure, a greater toxicity was found from nanoparticles than from their larger counterparts. This led to the conclusion that the inflammatory effect may be dependent on the surface area of nanoparticles, suggesting a need for changes in definitions and regulations related to dose and exposure limits. Indeed, smaller nanoparticles have higher surface area and particle number per unit mass compared to larger particles. The body will react differently to the same mass dose consisting of billions of nanoparticles compared to several microparticles. Larger surface area leads to increased reactivity [27] and is an increased source of reactive oxygen species, as demonstrated by *in vitro* experiments [49].

Intratracheal instillation studies on mice with titanium dioxide anatase show that small nanoparticles (20 nm) induce a much greater inflammatory response than larger nanoparticles (250 nm) for the same mass dose [20]. If instilled at the same surface area dose, they generated similar toxicity, fitting the same curve [20].

The higher surface area of nanoparticles causes a dose dependent increase in oxidation [49] and DNA damage [46], much higher than larger particles with the same mass dose [49]. Giving an example for the dose, high levels of oxidative DNA have been observed in cell culture experiments at 25 $\mu$g per well, with surface area of wells of 9.6 $cm^2$ [46]. In a simplified calculation, for a total surface area of the human lung alveolar region of 75 $m^2$, from which 3% are type II epithelial cells (target for cancer development), this dose is equivalent to about 4 years of exposure at the highest ambient particle concentration [46]. However, mathematical modeling of particle deposition in the airways indicate that some cells may receive 100-fold more particles depending on their orientation geometry [297]. Other studies suggested a threshold of 20 $cm^2$ surface area of instilled nanoparticles below which there is no significant inflammatory response in mice [295]. Extrapolating these findings to humans and environmental pollution, the critical surface area of nanoparticles becomes 30,000 $cm^2$ [295]. In a busy urban area with nanoparticles concentrations of up to 10 $\mu$g/$m^3$, with specific surface area of 110$m^2$/g, deposition efficiency of 70%, the lung burden results in 150 $cm^2$/day. If deposited particles accumulate in the lungs, the surface threshold for significant inflammatory effects is reached in about half a year [295]. However, subjects with respiratory or cardiovascular diseases may have a lower threshold. In addition, cardiovascular consequences may appear at a lower pollution threshold. We must emphasize that epidemiological studies do not indicate the existence of a threshold below which there are no adverse health effects [295].

Attempts have been made to contradict surface-area dependent toxicity [298]. One study claims that they tested toxicity of smaller nanoparticles against larger nanoparticles of similar composition and their findings show that they generate similar cytotoxicity or inflammatory reaction within the lung [298]. However, they used two different forms of titanium dioxide: rutile and anatase, which seems to have different toxicity levels regarding generation of oxidative compounds [142]. Similar composition does not necessary implies similar chemistry, chemical bonds. The best example is carbon, whose allotropes are: graphite, diamond, carbon nanotubes, and fullerenes, each with distinct physical and biological characteristics.

### 5.4. Concentration-dependent toxicity

There are many contradictory results related to the toxic effects of nanoparticles at different concentrations. Some studies show that certain materials are not as toxic as was observed by other studies. When comparing the results of different studies one must take into account that there are differences in the aggregation properties of nanoparticles in air and water, resulting in inherent discrepancies between inhalation studies and instillation or in vitro experiments. The aggregation may depend on surface charge, material type, size, among others.

One must stress the fact that aggregation of nanoparticles is essential in determining their toxicity, due to a more effective macrophage clearance for larger particles compared to smaller ones (that seem to evade easier this defense mechanism), leading to reduced toxicity of nanoparticle aggregates larger than 100-200 nm [20], [145]. It has been demonstrated that a high concentration of nanoparticles would promote particle aggregation [142], [299], and therefore reduce toxic effects compared to lower concentrations [145]. Most aggregates are observed to be larger than 100 nm, a size that seems to be a threshold for many of the adverse health effects of small particles. Therefore, experiments performed with high concentrations of nanoparticles will lead to the formation of nanoparticle aggregates that may not be as toxic as lower concentrations of the same nanoparticles.

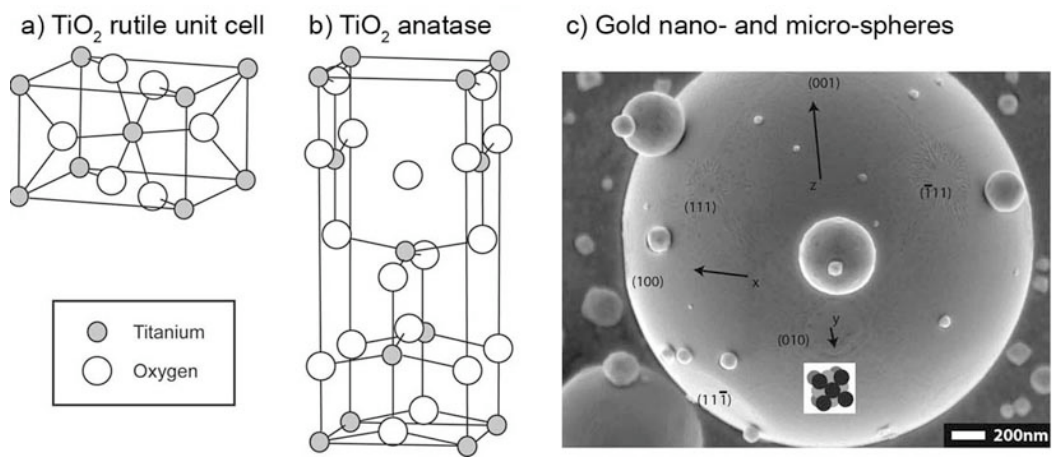

**Figure 40.** *(a) Unit cells of rutile, and (b) anatase, both crystalline forms of titanium dioxide; (c) gold micro- and nanoparticles formed by vacuum evaporation and vapor-phase condensation (Kevin Robbie, unpublished).*

### 5.5. Particle chemistry and crystalline structure dependent toxicity

Although there have been suggestions that size may be more important than chemical composition in deciding nanoparticles toxicity [46], one can not generally extrapolate the results of studies showing similar extent of inflammation for different nanoparticles chemistries. Particle chemistry is critical in determining nanoparticles toxicity. Particle chemistry is especially relevant from the point of view of cell molecular chemistry and oxidative stress. Namely, depending on their chemistry, nanoparticles can show different cellular uptake, subcellular localization, and ability to catalyze the production of reactive oxygen species [228].

One must make the distinction between composition and chemistry. Though particles may have the same composition, they may have different chemical or crystalline structure. The toxicity of a material depends on its type of crystalline form [142]. Let's take for example rutile and anatase, shown in Figure 40 a, b, both allotropes of titanium dioxide, i.e. polymorphs with the same chemical composition, but different crystalline structure, and hence different chemical and physical properties. Rutile nanoparticles (200 nm) were found to induce oxidative DNA damage in the absence of light, but anatase nanoparticles of the same size did not [142].

Nanoparticles can change crystal structure after interaction with water or liquids. For example, it is reported that zinc sulphide ZnS nanoparticles (3 nm across containing around 700 atoms) rearrange their crystal structure in the presence of water and become more ordered, closer to the structure of a bulk piece of solid ZnS [300]. Nanoparticles often exhibit unexpected crystal structures due to surface effects (Figure 40 c). The collection of gold nano- and microparticles shown in Figure 40 c was made by evaporting gold by heating it with an electron beam, and allowing the vapourized atoms sufficient time and density to condense into clusters before collection on a substrate. Condensation dynamics dictate that gold under these conditions will form these crystalline particles, which form equilibrium-seeking quasi-spheres as the condensing atoms jostle each other in random walks on the surface towards final resting places within the crystal. The effects of crystallinity on condensation are clearly observed in the faceting, and fine (nano) structure of the crystal faces. Incidentally interesting is the dendritic patterns on the (111) faces where the condensation forms a classic diffusion-limited aggregation structure. These nanoparticles are similar to the engineered nanoparticles produced in many industrial processes - they are engineered or designed by developing unique recipes that yield materials with beneficial characteristics. Finally, note the size of the largest gold particle in Figure 40 c, and that of the two progressively smaller particles stacked one upon the other. The largest is 2.5 μm in diameter with approximately $10^{11}$ atoms, the middle is 450 nm with $10^9$ atoms, and the smallest on top is 80 nm with $10^7$ atoms. The smallest nanoparticle in the image, just below the 'x' arrow, is only 25 nm in diameter, and contains roughly half a million atoms. Unique behavior emerges from these, and other nanomaterials, when small clusters of atoms form and manifest quantum effects.

### 5.6. Aspect ratio dependent toxicity

It was found that the higher the aspect ratio, the more toxic the particle is [201]. More exactly, lung cancer was associated with the presence of asbestos fibers longer than 10 microns in the lungs, mesothelioma with fibers longer than 5 microns, asbestosis with fibers longer than 2 microns [201]. All of these fibers had a minimum thickness of about 150 nm [201]. Long fibres (longer than 20 microns for humans) will not be effectively cleared from the respiratory tract due to the inability of macrophages to phagocytize them [30]. Alveolar macrophages were measured to have average diameters of 14-21 μm [38]. The biopersistence of these long aspect ratio fibers leads to long-term carcinogenic effects, as shown in Figure 41 [38].

The toxicity of long aspect ratio fibers is closely related to their bio-durability. The bio-durability of a fiber depends on its dissolution and mechanical properties (breaking). Longer fibers that break perpendicular to their long axis, become shorter and can be removed by macrophages. Asbestos fibers break longitudinally, resulting in more fibers with smaller diameter, being harder to clear [30]. If the lung clearance is slow, the longer the time these fibers will stay in the lung and the higher the probability of an adverse response. Fibers that are sufficiently soluble in lung fluid can disappear in a matter of months, while the insoluble fibers are likely to remain in the lungs indefinitely. Even short insoluble fibers that are efficiently phagocytized by alveolar macrophages may induce biochemical reactions (release of cytokines, reactive oxygen species, and other mediators).

Long aspect ratio engineered nanoparticles, such as carbon nanotubes (CNTs), are new materials of emerging technological relevance and have recently attracted a lot of attention due to their possible negative health effects [159], [274], [301-308], as suggested by their morphological similarities with asbestos. However, there is no consensus in the characterization of CNTs toxicity.

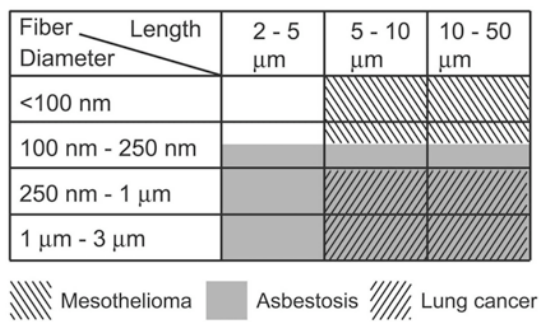

**Figure 41.** *Fiber health indices describing diseases associated to fibers of different size, after [201].*

The contradictory reports on CNTs toxicity could be associated with the multitude of morphologies, size, and chemical functionalization of their surface or ends. Carbon nanotubes can be single walled (SWCNTs) or multiple walled (MWCNTs), with varying diameter and length, with closed capped sections or open ends [309]. In addition to the many forms of nanotubes, they can also be chemically modified. The diameter of CNTs varies between 0.4 nm and 100 nm. Their lengths can range between several nanometers to centimeters [309]. Due to their hydrophobicity and tendency to aggregate, they are harmful to living cells in culture [274], [307]. For many applications, CNTs are oxidized to create hydroxyl and carboxyl groups, especially in their ends, which makes them more readily dispersed in aqueous solutions [310].

The conclusions of research on carbon nanotube cytotoxicity are that in general CNTs are very toxic, inducing cell death at sufficiently high doses of 400 µg/ml on human T cells [310], and 3.06 µg/cm$^2$ on alveolar macrophages [308]. Cell cultures with added SWCNTs at much lower doses of 3.8 µg/ml did not show cytotoxicity [306]. However, dose related inflammation or cell death is not in agreement between various studies. It was found that cells actively respond to SWCNTs by secreting proteins to aggregate and wrap them [306]. At the same time, SWCNTs induce up-regulation of apoptosis-associated genes [306].

Long-aspect ratio particles (SWCNTs) were reported to produce significant pulmonary toxicity compared to spherical particles (amorphous carbon black) [301], [303], [310]. Pharyngeal introduction of SWCNTs resulted in acute inflammation with onset of progressive fibrosis and granulomas in rats [301], [303]. For comparison, equal doses of carbon black or silica nanoparticles did not induce granulomas, alveolar wall thickening, causing only a weak inflammation and limited damage [301]. The enhanced toxicity was attributed to physicochemical properties and fibrous nature. Carbon nanotubes are not eliminated from the lungs or very slowly eliminated, 81% are found in rat lungs after 60 days after exposure [159].

### 5.7. Surface coating and functionalization

Due to the possibility of chemical interactions, the combined effects of inhalation, ingestion, or dermal application of nanoparticles with other nanoparticles, chemicals, and gases are largely unknown. The estimated risk of two or more pollutants is not a simple additive process. Particle surface plays a critical role in toxicity as it makes contact with cells and biological material. Surfactants can drastically change the physicochemical properties of nanoparticles, such as magnetic, electric, optical properties and chemical reactivity [20], [311], [312], affecting their cytotoxicity. Surface coatings can render noxious particles non-toxic while less harmful particles can be made highly toxic. The presence of oxygen, ozone [46], oxygen radicals [313], and transition metals [49] on nanoparticle surfaces leads to the creation of reactive oxygen species and the induction of inflammation. For example, specific cytotoxicity of silica is strongly associated with the occurrence of surface radicals and reactive oxygen species [30]. Experiments performed on hamsters showed that the formation of blood clots is more prominent when the surface of polystyrene nanoparticles is aminated [238]. Diesel exhaust particles interacting with ozone cause increased inflammation in the lungs of rats compared to diesel particles alone [46]. Nickel ferrite particles, with and without surface oleic acid, show different cytotoxicity [311]. The cytotoxicity of $C_{60}$ molecules systematically correlates with their chemical functionality in human (skin and liver) carcinoma cells with cell death occurring due to lipid oxidation caused by the generation of oxygen radicals [313]. Spherical gold nanoparticles with various surface coatings are not toxic to human cells, despite the fact that they are internalized [58], [59]. Quantum dots of CdSe can be rendered nontoxic when appropriately coated [60].

### 5.8. Adaptability to nanomaterials inhalation

Recent studies suggest that pre-exposure to lower concentrations of nanoparticles or shorter exposure times stimulates the phagocytic activity of cells, while high concentration of nanoparticles impairs this activity [177], [224], [225]. As a result, pulmonary inflammation is drastically reduced by several previous shorter exposure times to the same nanomaterials [177]. The severe pulmonary inflammatory response observed after only 15 minutes of rats exposed to $50 \mu g/m^3$ Teflon fume particles (with diameter of about 16 nm) can be prevented by three preceding daily 5 minutes exposure to the fumes [177]. During the three days of adaptation, the animals did not show clinical symptoms of respiratory effects, in contrast to the nonadapted group rats that were severely affected, showing difficulty breathing starting 1 h after exposure. The number of alveolar macrophages was significantly lower in nonadapted group.

### 5.9. Comparison studies

In order to assess the toxicity of various nanomaterials one must compare their toxic effects with those of known toxic particles. Several studies have pioneered this initiative [113], [159], [302], [303], [314], [315]. However, the database of studied materials is limited. The conclusions of these studies indicate that CNTs are extremely toxic, producing more damage to the lungs than carbon black or silica [159]. Varieties of CNT aggregates, and some carbon blacks, were shown to be as cytotoxic as asbestos (see Table 4) [113]. Silver nanoparticle aggregates were found to be more toxic than asbestos, while titanium oxide, alumina, iron oxide, zirconium oxide were found to be less toxic [113].

**Table 4.** *Nanomaterials, their morphologies, and their relative cytotoxicity index (RCI) on murine macrophage cells [113].*

| Material | Mean aggregate size (μm) | Mean particle size (nm) | RCI (at 5 μg/ml) | RCI (at 10 μg/ml) |
|---|---|---|---|---|
| Ag | 1 | 30 | 1.5 | 0.8 |
| Ag | 0.4 | 30 | 1.8 | 0.1 |
| $Al_2O_3$ | 0.7 | 50 | 0.7 | 0.4 |
| $Fe_2O_3$ | 0.7 | 50 | 0.9 | 0.1 |
| $ZrO_2$ | 0.7 | 20 | 0.7 | 0.6 |
| $TiO_2$ (rutile) | 1 | short fibers 5-15 nm diam. | 0.3 | 0.05 |
| $TiO_2$ (anatase) | 2.5 | 20 nm | 0.4 | 0.1 |
| $Si_3N_4$ | 1 | 60 | 0.4 | 0.06 |
| Asbestos Chrysotile | 7 | Fibers 20 nm diam., up to 500 aspect ratio | 1 | 1 |
| Carbon black | 0.5 | 20 | 0.8 | 0.6 |
| SWCNT | 10 | 100 nm diam. | 1.1 | 0.9 |
| MWCNT | 2 | 15 nm diam. | 0.9 | 0.8 |

**6. Applications of nanoparticles**

In this chapter we will outline several of the many applications of nanomaterials, both current and anticipated. To our knowledge, there is no comprehensive review of nanotechnology applications, likely due to the rapid development of this field. We feel that this chapter is necessary in order to broaden understanding of the importance that nanomaterials have and will play in our future, improving the quality of life through nanomedicine, electronics, and other nano-fields. Among the established applications of nanomaterials, we give as examples: microelectronics, synthetic rubber, catalytic compounds, photographic supplies, inks and pigments, coatings and adhesives, ultrafine polishing compounds, UV absorbers for sun screens, synthetic bone, ferrofluids, optical fiber cladding, and cosmetics. Applications currently entering widespread use include: fabrics and their treatments, filtration, dental materials, surface disinfectants, diesel and fuel additives, hazardous chemical neutralizers, automotive components, electronics, scientific instruments, sports equipment, flat panel displays, drug delivery systems, and pharmaceutics.

The unique properties of nanomaterials encourage belief that they can be applied in a wide range of fields, from medical applications to environmental sciences. Studies conducted by nanotechnology experts mapping the risks and opportunities of nanotechnology have revealed enormous prospects for progress in both life sciences and information technology [19]. Medical applications are expected to increase our quality of life through early diagnosis and treatment of diseases, and prosthetics, among others. Ecological applications include removal of persistent pollutants from soil and water supplies. Nanotechnology has become a top research priority in most of the industrialized world, including the USA, the EU and Japan. In the U.S.A. nanotechnology is now at the level of a federal program [316]. Since 2000, around 60 countries have initiated nanotechnology based initiatives at a national level [317].

*6.1. Electronics*

**Microelectronics**. Many of the current microelectronics applications are already at a nanoscale [318]. During the last four decades, the smallest feature of a transistor shrunk from 10 μm down to 30 nm [318]. The ultimate objective of microelectronics fabrication is to make electronic circuit elements that are nanoscopic. For example, by achieving a significant reduction in the size of circuit elements, the microprocessors (or better said, nanoprocessors) that contain these components could run faster and incorporate more logic gates, thereby enabling computations at far higher speeds. CNTs are exciting alternatives to conventional doped semiconductor crystals due to their varied electronic properties, ranging from metallic, to semiconducting [319], to superconducting [320].

**Displays.** The resolution of a television or a monitor improves with reduction of pixel size. The use of nanocrystalline materials can greatly enhance resolution and may significantly reduce cost. Also, flat-panel displays constructed with nanomaterials may possess much higher brightness and contrast than conventional displays owing to the enhanced electrical and optical properties of the new materials. CNTs are being investigated for low voltage field-emission displays [321]. Their combination of mechanical and electrical properties makes them potentially very attractive for long-life emitters.

**Data storage.** Devices, such as computer hard-disks function based on their ability to magnetize a small area of a spinning disk to record information, are established nano-applications. Discs and tapes containing engineered nanomaterials can store large amounts of information. Future avenues for magnetic recording that will drastically increase the capability of data storage include spintronics and nanowires.

**High energy density batteries.** New nanomaterials show promising properties as anode and cathode materials in lithium-ion batteries, having higher capacity and better cycle life than their larger-particle equivalents [322]. Among them are: aerogel intercalation electrode materials, nanocrystalline alloys, nanosized composite materials, carbon nanotubes, and nanosized transition-metal oxides [322].

**High-sensitivity sensors.** Due to their high surface area and increased reactivity, nanomaterials could be employed as sensors for detecting various parameters, such as electrical resistivity, chemical activity, magnetic permeability, thermal conductivity, and capacitance.

### 6.2. Transportation and telecommunication

**Car tires.** Nanoparticles of carbon black ranging between 10 nm - 500 nm act as a filler in the polymer matrix of tires, and are used for mechanical reinforcement.

**Car bumpers.** Clay particle based composites containing plastics and nano-sized clay are used to make car exteriors that are lighter and twice as resistant to scratches than usual materials [323].

### 6.3. Imaging

**Scanning microscope imaging.** SWCNTs have been used as probe tips for atomic-force microscopy imaging of antibodies, DNA, etc. [324]. Nanotubes are ideal probe tips for scanning microscopy due to their small diameter (which maximizes resolution), high aspect ratio, and stiffness.

**Molecular-recognition AFM tips.** SWCNTs with attached biomolecules are attached to AFM tips, and used for "molecular-recognition" in order to study chemical forces between molecules [324].

### 6.4. Biomedical applications

**Nanoscaffolds.** Nanofiber scaffolds can be used to regenerate central nervous system cells and possible other organs. Experiments performed on a hamster with severed optic tract demonstrated the regeneration of axonal tissue initiated by a peptide nanofibers scaffold [325].

**Antimicrobial nanopowders and coatings.** Certain nanopowders, possess antimicrobial properties [61], [326]. When these powders contact cells of *E. coli*, or other bacteria species and viruses, over 90% are killed within a few minutes. Due to their antimicrobial effect, nanoparticle of silver and titanium dioxide (<100nm) are assessed as coatings for surgical masks [291].

**Bioseparation.** Nanotube membranes can act as channels for highly selective transport of molecules and ions between solutions that are present on both side of the membrane [327]. For example, membranes containing nanotubes with inside diameters of molecular dimensions (less than 1 nm) separate small molecules on the basis of molecular size, while nanotubes with larger inside diameters (20–60 nm) can be used to separate proteins [328].

**Drug delivery.** The ability of nanoparticles to target and penetrate specific organs and cells contributes to their toxicity, however, this ability may be exploited in nanomedicine. Nanospheres composed of biodegradable polymers can incorporated drugs, allowing the timed release of the drug as the polymer degrades [329]. When particles are set to degrade in an acid microenvironment, such as tumor cells or around inflammation sites, this allows site-specific or targeted drug delivery.

**Gene transfection.** Surface-functionalized nanoparticles can be used to permeate cell membranes at a much higher level than nanoparticles without a functionalized surface [330]. This property can be used to deliver genetic material into living cells, a process called transfection. For example, silica nanospheres labeled on their outer surfaces with cationic ammonium groups can bind DNA (a polyanion) through electrostatic interactions [331]. Then nanoparticles deliver the DNA into cells.

**Medical imaging.** A variety of techniques currently called "non-invasive" have been used for more than a quarter of a century in medical imaging, for example superparamagnetic magnetite particles coated with dextran are used as image-enhancement agents in magnetic resonance imaging [332]. Intracellular imaging is also possible through attachment of quantum dots to selected molecules, which allows intracellular processes to be observed directly.

**Nasal vaccination.** Nanospheres carriers for vaccines are in development. Antigen-coated polystyrene nanospheres, used as vaccine carriers targeting human dendritic cells, have been researched for nasal vaccination [333]. Nanospheres had a direct effect on human dendritic cells, inducing transcription of genes important for, e.g., phagocytosis as well as an immune response.

**Nucleic acid sequence and protein detection.** Targeting and identifying various diseases could be made possible by detecting nucleic acid sequences unique to specific bacteria and viruses, or to specific diseases, or abnormal concentration of certain proteins that signal the presence of various cancers and diseases [334]. Nanomaterials-based assays are currently evaluated as well as more sensitive proteins detections methods. Nucleic acid sequences are currently detected with polymerase chain reaction (PCR) coupled with molecular fluorophore assays. Despite high sensitivity, PCR has significant drawbacks, such as: complexity, sensitivity to contamination, cost, and lack of portability [334]. Current protein detection methods, such as enzyme-linked immunoabsorbent assay (ELISA), allow the detection of proteins concentrations at which the disease is often advanced. More sensitive methods based on nanomaterials would revolutionize physical treatment of many cancer types and diseases [334].

**Smart nanophase extractors.** Differentially functionalized nanotubes are used as smart nanophase extractors, with molecular-recognition capabilities, to remove specific molecules from solutions [328].

**Treatment for local anesthetic toxicity.** Local anesthetic can be sometimes very toxic, ranging from local neurotoxicity to cardiovascular collapse and coma. In addition to conventional therapies, drug-scavenging nanoparticles have shown to increase survival rate from no animals in the control group to all animals in the treated group [335], [336].

## 6.5. Pollution remediation

Although research on environmental applications of nanoparticles is still a new area, it is growing rapidly. The potential of nanoparticles to react with pollutants in the air, soil, and water and transform them into harmless compounds is currently being researched. Nanotechnology could be applied at both ends of the environmental spectrum, to clean up existing pollution and to decrease or prevent its generation (see below).

**Elimination of pollutants.** Due to their enhanced chemical activity, nanomaterials can be used as catalysts to react with toxic gases (such as carbon monoxide and nitrogen oxide) in automobile catalytic converters and power generation equipment. This could prevent gaseous environmental pollution arising from burning gasoline and coal. Paints that absorb noxious gases from vehicle exhaust have already been developed [337]. They contain 30 nm spherical nanoparticles of titanium oxide and calcium carbonate mixed in a silicon-based polymer, polysiloxane, and absorb nitrogen oxide gases from vehicle exhausts, a pollution source that can cause smog and respiratory problems. The porous polysiloxane lets the nitrogen oxide gases diffuse and adhere to the titanium dioxide particles. UV radiation from sunlight converts nitrogen oxide to nitric acid, which is then neutralized by the calcium carbonate. The lifetime of the paint is said to be up to 5 years [337].

**Water Remediation.** Iron nanoparticles with a small content of palladium are tested to transform harmful products in groundwater into less harmful end products [338]. The nanoparticles are able to remove organic chlorine (a carcinogen) from water and soil contaminated with the chlorine-based organic solvents (used in dry cleaners) and convert the solvents to benign hydrocarbons.

### 6.6. Cosmetics

Titanium dioxide and zinc oxide become transparent to visible light when formed at the nanoscale, however are able to absorb and reflect UV light, being currently used in sunscreens and in the cosmetic industry. More cosmetics products containing nanoparticles are discussed in chapter 3.2.5.

### 6.7. Coatings

Nanomaterials have been used for very thin coatings for decades, if not centuries. Today thin coatings are used in a vast range of applications, including architectural glass, microelectronics, anticounterfeit devices, optoelectronic devices, and catalytically active surfaces. Structured coatings with nanometer-scale features in more than one dimension promise to be an important foundational technology for the future.

**Self-cleaning windows.** Self-cleaning windows have been demonstrated that are coated in highly hydrophobic titanium dioxide. The titanium dioxide nanoparticles speed up, in the presence of water and sunlight, the breakdown of dirt and bacteria that can then be washed off the glass more easily.

**Scratch resistant materials.** Nanoscale intermediate layers between the hard outer layer and the substrate material significantly improve wear and scratch resistant coatings. The intermediate layers are designed to give a good bonding and graded matching of mechanical and thermal properties, leading to improved adhesion.

**Textiles.** Nanoparticles have already been used in coating textiles such as nylon, to provide antimicrobial characteristics [339], Also the control of porosity at the nanoscale and surface roughness in a variety of polymers and inorganic materials led to ultrahydrophobic - waterproof and stain resistant fabrics.

### 6.8. Materials

**Insulation materials.** Nanocrystalline materials synthesized by the sol-gel technique exhibit a foam-like structure called an "aerogel" [340]. Aerogels are composed of three-dimensional, continuous networks of particles and voids. Aerogels are porous, extremely lightweight, and have low thermal conductivity.

**Nanocomposites.** Composites are materials that combine two or more components and are designed to exhibit overall the best properties of each component (mechanical, biological, optical, electric, or magnetic). Nanocomposites containing CNT and polymers used to control their conductivity are interesting for a wide range of applications, such as supercapacitors, sensors, solar cells, etc. [341].

**Paints.** Nanoparticles confer enhanced desired mechanical properties to composites, such as scratch resistant paints based on encapsulated nanoparticles [342]. The wear resistance of the coatings is claimed to be ten times greater than that for conventional acrylic paints.

### *6.9. Mechanical engineering*

**Cutting tools** made of nanocrystalline materials (such as tungsten carbide, WC) are much harder than their conventional due to the fact that the microhardness of nanosized composites is increased compared to that of microsized composites [343].

**Lubricants.** Nanospheres of inorganic materials could be used as lubricants, acting as nanosized ball bearings [344].

## 7. Conclusions and future directions

Human exposure to nanoparticles from natural and anthropogenic sources has occurred since ancient times. Following the invention of combustion engines and the development of industry, however, significant levels of nanoparticle pollution have arisen in most major cities and even across large regions of our planet, with climatic and environmental effects that are generally unknown.

There is heightened concern today that the development of nanotechnology will negatively impact public health, and it is indisputable that engineered nanomaterials are a source of nanoparticle pollution when not safely manufactured, handled, and disposed of or recycled. A large body of research exists regarding nanoparticle toxicity, comprising epidemiological, animal, human, and cell culture studies. Compelling evidence that relates levels of particulate pollution to respiratory, cardiovascular disease, and mortality, has shifted attention to particles with smaller and smaller sizes (nanometer scale). Research on humans and animals indicates that some nanoparticles are able to enter the body, and rapidly migrate to the organs via the circulatory and lymphatic systems. Subjects with pre-existing diseases (such as asthma, diabetes, among others) may be more prone to the toxic effects of nanoparticles. Genetic factors may also play an important role in the response of an organism to nanoparticles exposure.

As shown in this review, it is clear that workers in nanotechnology related industries may be potentially exposed to uniquely engineered nanomaterials with new sizes, shapes and physicochemical properties. Exposure monitoring and control strategies are necessary. Indeed, there is a need for a new discipline - nanotoxicology - that would evaluate the health threats posed by nanoparticles, and would enable safe development of the emerging nanotechnology industry [19]. We emphasize that this field of study should include not only newly engineered nanomaterials, but also those generated by nature and pollution.

The ability of nanoparticles to enter cells and affect their biochemical function makes them important tools at the molecular level. The toxic properties of nanoparticles can in some instances be harnessed to improve human health through targeting cancer cells or harmful bacteria and viruses. These very properties that might be exploited as beneficial may also have secondary negative effects on health and the environment. For example, nanoparticles used to destroy cancer cells may cause harmful effects elsewhere in the body, or nanoparticles used for soil remediation may have an adverse impact upon entering the food chain via microorganisms, such as bacteria and protozoa.

In the following we highlight important questions and research directions that should be addressed in the near future by the scientific community involved in the study of nanoparticles sciences and by government agencies responsible for regulations and funding.

Advanced analysis of the physical and chemical characteristics of nanoparticles will continue to be essential in revealing the relationship between their size, composition, crystallinity, and morphology and their electromagnetic response properties, reactivity, aggregation, and kinetics. It is important to note that fundamental properties of nanoparticles are still being discovered, such as magnetism in nanoparticles made of materials that are non-magnetic in bulk form. A systematic scientific approach to the study of nanoparticle toxicity requires correlation of the physical and chemical characteristics of nanoparticles with their toxicity. Existing research on nanotoxicity has concentrated on empirical evaluation of the toxicity of various nanoparticles, with less regard given to the relationship between nanoparticle properties (such as exact composition, crystallinity, size, size dispersion, aggregation, ageing) and toxicity. This approach gives very limited information, and should not be considered adequate for developing predictions of toxicity of seemingly similar nanoparticle materials.

Further studies on kinetics and biochemical interactions of nanoparticles within organisms are imperative. These studies must include, at least, research on nanoparticles translocation pathways, accumulation, short- and long-term toxicity, their interactions with cells, the receptors and signaling pathways involved, cytotoxicity, and their surface functionalization for an effective phagocytosis. Existent knowledge on the effects of nanoparticle exposure on the lymphatic and immune systems, as well as various organs, is sparse. For example it is known that nanoparticle exposure is able to modulate the response of the immune system to different diseases, however much research is needed in order to better understand to what extent this occurs and the full implications of risk groups (age, genotype). In order to clarify the possible role of nanoparticles in diseases recently associated with them (such as Crohn's disease, neurodegenerative diseases, autoimmune diseases, and cancer), nanoscale characterization techniques should be used to a larger extent to identify nanoparticles at disease sites in affected organs or tissues, and to establish pertinent interaction mechanisms.

Other important research topics to be pursued include nanoparticle ageing, surface modifications, and change in aggregation state after interaction with bystander substances in the environment and with biomolecules and other chemicals within the organisms. How do these interactions modify the toxicity of nanoparticles? Do they render toxic nanoparticles less toxic? Or can they render benign nanoparticles more toxic? What about the beneficial properties of some nanoparticles? Do they change in the short- and long-term after undergoing chemical interactions? Research should also be directed toward finding ways to reduce nanoparticle toxicity (such as antioxidants provided from dietary sources and supplements, metals chelators, anti-inflammatory agents).

Understanding and rationally dealing with the potentially toxic effects of nanoparticles requires a multidisciplinary approach, necessitating a dialogue between those involved in the disparate aspects of nanoparticle fabrication and their effects, including but not limited to nanomaterial fabrication scientists, chemists, toxicologists, epidemiologists, environmental scientists, industry, and policy makers. In order to achieve an interdisciplinary dialogue, systematic summaries should be prepared, discussing current knowledge in the various nano-fields, and using a common vocabulary. This will help bring together scientists in different fields, as well as policy makers and society at large. These summaries should include periodic written reviews, conferences, and accessible databases that contain the collected knowledge of nanoparticle synthesis, characterization, properties, and toxicity, in a format easily comprehensible to a wide audience of scientists. A database initiative has already begun, led by the National Institute for Occupational Safety and Health, as the "Nanoparticles Information Library".

We also suggest several directions for minimizing human exposure to nanoparticles, and thereby reducing associated adverse health effects. National governments and international organizations should enact stringent air quality policies, with standardized testing methods and low exposure limits. With such compelling existing evidence of the correlation between particle pollution levels, mortality, and a wide range of diseases (comprising cardiovascular, respiratory diseases, and malignant tumors), the primary source of atmospheric nanoparticles in urban areas – combustion-based vehicles - should be mandated to have lower nanoparticle emission levels. In the light of their potential toxicity, the commercialization of dietary and cosmetic nanoparticles, as well as other consumer products incorporating nanoparticles, must be strictly regulated. In particular they must be regulated as distinct materials from their bulk constituents. Before using these nanoparticles several questions should be answered: Are they biocompatible? Do they translocate and accumulate in the body (including skin)? What are the long-term effects of uptake and accumulation? In general, consumer products containing nanomaterials should be recycled. A model initiative began in 2001 in Japan for electrical appliances, where the retailers, manufacturers, and importers are now responsible for recycling the goods they produce or sell.

There is limited existing research regarding ecological and environmental implications of natural and anthropogenic nanoparticles pollution, though the role of nanoparticles in some forms of environmental degradation is well known, e.g. atmospheric nanoparticles play a central role in ozone depletion. Nanoparticulate pollution is likely to play an important role in global climate balance, despite the fact that current anthropogenic climate changes are attributed solely to greenhouse gases. This is dangerous as it encourages the misconception that wood burning does not contribute to pollution and/or climate change. In a simple calculation of carbon liberation and fixation, it appears that wood burning, as a so-called "renewable" source of energy, is benign to the environment. A proper accounting of nanoparticle pollution in addition to $CO_2$ reveals the naivety of this analysis.

Advances in nanotechnology are driven by rapid commercialization of products containing nanostructures and nanoparticles with remarkable properties. This is reflected in the enormous number of publications on nanotechnology. In comparison, the number of publications on nanoparticle toxicity is much smaller, as the funding available for toxicity studies are mostly government related. One way of increasing funding for nanotoxicity research might be via international regulations requiring that a fraction of the revenues of each company involved in their production and commercialization to be dedicated to this field of research. Without this level of commitment it is likely that a current or future industrial nanoparticle product, with non-obvious or delayed toxicity, will cause significant human suffering and/or environmental damage. The field of nanotechnology has yet to have a significant public health hazard, but it is a real possibility that can and should be prevented.

We conclude that the development of nanotechnology and the study of nanotoxicology have increased our awareness of environmental particulate pollution generated from natural and anthropogenic sources, and hope that this new awareness will lead to significant reductions in human exposure to these potentially toxic materials. With increased knowledge, and ongoing study, we are more likely to find cures for diseases associated with nanoparticle exposure, as we will understand their causes and mechanisms. We foresee a future with better-informed, and hopefully more cautious manipulation of engineered nanomaterials, as well as the development of laws and policies for safely managing all aspects of nanomaterial manufacturing, industrial and commercial use, and recycling.

## Acknowledgments

We gratefully acknowledge financial support from the Natural Sciences and Engineering Council of Canada (NSERC), the Canada Research Chairs Program (CRC), the Canadian Institute for Photonic Innovations (CIPI), and the Ontario Photonic Consortium (OPC). We also thank Jennifer K. Gregg and Prof. Alex Braginski for a critical reading of the manuscript.

# References

[1] Jarrett R F 2006 Viruses and lymphoma/leukaemia *J. Pathol.* **208** 176–186

[2] DiMaio D, Liao J B 2006 Human papillomaviruses and cervical cancer *Adv. Virus. Res.* **66** 125-159

[3] Levrero M 2006 Viral hepatitis and liver cancer: the case of hepatitis C *Oncogene* **25** 3834-3847

[4] Kusters J G, van Vliet A H, Kuipers E J 2006 Pathogenesis of Helicobacter pylori infection *Clin. Microbiol. Rev.* **19** 449-490

[5] Burgos J S 2005 Involvement of the Epstein-Barr virus in the nasopharyngeal carcinoma pathogenesis *Med. Oncol.* **22** 113-121

[6] Kajander E O and Çiftçioglu N 1998 Nanobacteria: an alternative mechanism for pathogenic intra- and extracellular calcification and stone formation *Proc. Natl. Acad. Sci.* **95** 8274-8279

[7] Kahn J S 2006 The widening scope of coronaviruses *Curr. Opin. Pediatr.* **18** 42-47

[8] Kol A, Santini M 2004 Infectious agents and atherosclerosis: current perspectives and unsolved issues *Ital. Heart J.* **5** 350-357

[9] Fohlman J, Friman G 1993 Is juvenile diabetes a viral disease? *Ann. Med.* **25** 569-574

[10] Itzhaki R F, Wozniak M A, Appelt D M, Balin B J 2004 Infiltration of the brain by pathogens causes Alzheimer's disease *Neurobiol. Aging.* **25** 619-627

[11] Lynch N E, Deiratany S, Webb D W, McMenamin J B 2006 PANDAS (Paediatric Autoimmune Neuropsychiatric Disorder Associated with Streptococcal Infection *Ir. Med. J.* **100** 155

[12] Miranda HC, Nunes S O, Calvo E S, Suzart S, Itano E N, Watanabe M A 2006 Detection of Borna disease virus p24 RNA in peripheral blood cells from Brazilian mood and psychotic disorder patients *J. Affect. Disord.* **90** 43-47

[13] Janka J, Maldarelli F 2004 Prion Diseases: Update on Mad Cow Disease, Variant Creutzfeldt-Jakob Disease, and the Transmissible Spongiform Encephalopathies *Curr. Infect. Dis. Rep.* **6** 305-315

[14] A Nanotechnology Consumer Products Inventory, http://www.nanotechproject.org/index.php?id=44

[15] Nanotechnology. Untold promise, unknown risk, Consumer Reports, July 2007, 40-45, www.ConsumerReports.org.

[16] Feynman R P 1959 There's Plenty of Room at the Bottom, An Invitation to Enter a New Field of Physics, Annual meeting of the American Physical Society, California Institute of Technology, Dec. 29; or http://www.zyvex.com/nanotech/feynman.html.

[17] Kittelson D B 2001 Recent measurements of nanoparticle emission from engines *Current Research on Diesel Exhaust Particles*, Japan Association of Aerosol Science and Technology, 9 January (Tokyo, Japan) and references therein

[18] Borm P J A, Robbins D, Haubold S, Kuhlbusch T, Fissan H, Donaldson K, Schins R P F, Stone V, Kreyling W, Lademann J, Krutmann J, Warheit D, Oberdorster E 2006 The potential risks of nanomaterials: a review carried out for ECETOC (review) *Part. Fibre Toxicol.* **3** 11

[19] Donaldson K, Stone V, Tran C, Kreyling W, Borm P J A 2004 Nanotoxicology *Occup. Environ. Med.* **61** 727-728 and references therein

[20] Oberdörster G, Oberdörster, E, Oberdörster J 2005 Nanotoxicology: an emerging discipline evolving from studies of ultrafine particles *Environ. Health. Perspect.* **113** 823-839 and supplemental material found at http://www.ehponline.org/members/2005/7339/7339.html

[21] Public Health Image Library, http://phil.cdc.gov/Phil/home.asp

[22] Jim Ekstrom, http://science.exeter.edu/jekstrom/SEM/SEM.html

[23] Hansen W R, Autumn K 2005 Evidence for self-cleaning in gecko setae *Proc. Nat. Acad. Sci.* **102** 385-

389

[24] Sioutas C, Delfino R J, Singh M 2005 Exposure Assessment for Atmospheric Ultrafine Particles (UFPs) and Implications in Epidemiologic Research *Environ. Health Res*. **113** 947-955 and references therein

[25] www.visualsunlimited.com

[26] https://histo.life.uiuc.edu/histo/atlas/index.php, University of Illinois at Urbana-Champaign, College of Medicine

[27] Roduner E 2006 Size matters: why nanomaterials are different *Chem. Soc. Rev*. **35** 583-592

[28] Kouwenhoven L P, Austing D G, Tarucha S 2001 Few-electron quantum dots *Rep. Prog. Phys*. **64** 701-736

[29] ISI web of science http://portal.isiknowledge.com/

[30] Hoet P H M, Bruske-Hohlfeld I, Salata O V 2004 Nanoparticles - known and unkown health risks *J. Nanobiotechnol*. **2** 12-27 and references therein

[31] Maynard A D, Kuempel E D 2005 Airborne nanostructured particles and occupational health *J. Nanopart. Res*. **7** 587-614

[32] Nel A, Xia T, Madler L, Li N 2006 Toxic potential of materials at the nanolevel *Science* **311** 622-627

[33] Kunzli N, Tager I B 2005 Air pollution: from lung to heart *Swiss Med. Wkly*. **135** 697-702

[34] Englert N 2004 Fine particles and human health – a review of epidemiological studies *Toxicol. Lett*. **149** 235-242

[35] Delfino R J, Sioutas C, Malik S 2005 Potential role of ultrafine particles in association between airborne particle mass and cardiovascular health *Environ. Health Persp*. **113** 934-946

[36] Ferin J 2004 Pulmonary retention and clearance of particles *Toxicol. Lett*. **72** 121-125 and references therein

[37] Oberdorster G 2001 Pulmonary effects of inhaled ultrafine particles *Int. Arch. Occup. Environ. Health* **74** 1-8

[38] Oberdörster G 2002 Toxicokinetics and effects of fibrous and nonfibrous particles *Inhalation Toxicol*. **14** 29-56 and references therein

[39] Borm P J A, Schins R P F, Albrecht C 2004 Inhaled particles and lung cancer, Part B: paradigms and risk assessment *Int. J. Cancer* **110** 3-14 and references therein

[40] Knaapen A M, Borm P J A, Albrecht C, Schins R P F 2004 Inhaled particles and lung cancer. Part A: mechanisms *Int. J. Cancer*. **109** 799-809

[41] Donaldson K, Tran L, Jimenez L A, Duffin R, Newby D E, Mills N, MacNee W, Stone V 2005 Combustion-derived nanoparticles: a review of their toxicology following inhalation exposure *part. Fibre Toxicol*. **2** 10

[42] Brown R C, Lockwook A H, Sonowana B R 2005 Neurodegenerative diseases: an overview of environmental risk factors *Environ. Health Perspect*. **113** 1250-1256 and references therein

[43] Churg A 2003 Interactions of exogeneous or evoked agents and particles: the role of reactive oxygen species *Free Radical Biol. & Med*. **34** 1230–1235

[44] Donaldson K, Stone V, Borm P J A, Jimenez L A, Gilmour P S, Schins R P F, Knaapen A M, Rahman I, Faux S P, Brown D M, MacNee W 2003 Oxidative stress and calcium signaling in the adverse effects of environmental particles PM10 *Free Radical Biol. & Med*. **34** 1369-1382

[45] Gonzales-Flecha B 2004 Oxidant mechanisms in response to ambient air particles *Mol. Aspects Med*. **25** 169-182

[46] Risom L, Moller P, Loft S 2005 Oxidative stress-induced DNA damage by particulate air pollution *Mutat. Res*. **592** 119-137 and references therein

[47] Tao F, Gonzales-Flecha B, Kobzik L 2003 Reactive oxygen species in pulmonary inflammation by ambient particulates *Free Radical Biol. & Med*. **35** 327-340

[48] Peters A 2005 Particulate matter and heart disease: evidence from epidemiological studies *Toxicol. Appl. Pharmacol*. **207** S477-S482 and references therein

[49] Donaldson K, Stone V 2003 Current hypotheses on the mechanisms of toxicity of ultrafine particles *Ann. Ist. Super Sanita* **39** 405-410 and references therein

[50] Fubini B, Hubbard A 2003 Serial review: role of reactive oxygen and nitrogen species (ROS/RNS) in lung injury and diseases *Free Rad. Biol. Med*. **34** 1507-1516

[51] Hussain N, Jaitley V, Florence A T 2001 Recent advances in the understanding of uptake of

microparticulate across the gastrointestinal lymphatics *Adv. Drug Delivery Rev.* **50** 107-142

[52] Moghimi S M, Hunter A C, Murray J C 2005 Nanomedicine: current status and future prospects *FASEB J.* **19** 311-330

[53] Powers K W, Brown S C, Krishna V B, Wasdo S C, Moudgil B M, Roberts S M 2006 Research strategies for safety evaluation of nanomaterials. Part VI. Characterization of nanoscale particles for toxicological evaluation *Toxicol. Sci.* **90** 296-303

[54] Oberdörster G, Maynard A, Donaldson K, Castranova V, Fitzpatrick J, Ausman K, Carter J, Karn B, Kreyling W, Lai D, Olin S, Monteiro-Riviere N, Warheit D, Yang H 2005 Principles for characterizing the potential human health effects from exposure to nanomaterials: elements of a screening strategy *Part. Fibre Toxicol.* **2** 8

[55] Thomas T, Thomas K, Sadrieh N, Savage N, Adair P, Bronaugh R 2006 Research strategies for safety evaluation of nanomaterials, part VII: evaluating consumer exposure to nanoscale materials *Toxicol. Sci.* **91** 14-19

[56] Colvin V L 2003 The potential environmental impact of engineered nanomaterials *Nature Biotechnol.* **21** 1166-1170

[57] Li N, Sioutas C, Cho A, Schmitz D, Misra C, Sempf J, Wang M, Oberley T, Froines J, Nel A 2003 Ultrafine Particulate Pollutants Induce Oxidative Stress and Mitochondrial Damage *Environ. Health Persp.* **111** 455-460

[58] Connor E E, Mwamuka J, Gole A, Murphy C J, Wyatt M D 2005 Gold nanoparticles are taken up by human cells but do not cause acute cytotoxicity *Small* **1** 325-327

[59] Goodman C M, McCusker C D, Yilmaz T, Rotello V 2004 Toxicity of gold nanoparticles functionalized with cationic and anionic side chains *Bioconjugate Chem.* **15** 897-900

[60] Derfus A M, Chan W C W, Bhatia S N 2004 Probing the cytotoxicity of semiconductor quantum dots *Nano Lett.* **4** 11-18

[61] Bosi S, da Ros T, Spalluto G, Prato M 2003 Fullerene derivatives: an attractive tool for biological applications *Eur. J. Med. Chem.* **38** 913-923 and references therein

[62] Schubert D, Dargusch R, Raitano J, Chan S W 2006 Cerium and yttrium oxide nanoparticles are neuroprotective *Biochem. Biophys. Res. Commun.* **342** 86-91

[63] Seshan K, Ed 2002 Handbook of Thin-Film Deposition Processes and Techniques - Principles, Methods, Equipment and Applications *William Andrew Publishing/Noyes* pp. 1-657

[64] Robbie K, Yang J, Elliott C, Dariani R, Buzea C 2007 Designed Nanoparticles, unpublished.

[65] Taylor D A 2002 Dust in the wind *Environ. Health Perspect.* **110** A80-A87

[66] Houghton J 2005 Global warming *Rep. Prog. Phys.* **68** 1343–1403

[67] Buseck P R, Pósfai M 1999 Airborne minerals and related aerosol particles: Effects on climate and the environment Proc. Nat. Acad. Sci. 96 3372-3379 and references therein

[68] Shi Z, Shao L, Jones T P, Lu S 2005 Microscopy and mineralogy of airborne particles collected during severe dust storm episodes in Beijing, China *J. Geophys. Res.* **110** D01303 and references therein

[69] d'Almeida G A, Schutz L 1983 Number, mass and volume distributions of mineral aerosol and soils of the Sahara *J. Clim. Appl. Meteor.* **22** 233-243

[70] Husar R B el al. 2001 The Asian Dust Events of April 1998 *J. Geophys. Res.* **106** 18317-18330

[71] McKendry I G, Hacker J P, Stull R, Sakiyama S, Mignacca D, Reid K 2001 Long-range transport of Asian dust to the Lower Fraser Valley, British Columbia, Canada *J. Geophys. Res.* Vol. **106** 18361-18370

[72] http://visibleearth.nasa.gov/, http://rapidfire.sci.gsfc.nasa.gov/firemaps/

[73] USA Today, Sept. 19 2005, http://www.usatoday.com/educate/college/healthscience/articles/20050925.htm

[74] http://www.lpi.usra.edu/expmoon/ Apollo17/A17_Experiments_SMI.html, Apollo 17 Soil Mechanics Investigation

[75] Taylor L A, Schmitt H H, Carrier W D III, Nakagawa M 2005 the lunardust problem: from liability to asset *First Space Exploration Conference: Contuinuing the Voyage of Discovery, 2005, Orlando, Florida* pdf.aiaa.org/preview/CDReadyMSEC05_1205/PV2005_2510.pdf

[76] http://science.nasa.gov/headlines/y2005/18mar_moonfirst.htm

[77] Watson T 2005 NASA to detail plans for trip to moon *USA Today* **18 September**

[78] Batsura I D, Kruglikov G G, Arutiunov V D 1981 Morphology of experimental pneumoconiosis

developing after exposure to lunar soil *Biull. Eksp. Biol. Med.* **92** 376-379

[79] Sapkota A, Symons J M, Kleissl J, Wang L, Parlange M B, Ondov J, Breysse P N, Diette G B, Eggleston P A, Buckley T J 2005 Impact of the 2002 Canadian Forest Fires on Particulate Matter Air Quality in Baltimore City *Environ. Sci. Technol.* **39** 24 –32

[80] http://maps.geog.umd.edu/

[81] Mott JA, Meyer P, Mannino D, Redd SC, Smith EM, Gotway-Crawford C, Chase E 2002 Wild land forest fire smoke: health effects and intervention evaluation, Hoopa, California, 1999. West J. Med. **176** 157-162

[82] Thibodeau GA, Patton KT, Anatomy and physiology, Fifth edition (2003) Mosby Inc., St. Luis

[83] National Aeronautics and Space Administration, Image ID: S73-15171, http://nix.larc.nasa.gov/

[84] Malin M C, Edgett K S, Carr M H, Danielson G E, Davies M E, Hartmann W K, Ingersoll A P, James P B, Masursky H, McEwen A S, Soderblom L A, Thomas P, Veverka J, Caplinger M A, Ravine M A, Soulanille T A, and Warren J L, NASA's Planetary Photojournal (http://photojournal.jpl.nasa.gov/), Dust Storms of 2001, MOC2-314, 7 May 2002; Devil-Streaked Plain, MOC2-1378, 19 February 2006

[85] Li J, Posfai M, Hobs P V, Buseck P R 2003 Individual aerosol particles from biomass burning in southern Africa: 2. Compositions and aging of inorganic particles *J. Geophys. Res.* **108** D13, 8484

[86] Yano E, Yokoyama Y, Higashi H, Nishii S, Maeda K, Koizumi A 1990 Health effects of volcanic ash: a repeat study *Archives Environ. Health* **45** 367-373

[87] Blundell G, Henderson W, Price E w 1989 Soil particles in the tissues of the foot in endemic elephantiasis of the lower legs *Ann. Trop. Med. Parasitol.* **83** 381-385

[88] Corachan M 1988 Endemic non-filarial elephantiasis of lower limbs - podoconiosis *Medicina Clinica* **91** 97-100

[89] Corachan M, Tura J M, Campo E, Soley M, Traveria A 1988 Podoconiosis in Aequatorial Guinea. Report of two cases from different geological environments *Trop. Geogr. Med.* 40 359-364

[90] Montella M, Franceschi S, Geddes da Filicaia M, De Macro M, Arniani S, Balzi D, Delfino M, Iannuzzo M, Buonanno M, Satriano RA 1997 Classical Kaposi sarcoma and volcanic soil in southern Italy: a case-control study *Epidemiol. Prev.* **21** 114-117

[91] Fuller L C 2005 Podoconiosis: endemic nonfilarial elephantiasis *Curr. Opin. Infect. Dis.* **18** 119-122

[92] Davey G, Hanna E G, Adeyemo A, Rotimi C, Newport M, Desta K 2006 Podoconiosis: a tropical model for gene—environment interactions? *Trans. R. Soc. Trop. Med. Hyg.* In print

[93] http://visibleearth.nasa.gov/view_rec.php?vev1id=25928

[94] Ballard S T, Parker J C, Hamm C R 2006 Restoration of mucociliary transport in the fluid-depleted trachea by surface-active instillates *Am. J. Respir. Cell. Mol. Biol.* **34** 500-5004

[95] http://eol.jsc.nasa.gov images id: ID. STS064-040-010; STS068-214-043; ISS005-E-19013, and ISS005-E-19024.

[96] Louisa Howard, louisa.howard@dartmouth.edu, http://remf.dartmouth.edu/imagesindex.html

[97] Simonart T 2004 Iron: a target for the management of Kaposi's sarcoma? *BMC Cancer* **4** 1 and references therein

[98] Ahmad A, Senapati S, Khan M I, Kumar R, Sastry M 2005 Extra-/intracellular biosynthesis of gold nanoparticles by an alkalotolerant fungus, Trichothecium sp. *J. Biomed. Technol.* **1** 47-53 and references therein

[99] Lunetta P, Penttila A, Hallfors G 1998 Scanning and transmission electron microscopical evidence of the capacity of diatoms to penetrate the alveolo-capillary barrier in drowning *Int. J. Legal Med.* **111** 229-237

[100] Ross Inman, Institute for Molecular Virology & Department of Biochemistry, University of Wisconsin-Madison, Wisconsin. USA, http://www.biochem.wisc.edu/inman/empics/virus.htm

[101] Kajander E O 2006 Nanobacteria – propagating calcifying nanoparticles *Lett. Appl. Microbiol.* **42** 549-552 and references therein

[102] Checkoway H, Heyer N J, Demers P A, Breslow N E 1993 Mortality among workers in the diatomaceous earth industry *Br. J. Ind. Med.* **50** 586-597

[103] Rawal B D, Pretorius A M 2005 "Nanobacterium sanguineum" is it a new life-form in search of human ailment or commensal: overview of its transmissibility and chemical means of intervention *Med. Hypotheses.* **65** 1062-1066

[104] Ciftcioglu N, Haddad R S, Golden D C, Morrison D R, McKay D S 2005 A potential cause for kidney

stone formation during space flights: enhanced growth of nanobacteria in microgravity *Kidney Int.* **67** 483-491

[105] Wood H M, Shoskes D A 2006 The role of nanobacteria in urologic disease *World J. Urol.* **24** 51-54

[106] Sommer A P, Miyake N, Chandra Wickramasinghe N, Narlikar J V, Al-Mufti S 2004 Functions and possible provenance of primordial proteins *J. Proteome Res.* **3** 12996-1299

[107] Rogers F, Arnott P, Zielinska B, Sagebiel J, Kelly K E, Wagner D, Lighty J S, Sarofim A F 2005 Real-time measurements of jet aircraft engine exhaust *J. Air Waste Manag. Assoc.* **55** 583-593

[108] Seames W S, Fernandez A, Wendt J O 2002 A study of fine particulate emissions from combustion of treated pulverized municipal sewage sludge *Environ. Sci. Technol.* **36** 2772-2776

[109] Linak W P, Miller C A, Wendt J O 2000 Comparison of particle size distribution and elemental partitioning from the combustion of pulverized coal and residual fuel oil *J. Waste Manag. Assoc.* **50** 1532-1544

[110] U.S. E.P.A. 2002, Health assessment for Diesel exhaust. Washington, dc, U.S. Environmental Protection Agency

[111] Westerdahl D, Fruin S, Sax T, Fine P M, Sioutas C 2005 Mobile platform measurements of ultrafine particles and associated pollutant concentrations on freeways and residential streets in Los Angeles *Atmos. Environ.* **39** 3597-3610

[112] Evelyn A, Mannick S, Sermon P A 2003 Unusual carbon-based nanofibers and chains among diesel-emitted particles *Nano. Lett.* **3** 63-64

[113] Soto K F, Carrasco A, Powell T G, Garza K M, Murr L E 2005 Comparative *in vitro* cytotoxicity assessment of some manufactured nanoparticulate materials characterized by transmission electron microscopy *J. Nanoparticle Res.* **7** 145-169 and references therein

[114] Kocbach A, Li Y, Yttri K E, Cassee F R, Schwarze P E, Namork E 2006 Physicochemical characterisation of combustion particles from vehicle exhaust and residential wood smoke *Part. Fibre Toxicol.* **3** 1

[115] Singh M, Phuleria H C, Bowers K, Sioutas C 2005 Seasonal and spatial trends in particle number and concentrations and size distributions at the children's health study sites in Southern California *J. Expo. Anal. Env. Epid.* 1-16 and references therein

[116] Penn A, Murphy G, Barker S, Henk W, Penn L 2005 Combustion-Derived Ultrafine Particles Transport Organic Toxicants to Target Respiratory Cells *Environ. Health Persp.* **113** 956-963

[117] Vermylen J, Nemmar A, Nemery B, Hoylaerts F 2005 Ambient air pollution and acute myocardial infarction *J. Thromb. Haemost.* **3** 1955-1961 and references therein

[118] Knox E G 2005 Oil combustion and childhood cancers *J. Epidemiol. Community Health* **59** 755-760

[119] Riediker M, Devlin R B, Griggs T R, Herbst M C, Bromberg P A, Williams R W, Cascio W E 2004 Cardiovascular effects in patrol officers are associated with fine particulate matter from brake wear and engine emissions *Part. Fibre Toxicol.***4** 2

[120] Garshick E, Schenker M B, Mumoz A, Segal T, Smith T J, Woskiw S, Hammond S K, Speizer F E 1988 A retrospective cohort study of lung cancer and diesel enhaust exposure in railroad workers *Am. J. Respir. Dis.* **137** 820-825

[121] Bigert C, Gustavsson P, Hallqvist J, Hogstedt C, Lewne M, Plato N, Reuterwall C, Scheele P 2003 Myocardial infarction among professional drivers *Epidemiology* **14** 333-339

[122] Hoek G, Brunekreef B, Goldbohm S, Fischer P, van den Brandt PA 2002 Association between mortality and indicators of traffic related air pollution in the Netherlands: a cohort study *Lancet* **360**1203-1209

[123] Environmental Protection Agency http://www.epa.gov/iaq/index.html, www.epa.gov/airtrends/pmreport03/pmunderstand_2405.pdf

[124] Afshari A, Matson U, Ekberg L E 2005 Characterization of indoor sources of fine and ultrafine particles: a study conducted in a full-scale chamber *Indoor Air* **15** 141-150

[125] See S W, Balasubramanian R 2006 Risk assessment of exposure to indoor aerosols associated with Chinese cooking *Environ. Res.* **102** 197-204

[126] World Health Organization http://www.who.int/heli/risks/indoorair/indoorair/en/index.html

[127] Lioy P J, Freeman N C G, Millette J R 2002 Dust: A Metric for Use in Residential and Building Exposure Assessment and Source Characterization *Environ. Health Persp.* **110** 969-983

[128] Ning Z, Cheung C S, Fu J, Liu M A, and Schnell M A 2006 Experimental study of environmental

tobacco smoke particles under actual indoor environment *Sci. Total Environ*. **367** 822-830

[129] Rushton L 2004 Health impact of environmental tobacco smoke in the home *Rev. Environ. Health* **19** 291-309 and references therein

[130] Husgavfel-Pursiainen K 2004 Genotoxicity of environmental tobacco smoke: a review *Mutat. Res*. **567** 427-445

[131] Wang X L, Wang J 2005 Smoking-gene interaction and disease development: relevance to pancreatic cancer and atherosclerosis *World J. Surg*. **29** 344-353

[132] Godtfredsen NS, Osler M, Vestbo J, Andersen I, Prescott E 2003 Smoking reduction, smoking cessation, and incidence of fatal and non-fatal myocardial infarction in Denmark 1976-1998: a pooled cohort study *J. Epidemiol. Community Health* **57** 412-416

[133] Stefani D, Wardman D, Lambert T 2005 The implosion of the Calgary General Hospital: ambient ait quality issues *J. Air Waste Manag. Assoc*. **55** 52-59

[134] see for example Fireman E M, Lerman Y, Ganor E, Greif J, Fireman-Shoresh S, Lioy P J, Banauch G I, Weiden M, Kelly K J, Prezant D J 2004 Induced sputum assessement in New York city firefighters exposed to World Trade Center dust *Environ. Health Perspect*. **112** 1564-1569 and references therein

[135] Lioy P J, Weisel C P, Millette J R, Eisenreich S, Vallero D, Offenberg J, Buckley B, Turpin B, Zhong M, Cohen M D, Prophete C, Yang III, Stiles R, Chee G, Johnson W, Porcja R, Alimokhtari S, Hale R C, Weschler C, Chen L C 2002 Characterization of the Dust/Smoke Aerosol that Settled East of the World Trade Center (WTC) in Lower Manhattan after the Collapse of the WTC 11 September 2001 *Environ. Health Perspect*. **110** 703-714

[136] McGee J K, Chen L C, Cohen M D, Chee G R, Prophete C M, Haykal-Coates N, Wasson S J, Conner T L, Costa D L, Gavett S H 2003 Chemical Analysis of World Trade Center Fine Particulate Matter for Use in Toxicologic Assessment *Environ. Health Perspect*. **111** 972-980

[137] Agera ® Rx Medical Formula http://www.agerarx.co.uk/whatisagera.asp

[138] Xiao L, Takada H, Maeda K, Haramoto M, Miwa N 2005 Antioxidant effects of water-soluble fullerene derivatives against ultraviolet ray or peroxylipid through their action of scavenging the reactive oxygen species in human skin keratinocytes *Biomed. Pharmacol*. **59** 351-358

[139] Nohynek G J, Lademann J, Ribaud C, Roberts M S 2007 Grey goo on the skin? Nanotechnology, cosmetic and sunscreen safety *Crit. Rev. Toxicol*. **37** 251-277 and references therein

[140] http://www.foe.org/camps/comm/nanotech/

[141] Brumfiel G 2006 Consumer products leap aboard the nano bandwagon *Nature* **440** 262

[142] Gurr J R, Wang A S S, Chen C H, Jan K Y 2005 Ultrafine titanium dioxide particles in the absence of photoactivation can induce oxidative damage to human bronchial epithelial cells *Toxicol*. **213** 66-73 and references therein

[143] Serpone N, Salinaro A, Emeline A 2001 Deleterious effects of sunscreen titanium dioxide nanoparticles on DNA: efforts to limit DNA damage by particle surface modification *Proc. SPIE* **4258** 86-98

[144] Shintani H, Kurosu S, Miki A, Hayashi F, Kato S 2006 Sterilization efficiency of the photocatalyst against environmental microorganisms in a health care facility *Biocontrol Sci*. **11** 17-26

[145] Takenaka S, Karg E, Roth C, Schulz H, Ziesenis A, Heinzmann U, Schramel P, Heyder J 2001 Pulmonary and systemic distribution of inhaled ultrafine silver particles in rats *Environ. Health Persp*. **109** (Suppl. 4) 547-551 and references therein

[146] Gatti A M, Montanari S, Monari E, Gambarelli A, Capitani F, Parisini B 2004 Detection of micro- and nano-sized biocompatible particles in the blood *J. Mater. Sci. Mater. Med*. **15** 469-472

[147] Gatti A M 2004 Biocompatibility of micro- and nano-particles in the colon. Part II *Biomater*. **25** 385-392

[148] Antonini J M, Santamaria A B, Jenkins N T, Albini E, Lucchini R 2006 Fate of manganese associated with the inhalation of welding fumes: potential neurological effects *NeuroToxicol*. **27** 304-310

[149] Aitken R J, Creely K S, Tran C L 2004 Nanoparticles: An occupational hygiene review. Prepared by the Institute of Occupational Medicine for the Health and Safety Executive. (Crown copyright 2004) www.hse.gov.uk/research/rrpdf/rr274.pdf

[150] Robbie K, Sit J C, Brett M J 1998 Advanced techniques for glancing angle deposition *J. Vac. Sci. Technol. B* **16** 1115-1112

[151] Robbie K, Beydaghyan G, Brown T, Dean C, Adams J, Buzea C 2004 Ultrahigh vacuum glancing angle

deposition system for thin films with controlled three-dimensional nanoscale structure *Rev. Sci. Instrum.* **75** 1089-1097

[152] Nanoparticles Information Library, The National Institute for Occupational Safety and Health, http://www2a.cdc.gov/niosh-nil/index.asp

[153] Kaminska K, Suzuki M, Kimura K, Taga Y, Robbie K 2004 Simulating structure and optical response of vacuum evaporated porous rugate filters *J. Appl. Phys.* **95** 3055-3062

[154] Kaminska K, Robbie K 2004 Birefringent omnidirectional reflector *Appl. Optics***43** 1570-1576

[155] Robbie K, Brett M J 1997 Sculptured thin films and glancing angle deposition: growth mechanics and applications *J. Vac. Sci. Technol. A* **15** 1460-1465

[156] Buzea C, Kaminska K, Beydaghyan G, Brown T, Elliott C, Dean C, Robbie K 2005 Thickness and density evaluation for nanostructured thin films by glancing angle deposition *J. Vac. Sci. Technol. B* **23** 2545-2552

[157] Buzea C, Beydaghyan G, Elliott C, Robbie K 2005 Control of powel law scaling in the growth of silicon nanocolumn pseudo-regular arrays deposited by glancing angle deposition *Nanotechnology* **16** 1986-1992

[158] Shah C P 1998 *Public Health and preventive medicine in Canada* (University of Toronto Press, Toronto, Canada)

[159] Muller J, Huaux F, Moreau N, Misson P, Heiliea J F, Delos M, Arras M, Fonseca A, Nagyb J B, Lison D 2005 Respiratory toxicity of multi-wall carbon nanotubes *Toxicol. Appl. Pharmacol.* **207** 221-231

[160] Wang H, Huff T B, Zweifel D A, He W, Low P S, Wei A, Cheng J X 2005 In vitro and in vivo two-photon luminescence imaging of single gold nanorods *Proc. Natl. Acad. Sci.* **102** 15752-15756

[161] Nemery B 1990 Metal toxicity and the respiratory tract *Eur. Respir. J.* **3** 202-219

[162] Maier LA 2002 Clinical approach to chronic beryllium disease and other nonpneumoconiotic interstitial lung diseases *J. Thorac. Imaging.* **17** 273-284

[163] Paris C, Bertrand O, de Abreu RR 2004 Lung cancer and occupational exposures (asbestos not considered) *Rev. Prat.* **54** 1660-1664

[164] Waalkes M P 2003 Cadmium carcinogenesis *Mutat. Res.* **533** 107-120

[165] McGrath K G 2003 An earlier age of breast cancer diagnosis related to more frequent use of antiperspirants/deodorants and underarm shaving *Eur. J. Cancer Prev.* **12** 479-485

[166] Guillard O, Fauconneau B, Olichon D, Dedieu G, Deloncle R 2004 Hyperaluminemia in a woman using an aluminum-containing antiperspirant for 4 years *Am. J. Med.* **117** 956-959

[167] Kawahara M 2005 Effects of aluminum on the nervous system and its possible link with neurodegenerative diseases *J. Alzheimers Dis.* **8** 171-182

[168] Joshi JG, Dhar M, Clauberg M, Chauthaiwale V 1994 Iron and aluminum homeostasis in neural disorders *Environ Health Perspect* **102** 207-213

[169] Deloncle R, Guillard O, Huguet F, Clanet F 1995 Modification of the blood-brain barrier through chronic intoxication by aluminum glutamate. Possible role in the etiology of Alzheimer's disease *Biol. Trace Elem. Res.* **47** 227-233

[170] Weiss B 2005 Economic implications of manganese neurotoxicity *NeuroToxicol.* **In** print

[171] Quintana C, Bellefqih S, Laval Y J, Guerquin-Kern J L, Wu T D, Avila J, Ferrer I, Arranz R, Patiño C 2006 Study of the localization of iron, ferritin, and hemosiderin in Alzheimer's disease hippocampus by analytical microscopy at the subcellular level *J. Struct Biol* **153** 42-54 and references therein

[172] Noonan C W, Pfau J C, Larson T C, Spence M R 2006 Nested case-control study of autoimmune disease in an asbestos-exposed population *Envir. Health Persp.* **114** 1243–1247

[173] Borm P J A and Tran L 2002 From quartz hazard to quartz risk: the coal mines revisited *Ann. Occup. Hyg.* **46** 25-32

[174] Gustavsson P, Gustavsonn A, Hogstedt C 1988 Excess of cancer in Swedish chimney sweeps *Br. J. Ind. Med.* **45** 777-781

[175] Gaafar R M, Eldin N H A 2005 Epidemic of mesothelioma in Egypt *Lung Cancer* **49** S17-S20

[176] Pfau J C, Sentissi J J, Weller G, Putnam E A 2005 Assessment of Autoimmune Responses Associated with Asbestos Exposure in Libby, Montana, USA *Environ. Health Perspect.* **113** 25–30

[177] Johnston C J, Finkelstein J N, Mercer P, Corson N, Gelein R, Oberdorster G 2000 Pulmonary effects induced by ultrafine PTFE particles *Toxicol. Appl. Pharmacol.* **168** 208-215 and references therein

[178] Ferin J, Oberdorster G 1992 Polymer degradation and ultrafine particles: potential inhalation hazards for astronauts *Acta Astronaut.* **27** 257-259

[179] Ostiguy C, Lapointe G, Menard L, Cloutier Y, Trottier M, Boutin M, Antoun M, Normans C 2006 Nanoparticles, actual knowledge about occupational health and safety risks and prevention measures. Studies and research projects report. R-470. www.irsst.qc.ca

[180] Tuch T H, Brand P, Wichmann H E, Heyder J 1997 Variation of particle number and mass concentration in various size ranges of ambient aerosols in eastern Germany *Atmos. Environ.* **31** 4187-4193

[181] Chung A, Herner J D, Kleeman M J 2001 Detection of alkaline uktrafine atmospheric particles at Bakersfield, California *Environ. Sci. Technol.* **35** 2184-2190

[182] http://www.airqualityontario.com/

[183] http://www.gsfc.nasa.gov/, Image ID: GL-2002-001716

[184] NASA Multi-angle Imaging SpectroRadiometer image ID: F09_0017, http://www-misr.jpl.nasa.gov/index.html

[185] Dockery D W, Luttmann-Gibson H, Rich D Q, Link M S, Mittleman M A, Gold D R, Koutrakis P, Schwartz J D, Verrier R L. 2005 Association of air pollution with increased incidence of ventricular tachyarrhythmias recorded by implanted cardioverter defibrillators *Environ. Health Persp.* **113** 670-674

[186] Lippmann M, Frampton M, Schwartz J, Dockery D, Schlesinger R, Koutrakis P, Froines J, Nel A, Finkelstein J, Godleski J, Kaufman J, Koenig J, Larson T, Luchtel D, Liu LJ, Oberdorster G, Peters A, Sarnat J, Sioutas C, Suh H, Sullivan J, Utell M, Wichmann E, Zelikoff J 2003 The U.S. Environmental Protection Agency Particulate Matter Health Effects Research Centers Program: a midcourse report of status, progress, and plans *Environ. Health Persp.* **111** 1074-1092

[187] Peters A, Wichmann HE, Tuch T, Heinrich J, Heyder J 1997 Respiratory effects are associated with the number of ultrafine particles *Am. J. Respir. Crit. Care Med.* **155** 1376-1383

[188] Peters A, Dockery D W, Muller J E, Mittleman M A 2001 Increased particulate air pollution and the triggering of myocardial infarction *Circulation* **103** 2810-2815

[189] Schwartz J, Morris R 1995 Air pollution and hospital admissions for cardiovascular disease in Detroit, Michigan *Am. J. Epidemiol.* **142** 23-35

[190] Pope C A, Burnett R T, Thun M J, Calle E E, Krewski D, Ito K, Thurston G D 2002 Lung cancer, cardiopulmonary mortality, and long-term exposure to fine particulate air pollution *JAMA* **286** 1132-1141

[191] Bates D, Sizto R 1987 Relationship between air pollution levels and hospital admissions in southern Ontario: acid summer haze effect *Environ. Res.* **47** 317-331

[192] Iwai K, Mizuno S, Miyasaka Y, Mori T 2005 Correlation between suspended particles in the environmental air and causes of disease among inhabitants: cross-sectional studies using the vital statistics and air pollution data in Japan *Environ. Res.* **99** 106-117

[193] Abey D E, Nishino N, McDonnel W F, Burchette R J, Knutsen S F, Beeson W L, Yang J X 1999 Long term inhalable particles and other air pollutants related to mortality in non-smokers *Am. J. Respir. Crit. Care Med.* **159** 373-382

[194] Nemmar A, Hoet P H, Vanquickenborne B, Dinsdale D, Thomeer M, Hoylaerts M F, Vanbilloen H, Mortelmans L, Nemery B 2002 Passage of inhaled particles into the blood circulation in humans *Circulation* **105** 411-414 and references therein

[195] Gilboa S M, Mendola P, Olshan A F, Langlois P H, Savitz D A, Loomis D, Herring A H, Fixler D E 2005 Relation between ambient air quality and selected birth defects, seven country study, Texas 1997-2000 *Am. J. Epidemiol.* **162** 238-252

[196] Sram R J, Binkova B, Dejmek J, Bobak M 2005 Ambient air pollution and pregnancy outcomes: a review of the literature *Environ. Health Persp.* **113** 375-382

[197] Kaiser R, Romieu I, Medina S, Schwartz J, Krzyzanowski M, Künzli N 2004 Air pollution attributable postneonatal infant mortality in U.S. metropolitan areas: a risk assessment study *Environ. Health: A Global Access Science Source* **3** 4 and references therein

[198] Schwartz J, Marcus A 1990 Mortality and air pollution in London: A time series analysis *Am. J. Epidemiol.* **131** 185-194

[199] Romieu I 2005 Nutrition and lung health *Int. J. Tuberc. Lung. Dis.* **9** 362-374

[200] Elder A, Gelein R, Silva V, Feikert T, Opanashuk L, Carter J, Potter R, Maynard A, Ito Y, Finkelstein J,

Oberdorster G 2006 Translocation of inhaled ultrafine manganese oxide particles to the central nervous system *Environ. Health Perspect.* **114** 1172-1178

[201] Lippmann M 1990 Effects of fiber characteristics on lung deposition, retention, and disease *Environ. Health Perspec.* **88** 311-317

[202] Wright D T, Cohn L A, Li H, Fischer B, Li C M, Adler K B 1994 Interactions of oxygen radicals with airway epithelium *Environ Health Perspect.* **102 Suppl. 10** 85-90

[203] Peters A, Veronesi B, Calderon-Garciduenas L, Gehr P, Chen L C, Geiser M, Reed W, Rothen-Rutishauer B, Schurch S, Schultz H 2006 Translocation and potential neurological effects of fine and ultrafine particles. A critical update *Part. Fibre Toxicol.* **3** 13

[204] Hopwood D, Spiers E M, Ross P E, Anderson J T, McCullough J B, Murray F E 1995 Endocytosis of fluorescent microspheres by human oesophageal epithelial cells: comparison between normal and inflamed tissue *Gut* **37** 598-602

[205] Thibodeau G. A., Patton K. T. 2003 Anatyomy and Physiology, Fifth Edition, St. Louis, Mosby Inc.

[206] Semmler M, Seitz J, Mayer P, Heyder J, Oberdörster G, Kreyling W G 2004 Long-term clearance kinetics of inhaled ultrafine insoluble iridium particles from the rat lung, including transient translocation into secondary organs *Inhalation Toxicol.* **16** 453-459

[207] Ng A W, Bidani A, Heming T A 2004 Innate host defense of the lung: effects of lung-lining fluid pH *Lung* **182** 297-317

[208] Garnet M C, Kallinteri P 2006 Nanomedicines and nanotoxicology: some physiological principles Occ. Med. 56 307-311

[209] Aderem A, Underhill D M 1999 Mechanisms of phagocytosis in macrophages *Annu. Rev. Immunol.* 17 593–623

[210] Palecanda A, Kobzik L 2000 Alveolar macrophage-environmental particle interaction: analysis by flow cytometry *Methods* **21** 241-247

[211] Kobzik L 1995 Lung macrophage uptake of unposonized environmental particulates. Role of scavenger-type receptors *J. Immunol.* **155** 367-376

[212] Inoue K I, Takano H, Yanagisawa R, Hirano S, Ichinose T, Shimada A, Yoshikawa T 2006 The role of toll-like receptor 4 in the airway inflammation induced by diesel exhaust particles *Arch. Toxicol.* **80** 275-279

[213] Park J B 2003 Phagocytosis induces superoxide formation and apoptosis in macrophages *Exp. Mol. Med.* **35** 325-335

[214] Porter A E, Muller K, Skepper J, Midgley P, Welland M 2006 Uptake of C60 by human monocyte macrophages, its localization and implications for toxicity: studied by high resolution electron microscopy and electron tomography *Acta Biomater.* **2** 409-419

[215] Brown D M, Donaldson K, Stone V 2004 Effects of PM10 in human peripheral blood monocytes and J774 macrophages *Respir. Res.* **5** 29

[216] Oberdörster G, Ferin J, Lehnert B E 1994 Correlation between particle size, *in vivo* particle persistence, and lung injury *Environ. Health Persp.* **102 Suppl 5** 173-179

[217] Oberdörster G 1988 Lung clearance of inhaled insoluble and soluble particles *J. Aerosol Med.* **1** 289-329

[218] Oberdorster G, Sharp Z, Atudorei V, Elder A, Gelein R, Kreyling W, Cox C 2004 Translocation of inhaled ultrafine particles to the brain *Inhal. Toxicol.* **16** 437-445

[219] Liu J, Wong H L, Moselhy J, Bowen B, Wu X Y, Johnston M R 2006 Targeting colloidal particulates to thoracic lymph nodes *Lung Cancer* **51** 377-386 and references therein

[220] Dasenbrock C, peters L, Creutzenberg O, Heinrich U 1996 The carcinogenic potency of carbon particles with and without PAH after repeated intratracheal administration in the rat *Toxicol. Lett.* **88** 15-21

[221] Driscoll K E, Carter J M, Howard B W, Hassenbein D G, Pepelko W, Baggs R B, Oberdoster G 1996 Pulmonary inflammation, chemokine, and mutagenic responses in rats after subchronic inhalation of carbon black *Toxicol. Appl. Pharmacol.* **136** 327-380

[222] Nikula K J, Snipes M B, barr E B, Griffith W C, Henderson R F, Mauderly J L 1995 Comparative pulmonary toxicities and carcinogenicities of chronically inhaled diesel exhaust and carbon black in F344 rats *Fundam. Appl. Toxicol.* **25** 80-94

[223] Lucarelli M, Gatti A M, Savarino G, Quatronni P, Martinelli L, Monari E, Boraschi D 2004 Innate defence functions of macrophages can be biased by nano-sized ceramics metallic particles *Eur. Cytokine Netw.* **15** 339-346

[224] Renwick L C, Donaldson K, Clouter A 2001 Impairment of alveolar macrophage phagocytosis by ultrafine particles *Toxicol. Appl. Pharmacol.* **172** 119-127

[225] Hoet P M H, Nemery B 2001 Stimulation of phagocytosis by ultrafine particles *Toxicol. Appl. Pharmacol.* **176** 203

[226] Wesselkamper S C, Chen L C, Gordon T 2005 Quantitative trait analysis of the development of pulmonary tolerance to inhaled zinc oxide in mice *Respir. Res.* **6** 73

[227] Sood N, Bennett W D, Zeman K, Brown J, Foy C, Boucher R C, Knowles M R 2003 Increasing concentration of inhaled saline with or without amiloride: effect on mucociliary clearance in normal subjects *Am. J. Respir. Crit. Care. Med.* **167** 158-163

[228] Xia T, Kovochich M, Brant J, Hotze M, Sempf J, Oberley T, Sioutas C, Yeh J I, Wiesner M R, Nel A E 2006 Comparison of the abilities of ambient and manufactured nanoparticles to induce cellular toxicity according to an oxidative stress paradigm *Nano Lett.* **6** 1794-1807

[229] Geiser M, Rothen-Rutishauser B, kapp N, Schurch S, Kreyling W, Schultz H, Semmler M, Im Hof V, Heyder J, Gehr P 2005 Ultrafine particles cross cellular membranes by nonphagocytotic mechanisms in lungs and in cultured cells *Environ. Health Perspect.* **113** 1555-1560

[230] Garcia-Garcia E, Andrieux K, Gil S, Kima H R, Le Doana T, Desmaele D, d'Angelo J, Taran F, Georgin D, Couvreur P 2005 A methodology to study intracellular distribution of nanoparticles in brain endothelial cells *Int. J. Pharm.* **298** 310–314

[231] Stone V, Shaw J, Brown D M, MacNee W, Faux S P, Donaldson K 1998 The role of oxidative stress in the prolonged inhibitory effect of ultrafine carbon black on epithelial cell function *Toxicol. In Vitro* **12** 649-659

[232] Singal M, Finkelstein J N 2005 Amourphous silica particles promote inflammatory gene expression through the redox sensitive transcription factor, AP-1, in alveolar epithelial cells *Exp. Lung Res.* **31** 581-597

[233] Kreyling W G, Semmler M, Erbe F, Mayer P, Takenata S, Oberdörster G, Ziesenis A 2002 Translocation of ultrafine insoluble iridium particles from lung epithelium to extrapulmonary organs is size dependent but very low *J. Toxicol. Environ. Health A* **65** 1513-1530

[234] Kukowska-Latallo J F, Candido K A, Cao Z, Nigavekar S S, Majoros I J, Thomas T P, Balogh L B, Khan M K, Baker J R Jr 2005 Nanoparticle targeting of anticancer drug improves therapeutic response in animal model of human epithelial cancer *Cancer Res.* **65** 5317-5324

[235] Kato T, Yashiro T, Murata Y, Herbert D C, Oshikawa K, Bando M, Ohno S, Sugiyama Y 2003 Evidence that exogenous substances can be phagocytized by alveolar epithelial cells and transported into blood capillaries *Cell Tissue Res.* **311** 47-51

[236] Juvin P, Fournier T, Boland S, Soler P, Marano F, Desmonts J M, Aubier M 2002 Diesel particles are taken up by alveolar type II tumor cells and alter cytokine secretion *Arch. Environ. Health* **57** 53-60

[237] Rothen-Rutishauser B M, Schurch S, Haenni B, Kapp N, Gehr P 2006 Interaction of fine particles and nanoparticles with red blood cells visualized with advanced microscopic techniques *Environ. Sci. Technol.* **40** 4353-4359

[238] Nemmar A, Hoylaerts M F, Hoet P H M, Dinsdale D, Smith T, Xu H, Vermylen J, Nemery B 2002 Ultrafine particles affect experimental thrombosis in an vivo hamster model *Am. J. Respir. Crit. Care Med.* **166** 998-1004

[239] Oberdörster G, Sharp Z, Atudorei V, Elder A, Gelein R, Lunts A 2002 Extrapulmonary translocation of ultrafine carbon particles following whole-body inhalation exposure of rats *J. Toxicol. Environ. Health A* **65** 1531-1543

[240] Brown D M, Donaldson K, Borm P J, Schins R P, Dehnhardt M, Gilmour P, Jimenez L A, Stone V 2004 Calcium and ROS-mediated activation of transcription factors and TNF-$\alpha$ cytokine gene expression in macrophages exposed to ultrafine particles Iam. *J. Physiol. Lung Cell. Mol. Physiol.* **286** L344-L353

[241] Long H, Shi T, Borm P J, Määttä J, Husgafvel-Pursiainen K, Savolainen K, Krombach F 2004 ROS-mediated TNF-$\alpha$ and MIP-2 gene expression in alveolar macrophages exposed to pine dust *Part. Fibre Toxicol.* **1** 3

[242] Wilson M R, Lightbody J H, Donaldson K, Sales J, Stone V 2002 Interactions between ultrafine particles and transition metals *in vivo* and *in vitro Toxicol*. *Appl*. *Pharmacol*. **184** 172-179

[243] Lim Y, Kim S H, Cho Y J, Kim K A, Oh M W, Lee K H 1997 Silica-induced oxygen radical generation in alveolar macrophages *Ind*. *Health* **35** 380-387

[244] Lomer M C E, Thompson R P H, Powell J J 2002 Fine and ultrafine particles of the diet: influence on the mucosal immune response and association with Crohn's disease *Proc*. *Nutrition Soc*. **61** 123-130 and references therein

[245] Liu G, Mena P, Harris P R L, Rolston R K, Perry G, Smith M A 2006 Nanoparticle iron chelators: A new therapeutic approach in Alzheimer disease and other neurologic disorders associated with trace metal imbalance *Neurosci*. *Lett*. **406** 189–193

[246] Bourrinet P, Bengele H H, Bennemain B, Dencausse A, Idee J M, Jacobs P M, Lewis J M 2006 Preclinical safety and pharmacokinetic profile of ferumoxtran-10, an ultrasmall superparamagnetic iron oxide magnetic resonance contrast agent *Investigative Radiol*. **41** 313-324

[247] Rabin O, Perez J M, Grimm J, Wojtkiewiccz G, Weissleder R 2006 An X-ray computed tomography imaging agent on long-circulating bismuth sulphide nanoparticles *Nature Mater*. **5** 118-122

[248] 293. Chung S M 2002 Safety issues in magnetic resonance imaging *J*. *Neuro-Ophtalmol*. **22** 35-39

[249] Lockman P R, Koziara J M, Mumper R J, Allen D D 2004 Nanoparticle syrface charges alter blood-brain barrier integrity and permeability *J*. *Drug Targeting* **12** 635-641

[250] Campbell A, Oldham M, Becaria A, Bondy S C, Meacher D, Sioutas C, Misra C, Mendez L B, Kleinman M 2005 Particulate matter in polluted air may increase biomarkers of inflammation in mouse brain *Neurotoxicol*. **26** 133-140

[251] Calderon-Garciduenas L, Maranpot R R, Torres-Jardon R, Henriquez-Roldan C, Schoonhoven R, Acuna-Ayala H, Villarreal-Calderon A, Nakamura J, Fernando R, Reed W, Azzarelli B, Swenberg J A 2003 DNA damage in nasal and brain tissues of canines exposed to air pollutants is associated with evidence of chronic brain inflammation and neurodegeneration *Toxicol*. *Pathol*. **31** 524-538

[252] Oberdörster E 2004 Manufactured nanomaterials (Fullerenes, C60) induce oxidative stress in the brain of juvenile largemouth bass *Environ*. *Health Persp*. **112** 1058-1062

[253] Mount C, Downton C 2006 Alzheimer disease: progress or profit? *Nat*. *Med*. **12** 780-784

[354] Shwe T T W, Yamamoto S, Kakeyama M, Kobayashi T, Fujimaki H 2005 Effect of intratracheally instillation of untrafine carbon black on proinflammatory cytokine and chemokine release and mRNA expression in lung and lymph nodes of mice *Toxicol*. *Appl*. *Pharmacol*. **209** 51-61

[255] Chen J, Tan M, Nemmar A, Song W, Dong M, Zhang G, Li Y 2006 Quantification of extrapulmonary translocation of intratrachealy-instilled particles in vivo in rats: effect of lipopolysaccharide *Toxicol*. **222** 195-201

[256] Brown JS, Zeman KL, Bennett WD. 2002 Ultrafine particle deposition and clearance in the healthy and obstructed lung *Am*. *J*. *Respir*. *Crit*. *Care Med*. **166** 1240-1247

[257] Mills N L, Amin N, Robinsion S D, Anand A, Davies J, Patel D, de la Fuente J M, Cassee F R, Boon N A, MacNee W, Millar A M, Donaldson K, Newby D E 2005 *Am*. *J*. *Respir*. *Crit*. *Care Med*. **173** 426-431

[258] Wiebert P, Sanchez-Crespo A, Seitz J, Falk R, Philipson K, Kreyling W G, Möller W, Sommerer K, Larsson S, Svartengren M. 2006 Neglijible clearance of ultrafine particles retained in healthy and affected human lungs *Eur*. *Respir*. *J*. **in** press

[259] Wiebert P, Sanchez-Crespo A, Falk R, Philipson K, Lundin A, Larsson S, Mooler M, Kreyling G, Svartengren M 2006 No significant translocation of inhaled 35-nm carbon particles to the circulation in humans *Inhal*. *Toxicol*. **18** 741-747

[260] Gatti A M, Montanari S, Gambarelli A, Capitani F, Salvatori R, 2005 In-vivo short- and long-term evaluation of the interaction material-blood *J*. *Mater*. *Sci*. *Mater*. *Med*. **16** 1213-1219

[261] Nemmar A, Nemery B, Hoylaerts M F, Vaermylen J 2002 Air pollution and thrombosis: an experimental approach *Pathophysiol*. *Haemost*. *Thromb*. **32** 349-350

[262] Schulz H, Hardewr V, Ibald-Mulkli A, Khandoga A, Koenig W, Krombach F, radykewicz R, Stampfl A, Thorand B, Peters A 2005 Cardiovascular effects of fine and ultrafine particles *J*. *Aerosol Med*. **18** 1-22

[263] Nemmar A, Vanbilloen H, Hoylaerts M F, Hoet P H, Verbruggen A, Nemery B 2001 Passage of intratracheally instilled ultrafine particles from the lung into the systemic circulation in hamster *Am*. *J*. *Respir*. *Crit*. *Care*. *Med*. **164** 1665-1668

[264] Schwab A J, Pang K S 2000 The multiple indicator dilution method and its utility in risk assessment *Environ. Health Perspect.* **108** 861-872

[265] Gatti A M, Rivasi F 2002 Biocompatibility of micro- and nanoparticles. Part I: in liver and kidney *Biomater.* **23** 2381-2387 and references therein

[266] Ballestri M, Baraldi A, Gatti A M, Furci L, Bagni A, Loria P, Rapaa M, Carulli N, Albertazzi A 2001 Liver and kidney foreign bodies granulomatosis in a patient with malocclusion, bruxism, and worn dental prostheses *Gastroenterology* **121** 1234-1238

[267] Cooper J R, Stradling G N, Smith H, Breadmore S E 1979 The reactions of 1.0 nanometre diameter plutonium-238 dioxide particles with rat lung fluid *Int. J. Radiat. Biol. Relat. Stud. Phys. Chem. Med.* **36** 453-466

[268] Lomer M C E, Hutchinson C, Volkert S, Greenfield S M, Catterall A, Thompson R P H and Powell J J 2004 Dietary sources of inorganic microparticles and their intake in healthy subjects and patients with Crohn's disease British J. Nutrition **92** 947-955

[269] Jani P, Halbert G W, Langridge J, Florence A T 1990 Nanoparticle uptake by the rat gastrointestinal mucosa: quantitation and particle size dependency *J. Pharm. Pharmacol.* **42** 821-826

[270] Rae C S, Khor I W, Wang Q, Destito G, Gonzalez M J, Singh P, Thomas D M, Estrada M N, Powell E, Finn M G, Manchester M 2005 Systemic trafficking of plant virus nanoparticles in mice via the oral route *Virology* in print

[271] Powell J J, Ainley C C, Harvey R S, Manson I M, Kendall M D, Sankey E A, Dhillon A P, Thompson R P 1996 Characterization of inorganic microparticles in pigment cells of human gut associated lymphoid tissue *Gut* **38** 390-395

[272] Toll R, Jacobi U, Richter H, Lademann J, Schaefer H, Blume-Peytavi U 2004 Penetration Profile of Microspheres in Follicular Targeting of Terminal Hair Follicles *J. Invest. Dermatol.* 123 168 –176

[273] Tinkle S S, Antonini J M, Rich B A, Roberts J R, Salmen R, DePree K, Adkins E J 2003 Skin as a route of exposure and sensitization in chronic beryllium disease *Environ. Health Perspect.* **111** 1202–1208

[374] Monteiro-Riviere N A, Nemanich R J, Inman A O, Wang Y Y, Riviere J E 2005 Multi-walled carbon nanotube interactions with human epidermal keratinocytes *Toxicol. Lett.* **155** 377-384

[275] Tsuji J S, Maynard A D, Howard P C, James J T, Lam C W, Warheit D B, and Santamaria A B 2006 Research strategies for safety evaluation of nanomaterials, Part IV: Risk assessment of nanoparticles *Toxicol. Sci.* **89** 42-50 and references therein

[276] Lademann J, Weigmann H, Rickmeyer C, Barthelmes H, Schaefer H, Mueller G, Sterry W 1999 Penetration of titanium dioxide microparticles in a sunscreen formulation into the horny layer and the follicular orifice *Skin Pharmacol. Appl. Skin Physiol.* **12** 247-256

[277] Kubota Y, Shuin T, Kawasaki C, Hosaka M, Kitamura H, Cai R, Sakai H, Hashimoto K, Fujishima A 1994 Photokilling of T-24 human bladder cancer cells with titanium dioxide *Br. J. Cancer* **70** 1107-11011

[278] Dunford R, Salinaro A, Cai L, Serpone N, Horikoshi S, Hidaka H, Knowland J 1997 Chemical oxidation and DNA damage catalysed by inorganic sunscreen ingredients *FEBS Lett.* **418** 87–90

[279] Rehn B, Seiler F, Rehn S, Bruch J, Maier M 2003 Investigations on the inflammatory and genotoxic lung effects of two types of titanium dioxide: untreated and surface treated *Toxicol. Appl. Pharmacol.* **189** 84–95

[280] Oberdörster G, Ferin J, Gelein R, Soderholm S C, Finkelstein J 1992 Role of the alveolar macrophage in lung injury: studies with ultrafine particles *Environ. Health Perspect.* **97** 193–199

[281] Peters K, Unger R E, Kirkpatrick C J, Gatti A M, Monari E 2004 Effects of nano-scaled particles on endothelial cell function *in vitro*: studies on viability, proliferation and inflammation *J. Mater. Sci. Mater. Med.* **15** 321–325

[282] Poon V K and Burd A 2004 In vitro cytotoxicity of silver: implication for clinical wound care *Burns* **30** 140-147 and references therein

[283] Dunn K and Edwards-Jones V 2004 The role of Acticoat with nanocrystalline silver in the management of burns *Burns* **30** S1-S9

[284] Araujo L, Lobenberg R, Kreuter J 1999 Influence of the surfactant concentration on the body distribution of nanoparticles *J. Drug Target.* **6** 373-385

[285] Labhasetwar V, Song C, Humphrey W, Shebuski R, Levy R J 1998 Arterial uptake of biodegradable nanoparticles: effect of surface modifications *J. Pharm. Sci.* **87** 1229-1234

[286] Reijnders L 2006 Cleaner nanotechnology and hazard reduction of manufactured nanoparticles *J. Cleaner Production* **14** 124-133 and references therein

[287] Visuri T I, Pukkala E, Pulkkinen P, Paavolainen P 2006 Cancer incidence and causes of death among total hip replacement patients: a review based on Nordic cohorts with a special emphasis on metal-on-metal bearings *Proc. ImechE. Part H: Eng. Med.* **220** 399-407

[288] Hallab N, Merritt K, Jacobs J J 2001 Metal sensitivity in patients with orthopaedic implants *J. Bone Joint Surg. Am.* **83** 428-436

[289] Milosev I, Trebse R, Kovac S, Cor A, Pisot V 2006 Survivorship and retrieval analysis of sikomet metal-on-metal total hip replacements at a mean of seven years *J. Bone Joint Surg. Am.* **88** 1173-1182

[290] Wang M L, Tuli R, Manner P A, Sharkey P F, Hall D J, Tuan R S 2003 Direct and indirect induction of apoptosis in human mesenchymal stem cells in response to titanium particles *J. Orthopaedic Res.* **21** 697-707

[291] Li Y, Leung P, Yao L, Song Q W, Newton E 2006 Antimicrobial effect of surgical masks coated with nanoparticles *J. Hosp. Infect.* **62** 58-63

[292] Brayner R, Ferrari-Iliou R, Brivois N, Djediat S, Benedetti M F, Fievet F 2006 Toxicological impact studies based on Esterichia coli bacteria in ultrafine ZnO nanoparticles colloidal medium *Nano Lett.* **6** 866-870

[293] Lok C N, Ho C M, Chen R, He Q Y, Yu W Y, Sun H, Tam P K, Chiu J F, and Che C M 2006 Proteomic analysis of the mode of antibacterial action of silver nanoparticles *J. Proteome Res.* **5** 916-924

[294] Elechiguerra J L, Burt J L, Morones J R, Camacho-Bragado A, Gao X, Lara H H, Yacaman M J 2005 Interaction of silver nanoparticles with HIV-1 *J. Nanobiotechnol.* **3** 6

[295] Stoeger T, Reinhard C, Takenaka S, Schroeppel A, Karg E, Ritter B, Heyder J, Schultz H 2006 Instillation of six different ultrafine carbon particles indicates a surface area threshold dose for acute lung inflammation in mice *Environ. Health. Perspect.* **114** 328-333

[296] Ferin J, Oberdörster G, Penney D P 1992 Pulmonary retention of ultrafine and fone particles in rats *Am. J. Respir. Cell Mol. Biol.* **6** 535-552

[297] Balashazy I, Hofmann W, Heistracher T 2003 Local particle deposition patterns may play a key role in the developing of lung cancer *J. Appl. Physiol.* **94** 1719-1725

[298] Warheit D B, Webb T R, Sayes C M, Colvin V L, Reed K L 2006 Pulmonary instillation studies with nanoscale TiO2 rods and dots in rats: toxicity is not dependent upon particle size and surface area *Toxicol. Sci.* **91** 227-236

[299] Churg A, Stevend B, Wright J L 1998 Comparison of the uptake of fine and ultrafine TiO2 in a tracheal explant system *Am. J. Physiol.* **274** L81-L86

[300] Zhang H, Gilbert B, Huang F, Banfield J F 2003 Water-driven structure transformation in nanoparticles at room temperature *Nature* **424** 1025-1029

[301] Shvedova A A, Kisin E R, Mercer R, Murray A R, Johnson V J, Potapovich A I, Tyurina Y Y, Gorelik O, Arepalli S, Schwegler-=Berry D, Hubs A F, Antonini J, Evans D E, Ku B K, Ramsey D, Maynard A, Kagan V E, Castranova V, Baron P 2005 Unusual inflammatory and fibrogenic pulmonary responses to single walled carbon nanotubes in mice *Am. J. Physiol. Cell Mol. Physiol.* June in press

[302] Warheit D B, Laurence B R, Reed K L, Roach D H, Reynolds G A M, Web T R 2004 Comparative pulmonary toxicity assessment of single-wall carbon nanotubes in rats *Toxicol. Sci.* **77** 117-125

[303] Lam C W, James J T, McCluskey R, Hunter R L 2004 Pulmonary toxicity of single-wall carbon nanotubes in mice 7 and 90 days after intratracheal instillation *Toxicol. Sci.* **77** 126-134

[304] Maynard A D, Baron P A, Foley M, Shvedova A a, Kisin E r, Castranova V 2004 Exposure to carbon nanotube material: aerosol release during the handling of unrefined single-walled carbon nanotube material *J. Toxicol. Environ. Health Part A* **67** 87-107

[305] Donaldson K, Tran C L 2004 An introduction to the short-term toxicology of respirable industrial fibres *Mutat Res.* **553** 5-9

[306] Cherukuri P, Bachilo S M, Litovsky S H, Weisman R B 2004 Near-infrared fluorescence microscopy of single-walled carbon nanotubes in phagocytic cells *J. Am. Chem. Soc.* **126** 15638-15639

[307] Cui D, Tian F, Ozkan C S, Wang M, Gao H 2005 Effect of single wall carbon nanotubes on human HEK293 cells *Toxicol. Lett.* **155** 73-85

[308] Jia G, Wang H, Yan L, Wang X, Pei R, Yan T, Zhao Y, Guo X 2005 Cytotoxicity of carbon

nanomaterials: single-wall nanotube, multi-wall nanotube, and fullerene *Environ. Sci. Technol.* **39** 1378-1383

[309] Dai H 2002 Carbon nanotubes: opportunities and challenges *Surf. Sci.* **500** 218-241

[310] Bottini M, Bruckner S, Nika K, Bottini N, Bellucci S, Magrini A, Bergamaschi A, Mustelin T 2005 Multi-walled carbon nanotubes induce T lymphocyte apoptosis *Toxicol. Lett.* **in print** and references therein

[311] Yin H, Too H P, Chow G M 2005 The effect of particle size and surface coating on the cytotoxicity of nickel ferrite *Biomaterials* **26** 5818-5826

[312] Gupta A K, Gupta M 2005 Cytotoxicity suppression and cellular uptake enhancement of surface modified magnetic nanoparticles *Biomater.* **26** 1565-1573

[313] Sayes C M, Fortner J D, Guo W, Lyon D, Boyd A M, Ausman K D, Tao Y J, Sitharaman B, Wilson L J, Highes J B, West J L, Colvin V L 2004 The differential cytotoxicity of water-soluble fullerenes *Nano Lett.* **4** 1881-1887

[314] Braydich-Stolle L, Hussain S, Schlager J, Hofmann M C 2005 *In vitro* cytotoxicity of nanoparticles in mammalian germ-line stem cells *Toxicol. Sci.* July 13

[315] Hussain S M, Hess K L, Gearhart J M, Geiss K T, Schlager J J 2005 *In vitro* toxicity of nanoparticles in BRL 3A rat liver cells *Toxicol. In Vitro* In press

[316] www.nano.gov

[317] Roco M C 2005 The emergence and policy implications of converging new technologies integrated from nanoscale *J. Nanopart. Res.* **7** 129-143

[318] Thompson S E, Parthasarathy S 2006 Moore's law: the future of Si microelectronics *Materials Today* **9** 20-25

[319] Jacoby M 2002 Nanoscale electronics *Chem. Eng. News* **80** 38–43

[320] Buzea C, Robbie K 2005 Assembling the puzzle of superconducting elements *Supercond. Sci. Technol.* **18** R1-R8

[321] Carey J D 2003 Engineering the next generation of large-area displays: prospects and pitfalls *Philos. Transact. A Math. Phys. Eng. Sci.* **361** 2891-2907

[322] Liu H K, Wang G X, Guo Z, Wang J, Konstantinov K 2006 Nanomaterials for lithium-ion rechargeable batteries *J. Nanosci. Nanotechnol.* **6** 1-15

[323] http://www.immnet.com/articles?article=1750

[324] Hafner J H, Cheung C L, Woolley A T, Lieber C M 2001 Structural and functional imaging with carbon nanotube AFM probes *Prog. Biophys. Mol. Biol.* **77** 73–110

[325] Ellis-Behnke R G, Liang Y X, You S W, Tay D K, Zhang S, So K F, Schneider G E 2006 Nano neuro knitting: peptide nanofibers scaffold for brain repair and axon regeneration with functional return of vision *Proc. Natl. Acad. Sci. USA* **103** 5054-5059

[326] Koper O B, Klabunde J S, Marchin G L, Klabunde K J, Stoimenov P, Bohra L 2002 Nanoscale powders and formulations with biocidal activity toward spores and vegetative cells of bacillus species, viruses, and toxins *Curr. Microbiol.* **44** 49-55

[327] Jirage K B, Hulteen J C, Martin C R 1997 Nanotubule based molecular-filtration membranes *Science* **278** 655–658

[328] Martin C R, Kohli P 2003 The emerging field of nanotube biotechnology, *Nature Rev. Drug Discovery* **2** 29-37

[329] Uhrich K E, Cannizzaro S M, Langer R S, Shakeshelf K M 1999 Polymeric systems for controlled drug release *Chem. Rev.* **99** 3181-3198

[330] Maïté L, Carlesso N, Tung C H, Tang X W, Cory D, Scadden D T, Weissleder R 2000 Tat peptide-derivatized magnetic nanoparticles allow *in vivo* tracking and recovery of progenitor cells *Nature Biotechnol.* **18** 410–414

[331] Kneuer C, Sameti M, Bakowsky U, Schiestel T, Schirra H, Schmidt H, Lehr C M 2000 A nonviral DNA delivery system based on surface modified silica-nanoparticles can efficiently transfect cells *in vitro* *Bioconj. Chem.* **11** 926–932

[332] Harisinghani M G, Barentsz J, Hahn P F, Deserno W M, Tabatabaei S, van de Kaa C H, de la Rosette J, Weissleder R 2003 Noninvasive detection of clinically occult lymph-node metastases in prostate cancer *N. Engl. J. Med.* **19** 2491-2499

[333] Matsusaki M, Larsson K, Akagi T, Lindstedt M, Akashi M, Borrebaeck C A K 2005 Nanosphere induced gene expression in human dendritic cells *Nano Lett.* to appear

[334] Rosi N L, Mirkin C A 2005 Nanostructures in biodiagnostics *Chem. Rev.* **105** 1547-1562

[335] Renehan E M, Enneking F K, Varshney M, Partch R, Dennis D M, Morey T E 2005 Scavenging nanoparticles: am emerging treatment for local anesthetic toxicity, *Region. Anesth. Pain Med.* **30** 380-384

[336] Weinberg G, Ripper R, Feinstein D L, Hoffman W 2003 Lipid emulsion infusion rescues dogs from bupivacaine-induced cardiac toxicity *Region. Anesth. Pain Med.* **28** 198-202

[337] Hogan J 2004 Smog-busting paint soaks up noxious gases *New Scientist* 4 February http://www.newscientist.com/

[338] He F, Zhao D Y 2005 Preparation and characterization of a new class of starch-stabilized bimetallic nanoparticles for degradation of chlorinated hydrocarbons in water *Environ. Sci. Technol.* **39** 3314-3320

[339] http://www.nanophase.com/applications/textile_fibers.asp

[340] Hrubesh L W, Poco J F 1995 Thin aerogel films for optical, thermal, acoustic and electronic applications *J. Noncrystalline Solids* **188** 46-53

[341] Baibarac M, Gomez-Romero P 2006 Nanocomposites based on conducting polymers and carbon nanotubes: from fancy materials to functional applications *J. Nanosci. Nanotechnol.* **6** 289-302

[342] Borup B, Leuchtenberger W 2002 Soft silanes for scratch-proof surfaces *Materials World* **10** 20-21

[343] Stiglich J J, Yu C C, Sudarshan T S 1996 Synthesis of nano WC/Co for tools and dies *Proc. 1995 3 Int. Conf. Tungsten Refract. Met.* 229-236

[344] Fleischer N, Genut M, Rapoport L, Tenne R 2003 New nanotechnology solid lubricants for superior dry lubrication *European Space Agency (Special Publication) ESA SP* **524** 65-66

[345] Matsudai M, and Hunt G 2005 Nanotechnology and public health *Nippon Koshu Eisei Zasshi* **52** 923-927